%% file: main.tex
\documentclass[aps,reprint,twocolumn,preprintnumbers,floatfix,nofootinbib,prb,nobibnotes]{revtex4-2}

\usepackage{amsfonts}
\usepackage[utf8]{inputenc}
\usepackage{graphicx}
\usepackage{subfig}
\usepackage{comment}
\usepackage{cancel}
\usepackage{hyperref}
\usepackage{physics}
\usepackage{slashed}
\usepackage{enumitem}
\usepackage{appendix}
\usepackage[usenames,dvipsnames]{xcolor}
\usepackage{url}
\usepackage[normalem]{ulem}

\newcommand{\keV}{\text{keV}}
\newcommand{\MeV}{\text{MeV}}
\newcommand{\GeV}{\text{GeV}}
\newcommand{\TeV}{\text{TeV}}

\newcommand{\phiend}{\Phi_{\text{end}}}
\newcommand{\rhoend}{\rho_{\text{end}}}
\newcommand{\Trh}{T_\text{rh}}

\newcommand{\Trhb}{T_{\text{rh},b}}
\newcommand{\Trhf}{T_{\text{rh},f}}
\newcommand{\arh}{a_\text{rh}}
\newcommand{\aend}{a_\text{end}}
\newcommand{\Tfi}{T_\text{fi}}
\newcommand{\zfi}{z_\text{fi}}
\newcommand{\Tmax}{T_\text{max}}
\newcommand{\Tmaxb}{T_{\text{max},b}}
\newcommand{\Tmaxf}{T_{\text{max},f}}
\newcommand{\amax}{a_\text{max}}
\newcommand{\yeff}{y_\text{eff}}
\newcommand{\mueff}{\mu_\text{eff}}
\newcommand{\Mp}{M_\text{Pl}}
\newcommand{\gs}{g_\star}
\newcommand{\gss}{g_{\star s}}

\newcommand{\mc}[1]{\mathcal{#1}}

\newcommand{\Lc}{\mathcal{L}}
\newcommand{\Xc}{\mathcal{X}}

\newcommand{\mdm}{m_\text{DM}}
\newcommand{\ndm}{n_\text{DM}}
\newcommand{\rhodm}{\rho_\text{DM}}
\newcommand{\Ydm}{Y_\text{DM}}
\newcommand{\omdm}{\Omega_\text{DM}h^2}
\newcommand{\Xdm}{\mc{X}_\text{DM}}
\newcommand{\ra}{\rightarrow}

\hypersetup{pdfstartview=FitV,colorlinks=true,linkcolor=blue,citecolor=red,filecolor=black,urlcolor=blue,breaklinks=true}

\usepackage[parfill]{parskip}
\begin{document}\sloppy 

\preprint{MITP-23-030}

\vspace*{1mm}

\title{Confronting Dark Matter Freeze-In during Reheating with Constraints from Inflation}
\author{Mathias Becker$^{a}$}
\email{bmathias@uni-mainz.de}
\author{Emanuele Copello$^{a}$}
\email{ecopello@uni-mainz.de}
\author{Julia Harz$^{a}$}
\email{julia.harz@uni-mainz.de}
\author{Jonas Lang$^{b}$}
\email{jonas.lang@tum.de}
\author{Yong Xu$^{a}$}
\email{yonxu@uni-mainz.de}
\vspace{0.5cm}

 \affiliation{$^a$ PRISMA$^+$ Cluster of Excellence \& Mainz Institute for Theoretical Physics,\\
 Johannes Gutenberg-Universit\"{a}t Mainz, 55099 Mainz, Germany\\} 
\affiliation{$^b$ Physik Department, Technische Universit\"{a}t M\"{u}nchen, 85748 Garching, Germany}

\date{\today}

\begin{abstract} 
We investigate the production of particle Dark Matter (DM) in a minimal freeze-in model considering a non-instantaneous reheating phase after inflation. 
We demonstrate that for low reheating temperatures, bosonic or fermionic reheating from monomial potentials can lead to a different evolution in the DM production and hence to distinct predictions for the parent particle lifetime and mass, constrained by long-lived particle (LLP) searches.
We highlight that such scenario predicts
parent particle decay lengths larger compared to using the instantaneous reheating approximation. Moreover, we demonstrate the importance of an accurate definition of the reheating temperature and emphasize its relevance for the correct interpretation of experimental constraints.
We explore different models of inflation, which can lead to the considered reheating potential. We find that the extent to which the standard DM freeze-in production can be modified crucially depends on the underlying inflationary model. Based on the latest CMB constraints, we derive lower limits on the decay length of the parent particle and confront these results with the corresponding reach of LLP searches.
Our findings underscore the impact of the specific dynamics of inflation on DM freeze-in production and highlight their importance for the interpretation of collider signatures. 
At the same time, our results indicate the potential for LLP searches to shed light on the underlying dynamics of reheating.

\end{abstract}

\maketitle

\setcounter{equation}{0}

\input{0Introduction.tex}
\input{2Cosmology.tex}
\input{3DMpheno.tex}
\input{4ConstraintsInflation.tex}
\input{5Conclusions.tex}

\begin{acknowledgments}
M.~B., E.~C. and J.~H. acknowledge support from the Deutsche Forschungsgemeinschaft (DFG, German Research
Foundation) through the Emmy Noether grant HA 8555/1-1.
Moreover, M.~B., E.~C., J.~H. and Y.~X. acknowledge support by the Cluster of Excellence “Precision Physics, Fundamental Interactions, and
Structure of Matter” (PRISMA$^+$ EXC 2118/1) funded by the Deutsche Forschungsgemeinschaft (DFG, German Research
Foundation) within
the German Excellence Strategy (Project No. 390831469).
\end{acknowledgments}

\onecolumngrid
\appendix

\setcounter{equation}{0}
\input{AModels}
\setcounter{equation}{0}
\input{AppendixA.tex}

\setcounter{equation}{0}
\input{AppendixC.tex}
\setcounter{equation}{0}
\input{AppendixB.tex}

\bibliographystyle{apsrev4-1}
\twocolumngrid
\bibliography{biblio} 

\end{document}

%% file: 0Introduction.tex
\section{\label{sec:Intro}Introduction}
The lack of observation of particle candidates within and beyond the Standard Model (SM, and BSM, respectively), which could explain the remarkably precise measurement of the dark matter (DM) abundance in the Universe \cite{Planck:2018parameters}, has created significant tensions for some of the most widely studied models of thermally-produced Weakly Interacting Massive Particles (WIMPs) \cite{Arcadi:2017kky,Roszkowski:2017nbc}.
While various possibilities for accommodating scenarios of WIMP DM may still be viable, in the last decade, there has been a growing interest in alternative production mechanisms (see, e.g., Refs.~\cite{Asadi:2022njl,Boveia:2022syt,Cooley:2022ufh} for recent reviews).

In particular, in this work, we focus on the hypothesis that dark matter is a Feebly Interacting Massive Particle (FIMP) created via the Freeze-In (FI) mechanism \cite{McDonald:2001vt,Hall:2009bx,Bernal:2017kxu} through decays of a heavier parent particle $P$ that carries SM gauge charges.
This minimal setup allows for testing models of FIMP DM on earth because a gauged-charged $P$ can be directly produced and leave identifiable tracks, such as long-lived particle (LLP) signatures at collider experiments \cite{Belanger:2018sti,Calibbi:2018fqf,Junius:2019dci,No:2019gvl,Calibbi:2021fld}.

We explore the possibility that the dark matter field $\chi$ interacts only via a trilinear interaction with $P$ and a SM fermion $f_{\text{SM}}$ of the form
\begin{align}
    y_{\text{DM}}\,P\,\chi\,f_\text{SM} \, .
    \label{eq:trilinear}
\end{align}
Due to gauge invariance, the parent particle needs to carry the same gauge quantum numbers of $f_\text{SM}$.
Moreover, in order for DM to be stable, we impose a $Z_2$ symmetry under which the dark sector fields are odd and the SM fields are even. We also consider DM to be the lightest dark sector particle.

The interaction in Eq.~\eqref{eq:trilinear} can be accommodated when, for example, DM is a Majorana fermion, singlet under the SM gauge group, with $P$ being a charged scalar.
Another possibility is given by exchanging the spin nature of the dark sector particles, with DM being a real scalar singlet and the bath particle $P$ a vector-like fermion.
Recently, Refs.~\cite{Belanger:2018sti,Calibbi:2021fld} have classified and constrained models that consider this realization, providing lifetimes and masses of the parent particles that could be detected in LLP searches for certain benchmark DM masses. 
For example, in the convention of \cite{Calibbi:2021fld}, these scenarios would correspond to the $\mc{S}_{\psi_\text{SM}\chi}$ and $\mc{F}_{\psi_\text{SM}\phi}$ models, respectively.
We refer the reader to App.~\ref{sec:Model} for a brief description of these two scenarios.

However, an important feature that was highlighted in \cite{Calibbi:2021fld} and neglected in \cite{Belanger:2018sti} concerns the cosmological expansion history during which DM production takes place. In the standard FI scenario \cite{Hall:2009bx}, the production of DM is formulated by assuming a standard radiation-dominated (RD) era right after inflation; 
this epoch would begin at the reheating temperature $\Trh$ and end at the temperature of matter-radiation equality $T_\text{eq}\sim\text{eV}$. 
This \emph{instantaneous reheating} assumption was made in Ref.~\cite{Belanger:2018sti} also in scenarios when the parent particle $P$ has a mass larger than $\Trh$ and therefore the underlying dynamics of reheating are expected to be relevant. 
For collider-friendly mass scales ($m_P\lesssim~\text{TeV}$), this would typically occur in a low reheating temperature scenario. 
However, this assumption leads to sensitivity solely to the Boltzmann-suppressed part of the $P$ distribution function and neglects possible critical effects from the non-standard cosmological expansion \cite{Calibbi:2021fld}.

In particular, during the period of coherent oscillations around the minimum of the inflaton potential during reheating, radiation and entropy are continuously injected into the (SM) bath while the inflaton dominates the energy density of the Universe: this has a strong impact on the production of DM \cite{Chung:1998rq,Giudice:2000ex,Co:2015pka,Allahverdi:2020bys}.
Moreover, even production of DM stops around $T\sim m_P$, it can still be significantly diluted due to the faster cosmic expansion by several orders of magnitude until the end of the reheating epoch, which is limited only by the onset of Big Bang Nucleosynthesis (BBN) with $T_\text{rh} > T_\text{BBN} \sim \mathrm{MeV}$ \cite{Kawasaki:2000en}.
In this sense, DM production would inevitably be affected by the non-standard expansion history for $m_P\gtrsim\Trh>\MeV$, rendering it sensitive to the physics of the very early Universe, including inflation.

In this work, we examine a realization of this scenario, by considering non-instantaneous reheating driven by inflaton potentials of the type $V(\Phi)\sim \Phi^k$ around their minimum.
The $k=2$ case reduces to the typical $\Phi$ condensate behaving like pressureless matter, while for $k>2$ the equation of state is stiffer, and the decay width of the inflaton exhibits also a time dependency \cite{Garcia:2020eof, Garcia:2020wiy}.
Potentials of these types, $V(\Phi) \sim \Phi^k$ with $k \geq 2$, are well-motivated in the context of supergravity-inspired inflation, such as the $\alpha$-attractor models \cite{Kallosh:2013hoa,Kallosh:2013maa,Starobinsky:1980te}, or polynomial inflation \cite{Hodges:1989dw,Nakayama:2013jka,Drees:2021wgd,Drees:2022aea}, which are not excluded by current cosmological observations \cite{Planck:2018inflation,BICEP:2021xfz} and could be probed in the near future \cite{Abazajian:2019eic}.
Depending on which interactions allow the inflaton to reheat the Universe, the duration of the reheating phase and the evolution of the energy densities and of the temperature of the Universe can substantially differ if $k>2$ \cite{Garcia:2020wiy}.
In particular, in this study, we investigate scenarios of bosonic and fermionic reheating, where the inflaton decays dominant into pairs of bosons or fermions.
Such cosmological histories have been recently considered in the context of DM freeze-in with (non-)renormalizable operators \cite{Garcia:2020eof,Garcia:2020wiy,Bernal:2020qyu,Ahmed:2021fvt,Barman:2022tzk,Ahmed:2022qeh,Ahmed:2022tfm,Bhattiprolu:2022sdd}, WIMPs \cite{Bernal:2022wck}, in the singlet-scalar DM model~\cite{Silva-Malpartida:2023yks}, CMB constraints on DM production \cite{Maity:2018dgy,Maity:2018exj,Haque:2020zco,Haque:2023yra}, and freeze-in Baryogenesis \cite{Dalianis:2023ixz}.

In the following, we show how collider-friendly minimal FIMP models are potentially sensitive to the physics of inflation and inflationary reheating, illustrating the dependence of the interpretation of collider limits on the cosmological history and the possible interplay of LLP searches with constraints on inflationary scenarios from CMB observations.
Specifically, our analysis focuses on leptophilic FIMP models and improves upon previous work \cite{Calibbi:2021fld,Belanger:2018sti} by demonstrating that: $i)$ interpreting collider constraints requires a consistent definition of the reheating temperature; $ii)$ in low reheating temperature scenarios, the phenomenology of FI DM at colliders is sensitive to reheating potentials with $k>2$, with  bosonic reheating scenarios generally predicting shorter $P$ lifetimes than fermionic reheating ones; $iii)$ CMB constraints derived for specific $\alpha$-attractor models of inflation have the potential of already excluding certain parameter space regions for the FIMP models considered.

In essence, we show how inflation and reheating model building can dramatically affect DM production in freeze-in models, impacting their phenomenological predictions and thus the interpretation of collider signatures:
on the one side, constraints on inflation are able to restrict if and to which extent a model of FIMP DM is altered by low reheating temperature scenarios; on the other side, a positive measurement of an LLP could give us hints about the dynamics that might have occurred during reheating.

The paper is outlined as follows: section~\ref{sec:Cosmo} reviews inflationary reheating and the different expansion histories; section~\ref{sec:DMproduction} analyzes DM production from decays of a heavy parent; section~\ref{sec:Constraints} presents the combined constraints on the models from LLP searches and cosmological observations. 
Finally, we summarize our results and conclude in section~\ref{sec:Conclusions}.

%% file: 2Cosmology.tex
\section{\label{sec:Cosmo}Reheating phase}
After the end of inflation, the inflaton oscillates around the minimum of its potential and \emph{reheats} the Universe by decaying into highly energetic particles, which eventually dominate the energy budget of the Universe, once the inflaton condensate has transferred all of its energy.
We assume that this minimum can be described by a monomial potential of the form
\begin{align}
    V(\Phi)=\lambda\dfrac{|\Phi|^k}{M^{k-4}}\,,
    \label{eq:PhiPotential}
\end{align}
with $k\geq2$ and where the dimensionless coupling $\lambda$ and the energy scale $M$ are fixed by the CMB normalization.
In what follows, without loss of generality, we will assume this scale to be the reduced Planck mass $\Mp=2.4~\times~10^{18}~\GeV$.

Potentials that exhibit this type of behavior around their minimum are ubiquitous in various inflationary models, such as the $\alpha$-attractor $T$-models and $E$-models \cite{Kallosh:2013hoa,Kallosh:2013maa} (including Starobinsky inflation \cite{Starobinsky:1980te}) or polynomial inflation \cite{Hodges:1989dw,Nakayama:2013jka,Drees:2021wgd,Drees:2022aea}, which are currently favoured by recent constraining observations \cite{Planck:2018inflation,BICEP:2021xfz,Kallosh:2021mnu}.
The exact form of the potential is irrelevant to the discussion at hand, which pertains to how the production of dark matter (DM) is affected during the reheating phase. Only Eq.~\eqref{eq:PhiPotential} is sufficient for our present purposes.
However, we will later make use of constraints on these models to demonstrate the interplay between cosmological and collider probes with respect to both reheating and the parameters of the DM model (cf. Sec.~\ref{sec:Constraints} and App.~\ref{sec:AppA}).

The equation of motion for the (homogeneous) inflaton field in the expanding Universe is \cite{Kolb:1990vq}
\begin{align}
    \Ddot{\Phi}+\left(3H + \Gamma_\Phi \right)\dot{\Phi}+V'(\Phi)=0\,,
    \label{eq:PhiEoM}
\end{align}
where dots indicate a derivative with respect to cosmic time, $V'(\Phi)=\partial_\Phi V(\Phi)$, $H=\dot{a}/a$ is the Hubble parameter, and $\Gamma_\Phi$ the decay width of the inflaton.
We could rewrite Eq.~\eqref{eq:PhiEoM} in terms of the inflaton energy density $\rho_\Phi=\frac{1}{2}\dot{\Phi}-V(\Phi)$ and isotropic pressure $P_\Phi=\frac{1}{2}\dot{\Phi}+V(\Phi)$, together with their equation of state $w_\Phi=P_\Phi/\rho_\phi$.
However, due to the oscillatory behavior, it is more useful to average these quantities over several periods of oscillation, such that, for the potential in Eq.~\eqref{eq:PhiPotential}, one obtains \cite{Garcia:2020eof,Garcia:2020wiy}
\begin{align}
    \expval{\rho_\Phi}&=\left(\frac{k}{2}+1\right)\lambda\frac{\expval{\Phi^k}}{\Mp^{k-4}}\,,\label{eq:avRho}\\
    \expval{P_\Phi}&=\left(\frac{k}{2}-1\right)\lambda\frac{\expval{\Phi^k}}{\Mp^{k-4}}\,,\label{eq:avPressure}\\
    \expval{w_\Phi}&=\frac{\expval{\rho_\Phi}}{\expval{P_\phi}}=\frac{k-2}{k+2}\,.
    \label{eq:avEoS}
\end{align}
Note that the $k=2$ case corresponds to the standard inflaton oscillating with a quadratic potential and behaving like pressureless matter, while for $k=4$, its equation of state describes a radiation-like condensate.

By combining Eqs.~\eqref{eq:avRho}-\eqref{eq:avEoS} with Eq.~\eqref{eq:PhiEoM} we obtain the equation for the time evolution of the inflaton energy density
\begin{equation}
    \dot{\rho}_\Phi+\frac{6k}{k+2}H\,\rho_\Phi=-\frac{2k}{k+2}\Gamma_\Phi\,\rho_\Phi\,.
    \label{eq:BoltzFriedPhi}
\end{equation}
The decay of the inflaton produces highly energetic particles behaving like a relativistic fluid, whose energy density $\rho_\text{R}$ evolves according to
\begin{equation}
    \dot{\rho}_R+4H\rho_R=\frac{2k}{k+2}\Gamma_\Phi\,\rho_\Phi\,.
    \label{eq:BoltzFriedRad}
\end{equation}
The system of equations for $\rho_\Phi(t)$, $\rho_R(t)$ and $a(t)$ is closed by the first Friedmann equation
\begin{equation}
    H^2=\frac{\rho_\Phi+\rho_R}{3 \Mp^2}\,,
    \label{eq:1stFriedmann}
\end{equation}
so that one is able to obtain a solution for the energy densities as functions of the scale factor $a$.
The evolution of energy densities in this work has been obtained by numerically solving the Eqs.~\eqref{eq:BoltzFriedPhi}-\eqref{eq:1stFriedmann}, including the effects of the inflaton decay width on its energy density at all times.

An analytic solution can be derived by firstly solving Eq.~\eqref{eq:BoltzFriedPhi} for $\rho_\Phi$ by assuming a negligible $\Gamma_\Phi$ when $\aend\ll a \ll a_\text{rh}$, where $\aend$ and $a_\text{rh}$ are the scale factors at the beginning and at the end of the reheating phase, respectively.
This is a well-motivated approximation since the inflaton decay width is negligible compared to the Hubble rate at the beginning of the reheating phase.
We thus obtain
\begin{align}
    \rho_\Phi(a)\simeq \rho_\Phi(a_\text{rh})\left(\frac{a_\text{rh}}{a}\right)^{\frac{6k}{k+2}}\quad\text{if}\quad \aend\ll a \ll a_\text{rh}\,.
    \label{eq:Rho_Phi}
\end{align}
In particular, $\rho_\Phi\propto a^{-3}$ for a quadratic reheating potential; and $\rho_\Phi\propto a^{-4}$ for a quartic. 
At later times, $a>a_\text{rh}$, the inflaton will decay away with an exponential factor, assuming that $\Gamma_\Phi$ is constant in time. 
This is the case for $k=2$, where the density of the inflaton, $\rho_\Phi(t)$, scales as $e^{-\Gamma_\phi t}$.
In contrast, for $k>2$, the decay width will acquire a dependence on time (cf. Eqs.~\eqref{eq:GammaBoson} and \eqref{eq:GammaFermion}) and the scaling of $\rho_\Phi(t)$ will be model-dependent in general.

By inserting this solution into Eq.~\eqref{eq:BoltzFriedRad}, we can solve for $\rho_R$ as \cite{Bernal:2022wck}
\begin{align}
    \rho_R(a)\simeq\frac{2\sqrt{3}\,k}{k+2}\frac{\Mp}{a^4}\int_{\aend}^a \dd a'\, \Gamma_\Phi(a')\,\rho^{1/2}_\Phi(a')\,(a')^3\,.
    \label{eq:RhoRad}
\end{align}
The temperature of the radiation thermal bath is then determined by
\begin{align}
    T(a)=\left(\dfrac{30\,\rho_R(a)}{\pi^2 \gs}\right)^{1/4}\,,
    \label{eq:Temp(a)}
\end{align}
where $\gs\sim\order{100}$ denotes the effective number of relativistic degrees of freedom that we assume to be constant throughout the reheating process, which is a good assumption for the considered temperatures of $T> 1~\text{GeV}$.
We define the scale factor $\arh$ and the reheating temperature $\Trh=T(\arh)$ as the point in time when the energy density of radiation becomes equal to that of the inflaton,
\begin{align}
    \rho_\Phi(\arh)=\rho_R(\arh)\,,
    \label{eq:T_RH_Implicit}
\end{align}
and starts to dominate. 
We should note that this definition of the reheating temperature is not equivalent to the definition used in the instantaneous reheating approximation, which is based on the relations $\Gamma_\Phi=H$ or $\Gamma_\Phi=t^{-1}=\frac{3}{2}H$ in the context of matter-dominated (MD) reheating \cite{Kolb:1990vq}.  Note that $\Trh$ marks the onset of the radiation phase, and in this sense, the definition in Eq.~\eqref{eq:T_RH_Implicit} is more accurate.

In section \ref{sec:Constraints}, we show that the definition of the reheating temperature, as well as the properties of the inflaton decay products, can have a substantial impact on the interpretation of experimental constraints. In the following, we consider two inflaton decay scenarios: inflaton decays to bosonic species and inflaton decays to fermionic species.

\subsection{Bosonic reheating}

Inflationary reheating could proceed via inflaton decays into pairs of scalars via a trilinear coupling of the type $\mu\Phi|X|^2$.
This eventuality can be naturally realized through the interaction $\mu_H\Phi |H|^2$, such that the inflaton can reheat the Universe by decaying into Higgs particles.
Additionally, other bosonic decay channels can be considered if more scalars are added to the SM. 
In this work, for example, we examine FIMP models with Majorana DM produced from the decays of heavier charged parent particle scalars $P$, as we will detail in Sec.~\ref{sec:DMproduction} and in App.~\ref{sec:Model}.
Thus, the inflaton could decay into pairs of parent particles $P$ via $\mu_P \Phi|P|^2$.
In any case, we can assume that the produced scalars thermalize fast enough via their gauge interactions so that we can treat them as being part of the radiation bath described by Eq.~\eqref{eq:BoltzFriedRad}.

We consider the bare masses of the decay products to be negligible compared to the inflaton mass.
Hence, we can write the inflaton width as follows \cite{Ichikawa:2008ne, Garcia:2020wiy,Bernal:2022wck}
\begin{align}
    &\Gamma_{\Phi\ra X X^{\dagger}}(t)=\dfrac{\mu_\text{eff}^2(k)}{8\pi m_\Phi (t)}\,,
    \label{eq:GammaBoson}
\end{align}
where
\begin{align}
    \mu_\text{eff}^2(k)=\mu^2\,\alpha_\mu(k,\mc{R})\,\frac{(k+2)(k-1)}{4}\sqrt{\dfrac{\pi\,k}{2(k-1)}}\dfrac{\Gamma(\frac{k+2}{2k})}{\Gamma(\frac{1}{k})}
    \label{eq:mu_eff}
\end{align}
is an effective coupling calculated from a (proper) average over inflaton oscillations.
Here, $\mu$ can be either $\mu_H$ or $\mu_P$; as long as the scalar daughter particles are part of the radiation bath, the details of the micro-physics involved are irrelevant to our discussion.
The function $\alpha_\mu(k,\mc{R})$ depends on a kinematic factor $\mc{R}$ which in essence quantifies if the inflaton-induced $P$ mass $m_\text{eff}^2(t)=2\mu\Phi$ is comparable to the inflaton mass $m_\Phi$ ($\mc{R}\gtrsim1$) or not ($\mc{R}<1$).
In our setup, we are mainly interested in low-reheating temperature scenarios, achieved by letting the inflaton slowly decay, such that $\mu$ is assumed to be very small (for instance relative to $M_P$) and thus $\mc{R}\ll 1$.
In this situation, $\alpha_\mu$ varies very slowly with increasing $k$, being $\alpha_\mu\simeq 1$ for $k=2$ and $\alpha_\mu\simeq0.9$ for $k=4$ \cite{Garcia:2020wiy}.

The mass of the inflaton can be derived from its potential (cf. Eq.~\eqref{eq:PhiPotential}), yielding
\begin{align}
    m_\Phi^2(t)&=V''(\Phi)=k (k-1) \lambda \dfrac{\Phi^{k-2}(t)}{\Mp^{k-4}}\nonumber\\
    &\simeq k(k-1)\lambda^{\frac{2}{k}}\Mp^{\frac{2(4-k)}{k}}\rho_\Phi^{\frac{k-2}{k}}(t)\,,
    \label{eq:mPhi}
\end{align}
where we used the ``envelope'' approximation for $\rho_\Phi$, defined by $\rho_\Phi(t)=V(\Phi_0(t))$, where $\Phi_0(t)$ encodes the effects of redshift and decay of the condensate $\Phi(t)~\simeq~\Phi_0(t)P(t)$ at large scales, while $P(t)$ describes its anharmonicity on short scales \cite{Garcia:2020eof,Garcia:2020wiy}. 
Notice that, in the standard $k=2$ case, $m_\Phi=\sqrt{2\lambda}\,\Mp$ and is constant in time, while for $k>2$ it decreases with $\rho_\Phi$ during reheating.
This has two consequences: firstly, the bosonic decay width is kinematically enhanced at later times, leading to a faster reheating phase; secondly, the inflaton mass could potentially become comparable to or even smaller than the product particle mass, breaking the perturbative reheating assumption and forcing to consider additional effects, such as back-reactions.
This scenario is avoided as long as $\mc{R}\ll1$, which needs to hold for the whole reheating phase.
However, it has been shown in Ref.~\cite{Garcia:2020wiy} for $k=4$ in an $\alpha$-attractor $T$-model, that $\mc{R} \gtrsim 1$ whenever $10^{-19}\lesssim\mu/\Mp\lesssim 10^{-9}$.
Nevertheless, since we are interested in low reheating temperature scenarios (i.e., $\Trh<10\,\TeV$), we checked that typical values for $\mu/\Mp$ lie below $10^{-19}$ (cf. Eq.~\eqref{eq:TRH_boson}) so that an eventual kinematic enhancement of the decay rate or non-perturbative particle production are expected to possibly play a role only in the latest stages of the reheating phase.
Moreover, Ref.~\cite{Garcia:2020wiy} found also non-perturbative effects for temperatures as large as $T_\text{max}$ in the case of reheating mediated by inflaton scatterings. This case, however, is not discussed in this article and would also not be relevant in the considered hierarchy $T_\text{max} \gg m_P$.
An analysis that would address how freeze-in production in minimal FIMP models would be affected by these effects lies beyond the scope of this work and is left for future improvements.
\begin{figure*}[t!]
    \centering
    \includegraphics[scale=0.4]{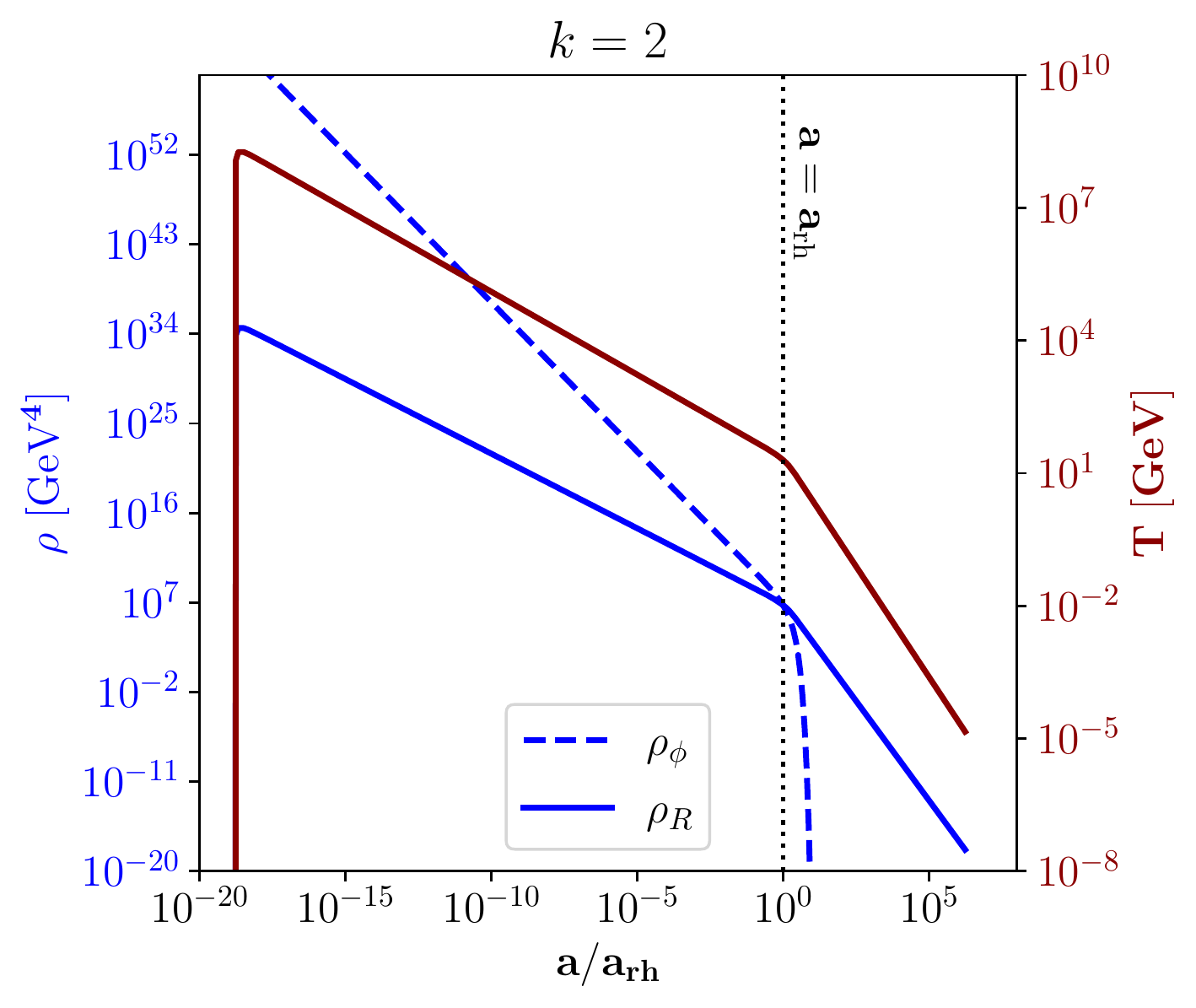}
    \includegraphics[scale=0.4]{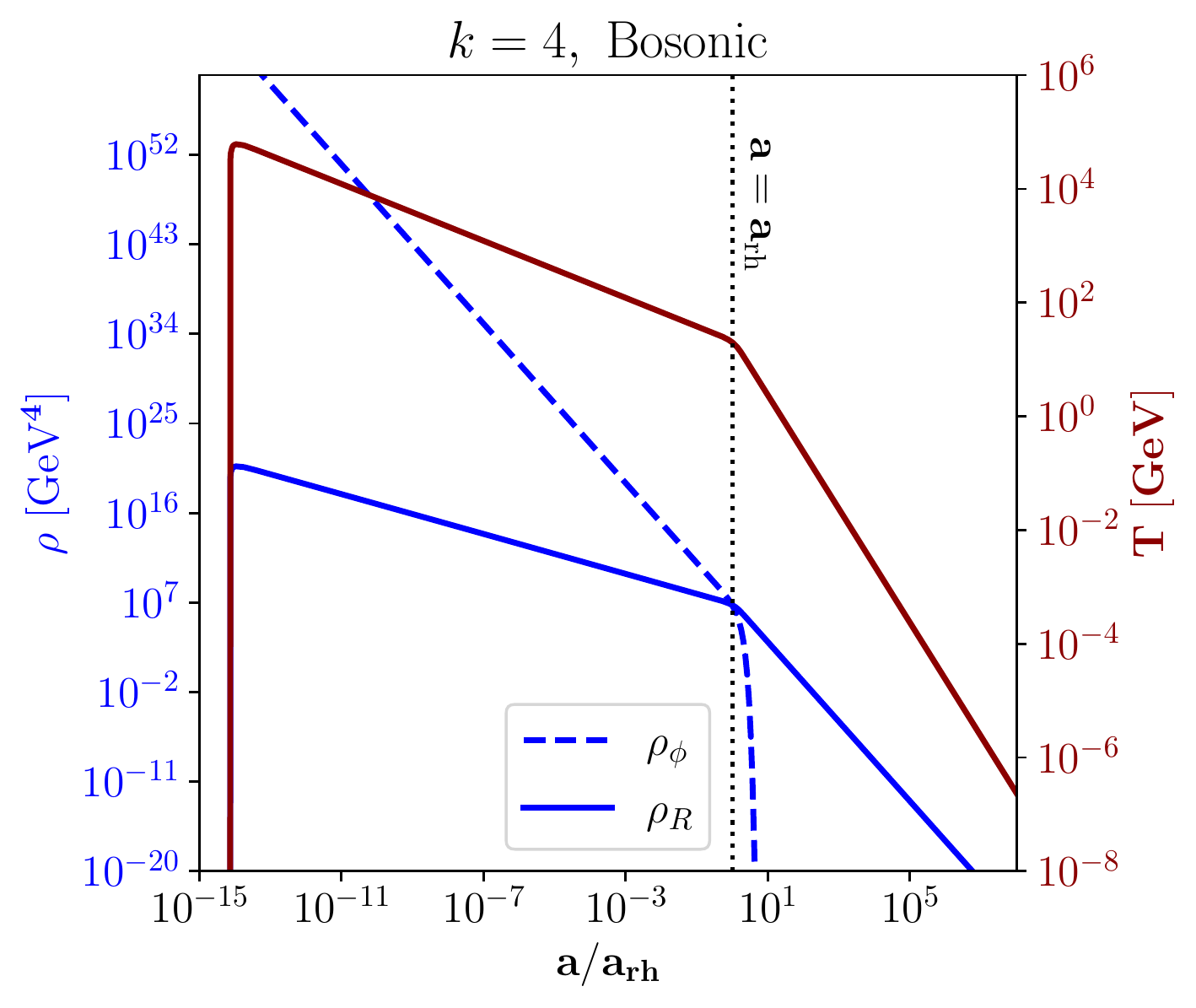}
    \includegraphics[scale=0.4]{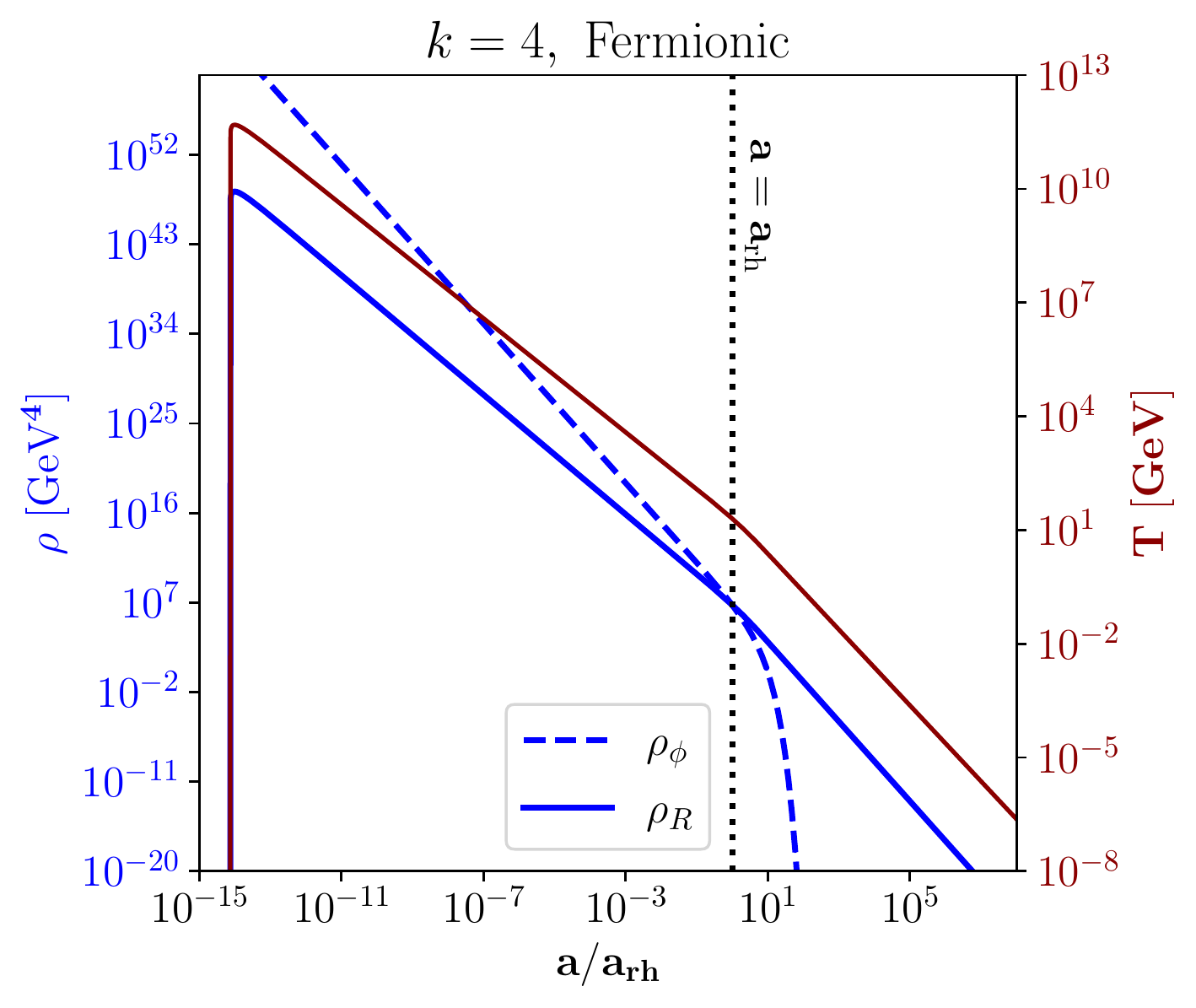}
    \caption{Evolution of the energy densities of the inflaton (blue dashed line) and of radiation (blue solid line), together with the radiation temperature (magenta solid), for $\Trh = 20~\text{GeV}$. The panels correspond to reheating potentials with $k=2$ (left), bosonic $k=4$ (middle), and fermionic $k=4$ (right).}
    \label{fig:rho}
\end{figure*}

With these assumptions, by inserting Eq.~\eqref{eq:GammaBoson} into Eq.~\eqref{eq:RhoRad}, one obtains
\begin{align}
    &\rho_R(a)\simeq \frac{\sqrt{3}}{8\pi}\frac{1}{1+2k}\sqrt{\frac{k}{k-1}}\frac{\mu_\text{eff}^2}{\lambda^{\frac{1}{k}}}M_P^{\frac{2k-4}{k}}\nonumber\\
    &\qquad\qquad\times\rho_\Phi^{\frac{1}{k}}(\arh)\left(\dfrac{\arh}{a}\right)^{\frac{6}{2+k}}\left[1-\left(\dfrac{\aend}{a}\right)^{\frac{2(1+2k)}{2+k}}\right]\,.
    \label{eq:RhoRadBoson}
\end{align}
This equation shows that the radiation energy density and the bath temperature are zero at the end of inflation $\rho_R(\aend)=0$ and $T(\aend)=0$, but that they quickly reach a maximum at $\amax\simeq \aend (\frac{2k+4}{3})^{\frac{k+2}{4k+2}}$.

Moreover, during reheating ($a\gg \aend$), one finds that the radiation energy density redshifts as $\rho_R(a)\propto a^{-\frac{6}{2+k}}$,
resulting in
\begin{align}
    T(a)\simeq \Trh\,\left(\dfrac{\arh}{a}\right)^{\frac{3}{4+2k}}\,.
    \label{eq:Temp_BR}
\end{align}
This implies that, in particular, for $k=2$ (matter-like reheating) $\rho_R\propto a^{-3/2}$ and $T\propto a^{-3/8}$, while for $k=4$ (radiation-like reheating) we have $\rho_R\propto a^{-1}$ and $T\propto a^{-1/4}$.
In Tab.~\ref{tab:reheating}, we provide a summary of the different scalings relevant to this work.
Notice that, these behaviors differ from the redshifting during an adiabatic cosmic expansion when entropy is conserved, where $\rho_R\propto~a^{-4}$ and $T\propto~a^{-1}$.

With this solution at hand, combining Eq.~\eqref{eq:Temp_BR} with Eq.~\eqref{eq:Rho_Phi} and Eq.~\eqref{eq:1stFriedmann} yields the Hubble rate during BR when $\Phi$ dominates,
\begin{align}
    H(T)\simeq\dfrac{\sqrt{\rho_\Phi(\Trh)}}{\sqrt{3}\Mp}\left(\frac{T}{\Trh}\right)^{2k}\,.
    \label{eq:HubbleBR}
\end{align}
The maximum and the reheating temperatures can be obtained from Eq.~\eqref{eq:Temp(a)} by employing the corresponding $\amax$ and by imposing $\rho_\Phi=\rho_{\rm R}$ at $\arh$, respectively.
Thus, we find
\begin{align}
    \Tmaxb^4\mkern-1mu=\mkern-1mu\frac{15}{4\pi^3\gs}\frac{k}{\sqrt{3k(k-1)}}\frac{M_P^{\frac{2k-4}{k}}}{\lambda^{\frac{1}{k}}}\mueff^2\rho_\text{end}^{\frac{1}{k}}\mkern-2mu\left(\mkern-1mu\dfrac{3}{2k+4}\mkern-1mu\right)^{\mkern-5mu\frac{2k+4}{2k+1}}
    \label{eq:Tmax_boson}
\end{align}
and
\begin{align}
    \Trhb^4=\dfrac{30}{\pi^2\gs}\left[\frac{\sqrt{3}}{8\pi\,(1+2k)}\sqrt{\frac{k}{k-1}}\lambda^{-\frac{1}{k}}\frac{\mu_\text{eff}^2}{\Mp^2}\right]^{\frac{k}{k-1}}\Mp^4\,.
    \label{eq:TRH_boson}
\end{align}
The evolution of $\rho_\Phi(a)$, $\rho_R(a)$, and $T(a)$ in BR obtained by numerically solving the Eqs.~\eqref{eq:BoltzFriedPhi}-\eqref{eq:1stFriedmann}, as explained in the previous section, are shown in the left panel ($k=2$) and the central panel ($k=4$) of Fig.~\ref{fig:rho}. 
\subsection{Fermionic reheating}
The inflaton could alternatively reheat the Universe by decaying into pair of fermions via a trilinear coupling of the form $y\Phi\bar{F}F$, in what is called ``fermionic reheating'' (FR).
This possibility is slightly more involved, as scalar inflaton fields cannot directly decay into SM fermion pairs, due to their chiral nature.
Therefore, one necessarily needs additional Majorana or vectorlike fermion species.
In this work, for instance, we study FIMP models that could be realized with a scalar singlet DM candidate produced by decays of a heavier vectorlike parent charged under the SM gauge group (cf. Sec.~\ref{sec:DMproduction} and App.~\ref{sec:Model}).
These could play the role of inflaton decay products and, once quickly thermalized via gauge interactions, be treated as part of the radiation bath.
Since we do not consider direct inflaton couplings to DM particles, models with only Majorana DM and charged scalar parents could not accommodate this scenario, unless an additional vectorlike fermion is added.
\setlength{\tabcolsep}{10pt} 
\renewcommand{\arraystretch}{1.5} 
\begin{table}[!t]
    \centering
    \begin{tabular}{l c c}
    \hline
      Type   &  $k=2$ &  $k=4$  \\
    \hline
      BR  &  \begin{tabular}{@{}l@{}}
              $\rho_R\propto a^{-3/2}$\\
              $T\propto a^{-3/8}$
             \end{tabular}  & 
             \begin{tabular}{@{}l@{}}
             $\rho_R\propto a^{-1}$\\
             $T\propto a^{-1/4}$
             \end{tabular}\\
     \hline
      FR  &  \begin{tabular}{@{}l@{}}
              $\rho_R\propto a^{-3/2}$\\
              $T\propto a^{-3/8}$
             \end{tabular}  & 
             \begin{tabular}{@{}l@{}}
             $\rho_R\propto a^{-3}$\\
             $T\propto a^{-3/4}$
             \end{tabular}\\
    \hline
    \end{tabular}
    \caption{Redshifting of the energy density and the temperature of the radiation bath during BR and FR for $\Phi^k$ reheating potentials with $k=2$ and $k=4$.}
    \label{tab:reheating}
\end{table}
The decay width for final state fermions with negligible masses is
\begin{align}
    \Gamma_{\Phi\ra F \Bar{F}}(t)=\dfrac{y^2_\text{eff}(k)}{8\pi}m_\Phi(t)\,,
    \label{eq:GammaFermion}
\end{align}
where
\begin{align}
    y_\text{eff}^2(k)=y^2\alpha_y(k,\mc{R})\sqrt{\frac{\pi k}{2(k-1)}}\frac{\Gamma(\frac{k+2}{2k})}{\Gamma(\frac{1}{k})}
    \label{eq:y_eff}
\end{align}
is the effective coupling from averaging the inflaton oscillations with the inflaton mass given by Eq.~\eqref{eq:mPhi}.
For the function $\alpha_y(k,\mc{R})$, a similar argument as for $\alpha_\mu(k,\mc{R})$ can be applied: the inflaton-induced fermion mass $y^2\Phi^2$ is negligible ($\mc{R}\ll1$) if $y$ is tiny enough, $y < \order{10^{-6}}$, so that $\alpha_y$ is approximately constant in $\mc{R}$, while it only varies between $\alpha_y=1$ at $k=2$ and $\alpha_y\simeq1.05$ at $k=4$ \cite{Garcia:2020wiy}.~\footnote{Notice that, differently from the bosonic case, the broad resonance regime ($\mc{R}\gg 1$) would be less impactful as effects from non-perturbative backreaction would be damped by Pauli blocking \cite{Greene:1998nh,Garcia:2021iag}. 
Nonetheless, the inflaton decay rate would also be reduced by a factor $\mc{R}^{-1/2}$, leading to a more prolonged reheating phase. These effects are however not of relevance as long as $y<10^{-5}$ \cite{Drewes:2017fmn}, a necessary condition in order to be in a low reheating temperature scenario.}

Inserting Eq.~\eqref{eq:GammaFermion} into Eq.~\eqref{eq:RhoRad} yields
\begin{align}
    &\rho_R(a)\simeq\dfrac{\sqrt{3}}{8\pi}\dfrac{k\sqrt{k(k-1)}}{7-k}y_\text{eff}^2\lambda^{\frac{1}{k}}\Mp^{\frac{4}{k}}\nonumber\\
    &\qquad\qquad\times \rho_\Phi^{\frac{k-1}{k}}(\arh)\left(\frac{\arh}{a}\right)^{\frac{6(k-1)}{2+k}}\left[1-\left(\frac{\aend}{a}\right)^{\frac{2(7-k)}{2+k}}\right]\,,
    \label{eq:RhoRadFermion}
\end{align}
with $\rho_R(\aend)=0$, but with the maximum attained at $\amax\simeq \aend (\frac{2k+4}{3k-3})^{\frac{k+2}{14-2k}}$, differently from the bosonic scenario.
For the values of $k$ of interest to us ($k<7$), during the reheating phase, one finds $\rho_R(a)~\propto~a^{-\frac{6(k-1)}{2+k}}$, with the SM temperature evolving as
\begin{align}
    T(a)\simeq \Trh \left(\frac{\arh}{a}\right)^{\frac{3k-3}{4+2k}}\,.
    \label{eq:Temp_FR}
\end{align}
As a consequence, for quadratic potentials ($k=2$), $\rho_R\propto a^{-3/2}$ and $T\propto a^{-3/8}$, likewise during BR, but, for quartic potentials ($k=4$) radiation redshifts as $\rho_R\propto a^{-3}$ and hence the temperature-scale factor relation is $T\propto a^{-3/4}$, which entails a faster cooling of the radiation bath with respect to BR. In Tab.~\ref{tab:reheating}, we provide a summary of the different behaviors relevant to this work.
Combining Eq.~\eqref{eq:Temp_FR} with Eq.~\eqref{eq:Rho_Phi} and Eq.~\eqref{eq:1stFriedmann} yields the Hubble rate during FR when $\Phi$ dominates,
\begin{align}
    H(T)\simeq\dfrac{\sqrt{\rho_\Phi(\Trh)}}{\sqrt{3}\Mp}\left(\frac{T}{\Trh}\right)^{\frac{2k}{k-1}}\,.
    \label{eq:HubbleFR}
\end{align}
The maximum and reheating temperature are in this case
\begin{align}
    \Tmaxf^4=\frac{15}{4\pi^3\gs}\frac{k^2}{\sqrt{3k(k-1)}}\,\lambda^{\frac{1}{k}}\Mp^{\frac{4}{k}}\yeff^2\,\rho_\text{end}^{\frac{k-1}{k}}\left(\frac{3k-3}{2k+4}\right)^{\frac{2k+4}{7-k}}
    \label{eq:Tmax_fermion}
\end{align}
and
\begin{align}
    \Trhf^4=\frac{30}{\pi^2\gs}\left[\frac{k\sqrt{3k(k-1)}}{7-k}\lambda^{\frac{1}{k}}\frac{y_\text{eff}^2}{8\pi}\right]^k \Mp^4\,.
    \label{eq:TRH_fermion}
\end{align}
The evolution of $\rho_\Phi(a)$, $\rho_R(a)$, and $T(a)$ in FR is shown in the left panel ($k=2$) and the right panel ($k=4$) of Fig.~\ref{fig:rho}. 
For $k=2$, the scaling behavior of $\rho_R(a)$ in the bosonic scenario exhibits identical behavior to that of the fermionic reheating, as depicted by the blue solid line in the left panel of Fig.~\ref{fig:rho}. 
However, as the parameter $k$ increases, the disparity between the two scenarios becomes more pronounced. Specifically, in the bosonic scenario, $\rho_R$ is suppressed compared to the fermionic case, owing to the characteristic time dependence of the decay rate. 
It is important to note that in bosonic reheating, the decay rate is proportional to $1/m_\Phi(t)$ (cf. Eq.~\eqref{eq:GammaBoson}) while, in the fermionic case, it is proportional to $m_\Phi(t)$ (cf. Eq.~\eqref{eq:GammaFermion}). 
This difference in the decay rate behavior elucidates the reason behind the less steep slope observed in the solid blue line of the bosonic case compared to that of the fermionic scenario, particularly when $k>2$.

%% file: 3DMpheno.tex
\section{\label{sec:DMproduction}Dark matter production}
To follow the evolution of the DM number density $n_\text{DM}$, we need to solve the Boltzmann equation
\begin{align}
    \dot{n}_\text{DM}+3Hn_\text{DM}=\mc{C}_\text{DM}\,,
    \label{eq:BEQ_ndm}
\end{align}
where the l.h.s describes the redshifting of the number of $\chi$ particles per comoving volume, whereas, in the r.h.s, the collision operator $\mc{C}_\text{DM}$ characterizes how much the reactions in which DM is involved contribute in its creation or depletion.

In this work, we mainly focus on DM freeze-in production from the decays\footnote{Notice that, for very light $\chi$, freeze-in contributes to the production of dark radiation. This scenario has been discussed for instance in \cite{Bernreuther:2022bdw}.} $P\rightarrow \chi\,f_\text{SM}$ of the parent particles $P$, assumed to be in thermal equilibrium with the SM bath.
Note that this assumption does not hold at times immediately after inflation where the particle bath is populated by non-thermalized decay products of the inflaton, a phase called ``pre-thermalization''. 
In App.~\ref{sec:AppB}, we estimate the maximal effects of pre-thermalization on the DM relic abundance and find them to be negligible if $m_P/\Trh \lesssim 10^3$. Thus, we neglect pre-thermalization effects and assume the parent particles are in equilibrium with the SM at all times\footnote{Note that, finite-temperature effects can substantially correct the relic abundance from freeze-in for small mass splittings between DM and the heavier parent. In our case, $m_P\gg \mdm$, such that the corrections are expected to be smaller than $\mathcal{O}(10\%)$ \cite{Biondini:2020ric}.}.
We also notice that the gravitational production is Planck-suppressed (cf. Refs.~\cite{Bernal:2018qlk,Mambrini:2021zpp}) and therefore irrelevant to our study. 
With these assumptions at hand, the collision operator assumes the following form \cite{Hall:2009bx}
\begin{align}
    \mc{C}_\text{DM}=\dfrac{g_P}{2\pi^2}\Gamma_Pm^2_P\,T K_1\left(\frac{m_P}{T}\right)\,,
    \label{eq:CollOp_DM}
\end{align}
where $g_P$ counts the internal degrees of freedom of $P$, $\Gamma_P$ is the parent decay width into DM, and $K_1$ is a modified Bessel function of the second kind.
Notice that, for vectorlike fermion parents (real scalar DM), the rate of production is twice as large as for charged scalar parents (Majorana DM). 
We will not specify the particular form of $\Gamma_P$, as our goal is to constrain $\Gamma_P$ directly from LLP searches such that we can actually relate it to the parent particle's decay length $c\tau=\Gamma_P^{-1}$.

For operators of this form, DM starts being created from an initial zero abundance and grows until it freezes at the temperature $\Tfi\simeq m_P/\zfi$, where $\zfi\simeq 4-10$, typically.
Therefore, the bulk of the production lies around the parent particle mass and we must differentiate between the scenarios when $\Tfi<\Trh$ and $\Tfi>\Trh$.
In the first case, DM production peaks well within the RD phase, such that one can assume the instantaneous reheating approximation ($\Trh=\Tmax$) and solve Eq.~\eqref{eq:BEQ_ndm} from $\Trh$ to the present day.
It is useful in this case to trade $n_\text{DM}$ for the \emph{comoving yield} $Y_\text{DM}=n_\text{DM}/s$ exploiting the conservation of entropy during RD.
As a result, using Eq.~\eqref{eq:CollOp_DM} with $H(T)~\sim~T^2/\Mp$, one can find the estimate \cite{Hall:2009bx}
\begin{align}
    &Y_{\text{DM}}^\text{RD}(T)\sim \frac{\Gamma(T)}{H(T)} \sim\Gamma_P \dfrac{m_P\,\Mp}{T^3}\\
    \implies &Y_\text{DM}^\text{RD}(\zfi)\sim\zfi^3\frac{\Gamma_P\,\Mp}{m_P^2}\,,
    \label{eq:Yield_prod}
\end{align}
for the produced DM yield.

However, for $m_P>\Trh$, the different expansion rates, together with the production of entropy during inflaton-dominated reheating, significantly modify the parametric dependence of the DM yield when it grows between $\Tmax$ and $\Tfi$.
This implies that the yield is lowered between $\Tfi$ and $\Trh$ by a dilution factor \cite{Co:2015pka}
\begin{align}
    D(T)&\simeq\frac{S(T)}{S(\Trh)}=\frac{s(T)a(T)^3}{s(\Trh)a(\Trh)^3}\nonumber\\
    &\simeq\begin{cases}
        \left(\dfrac{\Trh}{T}\right)^{1+2k}\,\text{BR}\\
        \left(\dfrac{\Trh}{T}\right)^{\frac{7-k}{k-1}}\;\text{FR}
    \end{cases}\,.
    \label{eq:D(T)}
\end{align}
Here, $S(T)=s(T)a(T)^3$ and $s(T)=\frac{2\pi}{45} \gss T^3$ are the total entropy and entropy density of the SM bath, respectively, with $\gss$ being the number of effective relativistic degrees of freedom for the entropy.
The dilution factor depends on the temperature~--~scale factor relation, which does not follow the usual $a\sim T^{-1}$ law of adiabatic expansion, but rather Eq.~\eqref{eq:Temp_BR} within bosonic reheating (BR), or Eq.~\eqref{eq:Temp_FR} within fermionic reheating (FR).  
By also taking into account the different Hubble rates as in Eqs.~\eqref{eq:HubbleBR} and~\eqref{eq:HubbleFR}, the yield during reheating reads now as  
\begin{align}
    &Y^\text{RH}_\text{DM}(T)\mkern-1mu\sim\mkern-1mu\frac{\Gamma \left( T \right)}{H \left( T \right)}D(T)\mkern-1mu\sim\mkern-1mu\frac{\Gamma_P\,m_P\Mp}{T^3}\mkern-2mu\times\mkern-2mu
    \begin{cases} 
        \left(\dfrac{\Trh}{T}\right)^{\mkern-5mu 4k-1}\mkern-5mu\text{BR}\\
        \left(\dfrac{\Trh}{T}\right)^{\mkern-5mu\frac{9-k}{k-1}}\text{FR}
    \end{cases}\mkern-22mu,    \label{eq:Ydm_T_dil}\\
    &Y_\text{DM}^\text{RH}\left(\frac{m_P}{\zfi}\right)\sim\zfi^3\frac{\Gamma_P\,\Mp}{m_P^2}\times
    \begin{cases} 
        \left(\dfrac{\zfi\Trh}{m_P}\right)^{4k-1}\,&\text{BR}\\
        \left(\dfrac{\zfi\Trh}{m_P}\right)^{\frac{9-k}{k-1}}\;&\text{FR}
    \end{cases}.
    \label{eq:Ydm_fi_dil}
\end{align}
We can see that, for low reheating temperature scenarios, the DM evolution highly depends on the expansion history during reheating and, in particular, on the type of reheating scenario considered (BR vs. FR).
As we can see, the value of $\Tfi$ plays an important role in estimating the dilution factor in Eq.~\eqref{eq:D(T)} and, as we will realize soon, typical values for $\zfi$ can vary between 4 and 10, depending on the cosmological history.
Given the magnitude of the exponents in Eq.~\eqref{eq:Ydm_fi_dil}, this can result in differences of several orders of magnitude.

In order to be able to solve Eq.~\eqref{eq:BEQ_ndm} for every choice of $m_P$, we need to describe the evolution of the energy and number densities in a consistent way throughout the entire expansion history, including reheating and RD.
Given the initial non-adiabatic expansion, it is better to utilize the following, more suited comoving number density \cite{Co:2015pka} 
\begin{align}
    \mathcal{X}(a)\equiv n_\text{DM}(a)\,\left(\frac{a}{\aend}\right)^3\,,
    \label{eq:Chi_yield}
\end{align}
such that Eq.~\eqref{eq:BEQ_ndm} turns into
\begin{align}
    \frac{\dd \Xc (a)}{\dd \ln a}=\left(\frac{a}{\aend}\right)^3\frac{\mc{C}_\text{DM}(a)}{H(a)}\,,
    \label{eq:BEQ_chi}
\end{align}
where
\begin{align}
    H(a)=\dfrac{\sqrt{\rho_\Phi(a)+\rho_\text{R}(a)}}{\sqrt{3}\,\Mp}\,,
    \label{eq:HubblePhiRad}
\end{align}
and $\mc{C}_\text{DM}(a)\equiv\mc{C}_\text{DM}(T(a))$ is given in Eq.\eqref{eq:CollOp_DM}.
Once we have obtained the solution for $\mc{X}(a)$, we can relate it to the more common $Y(T)$ simply by dividing by the total entropy, $Y(T)=\mc{X}(a(T))/S(T)$.
The DM relic abundance is then defined by
\begin{align}
    \omdm=\frac{\rho_\text{DM}}{\rho_c\,h^{-2}}=\frac{\mdm\,\Xdm(a_0)}{\rho_c\,h^{-2}}\left(\frac{\aend}{a_0}\right)^3\,,
    \label{eq:OmegaDM_def}
\end{align}
where $h=H_0/100~\text{Mpc}~\text{s}~\text{km}^{-1}$, $\rho_c$ is the critical energy density of the Universe, and where we used $\rhodm~=~\ndm~\mdm$.
From this definition, we can derive an approximate analytical expression for the relic abundance as
\begin{align}
    &\frac{\omdm}{0.12}\simeq\mkern-2mu\left(\frac{1.5\,\text{m}}{c\tau}\right)\mkern-5mu\left(\frac{106.75}{\gs}\right)^{\mkern-2mu3/2}\mkern-4mu\left(\frac{\mdm}{100\,\keV}\right)\mkern-5mu\left(\frac{200\,\GeV}{m_P}\right)^{\mkern-2mu2}\nonumber\\
    &\qquad\qquad\times
    \begin{cases}
        &\mkern-10mu\dfrac{2k+4}{3}\left(\dfrac{\Trh}{m_P}\right)^{\mkern-2mu 4k-1}\mc{I}_\text{rh,b}+\mc{I}_\text{RD}^0\quad\mkern-7mu\text{in BR}\\[10pt]
        &\mkern-10mu\dfrac{2k+4}{3k-3}\left(\dfrac{\Trh}{m_P}\right)^{\mkern-2mu\frac{9-k}{k-1}}\mc{I}_\text{rh,f}+\mc{I}_\text{RD}^0\quad\text{in FR}
    \end{cases}\,,
    \label{eq:OmegaDM_sol_main}
\end{align}
where
\begin{align} 
    \mc{I}_\text{rh,b}=\int_{z_\text{end}}^{z_\text{rh}}\dd z' z'^{2+4k}K_1(z')\,,\label{eq:int_rh_b}\\
    \mc{I}_\text{rh,f}=\int_{z_\text{end}}^{z_\text{rh}}\dd z' z'^{\frac{2k+6}{k-1}}K_1(z')\,,\label{eq:int_rh_f}\\
    \mc{I}_\text{RD}^0=\int_{z_\text{rh}}^{z_0}\dd z' z'^3 K_1(z')\,. \label{eq:int_RD}
\end{align}
Here, $z_\text{rh}=m_P/\Trh$, $z_\text{end}\simeq 0$, and $z_0=m_P/T_0~\simeq~\infty$. 
These integrals account for the contribution to the DM number density from the reheating phase (Eq.~\eqref{eq:int_rh_b} for BR and Eq.~\eqref{eq:int_rh_f} for FR) and from the RD epoch (Eq.~\eqref{eq:int_RD}).
Notice that whenever $z_\text{rh}\ll1$, the contribution from the reheating phase becomes negligible since the contribution of the integrals in Eqs.~\eqref{eq:int_rh_f}~and~\eqref{eq:int_rh_b} vanishes.
On the contrary, for low reheating temperature scenarios with $z_\text{rh}\gg1$, these contributions always add up to that one during RD.
Note also that the dependence on $\Trh/m_P$ in front of the reheating phase integrals in Eq.~\eqref{eq:OmegaDM_sol_main} matches with the estimates in Eq.~\eqref{eq:Ydm_T_dil}.
For a more detailed discussion about the calculation of the relic abundance in comparison with previous literature, we refer the reader to App.~\ref{sec:AppC}.

\begin{figure*}[t!]
    \def\sepf{0.45}
\centering
    \includegraphics[scale=0.45]{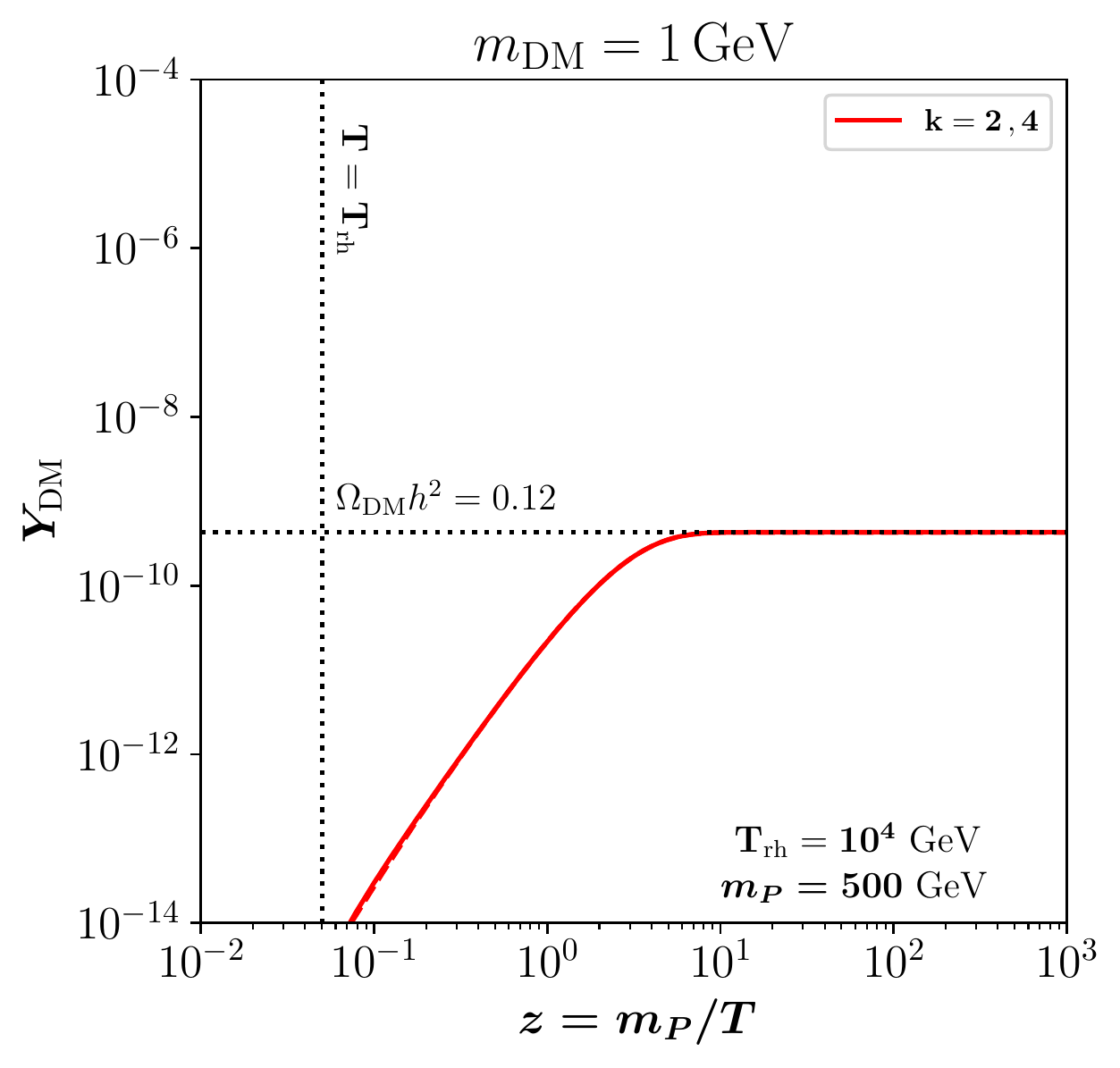}
     \includegraphics[scale=0.45]{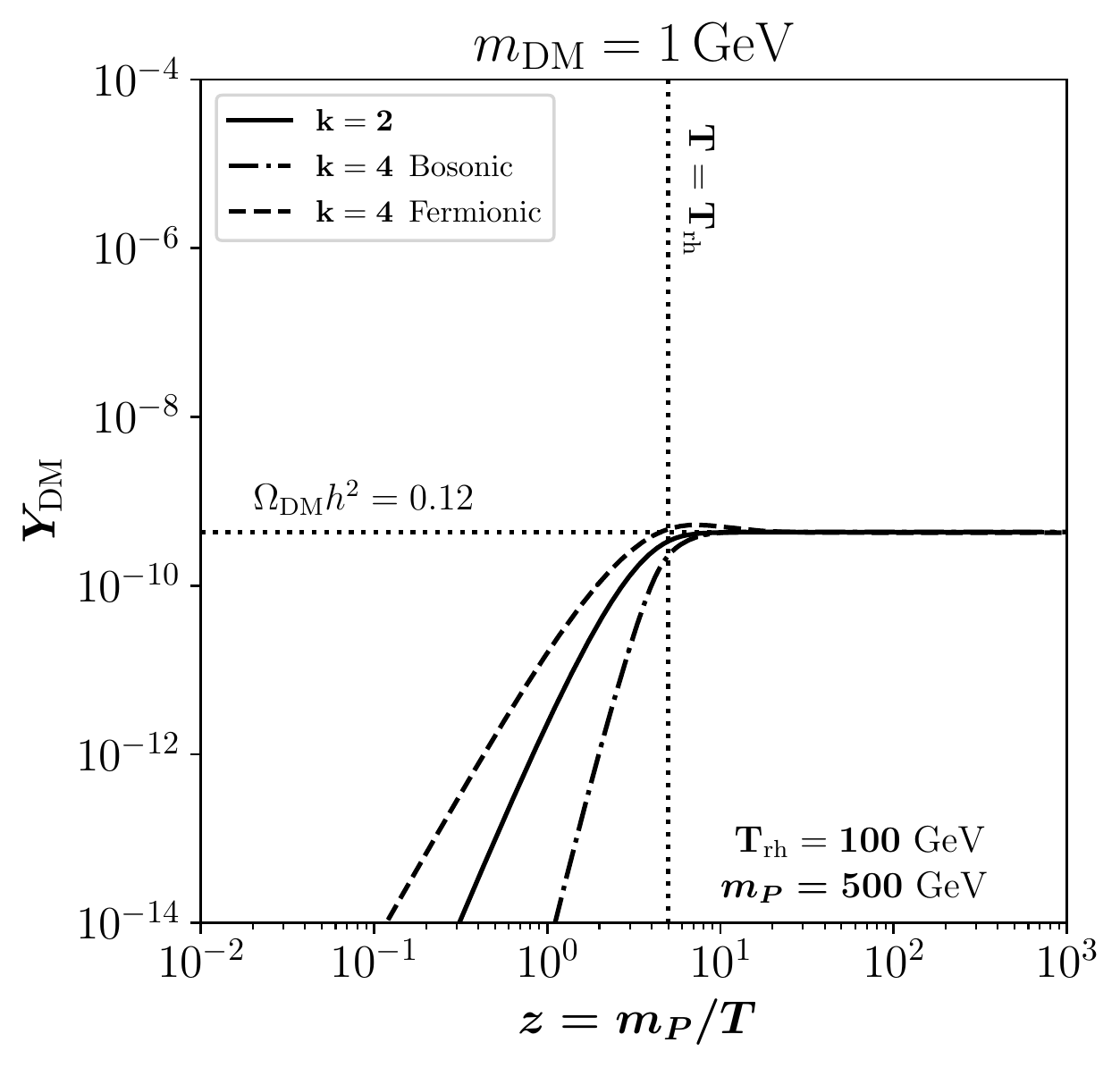}\\
    \includegraphics[scale=0.45]{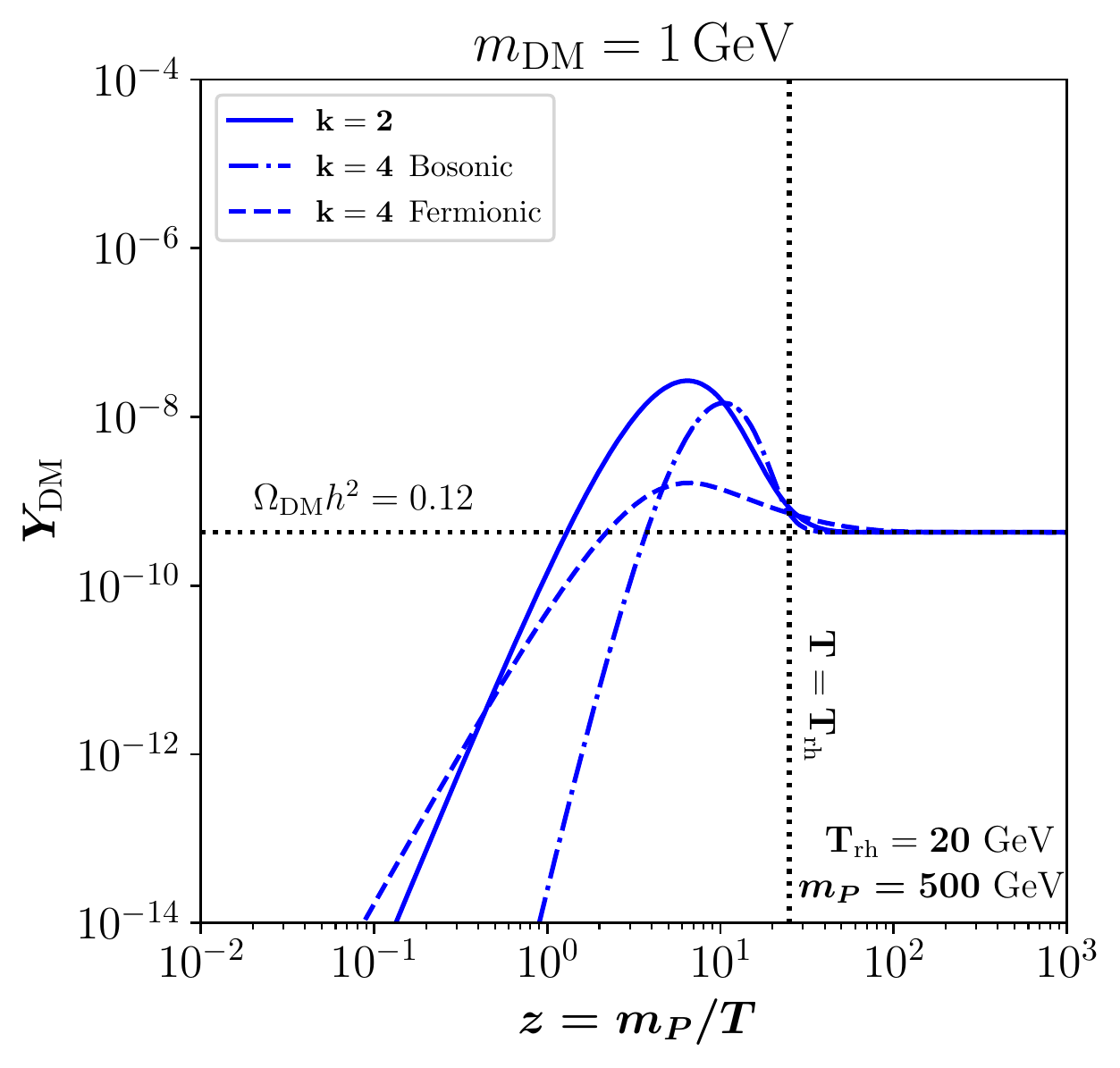}
     \includegraphics[scale=0.45]{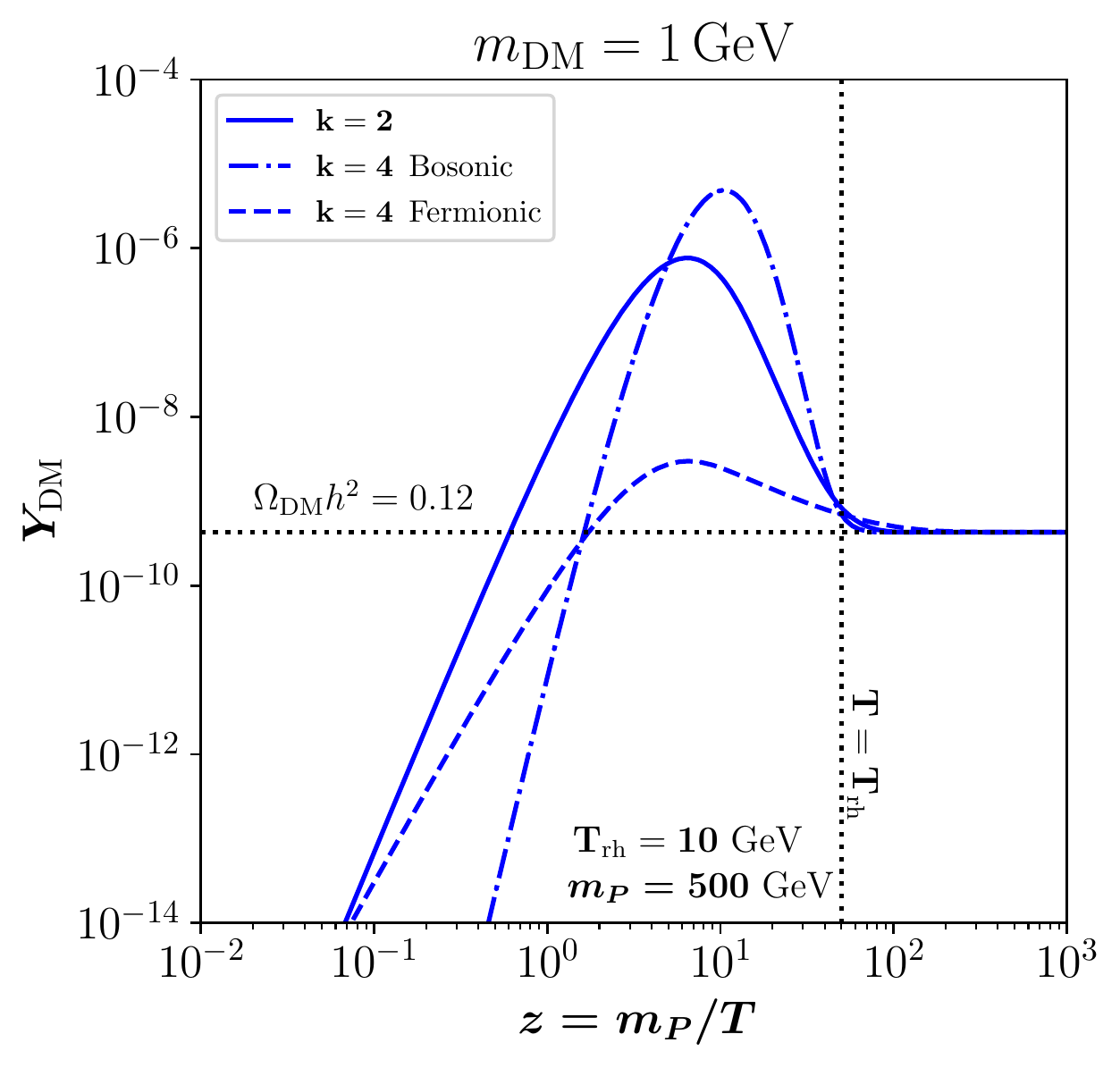}\\
    \caption{Time evolution of the DM yield $\Ydm$ for $\mdm=1\,\GeV$, and $m_P=500\,\GeV$. From upper-left to bottom-right, the reheating temperature is fixed at $10^4$, 100, 20, and 10 GeV. Solid lines correspond to the $k=2$ scenario, while dashed and dot-dashed lines show the FR and BR scenarios for $k=4$. The observed relic abundance $\omdm=0.12$ is obtained for the decay length values listed in Tab.~\ref{tab:ctau_values}.}
\label{fig:yield}
\end{figure*}

In Fig.~\ref{fig:yield}, we show the evolution of the DM yield with respect to the time variable $z=m_P/T$ obtained by numerically solving the system of differential equations consisting of Eq.~\eqref{eq:BoltzFriedPhi} and Eq.~\eqref{eq:BoltzFriedRad} (which determine the Hubble rate in Eq.~\eqref{eq:HubblePhiRad}), and the Boltzmann equation in Eq.~\eqref{eq:BEQ_chi}.
We fix the DM mass to $\mdm=1\,\GeV$, $m_P=500\,\GeV$, and for different $\Trh$, equal to $10^4\,\GeV$ (DM production during RD), $100\,\GeV$ (DM production around $\Trh$), $20\,\GeV$ and $10\,\GeV$ (DM production well within reheating).
The curves displayed consist of the quadratic ($k=2$, solid line) and the quartic ($k=4$) reheating potentials. In the second case, the nature of the reheating phase, i.e. whether BR (dot-dashed line) or FR (dashed line), crucially affects both the production and the dilution phases.
The observed relic abundance $\omdm=0.12$ is achieved for the parent particle decay lengths listed in Tab.~\ref{tab:ctau_values}.

\setlength{\tabcolsep}{10pt} 
\renewcommand{\arraystretch}{1.5} 
\begin{table}[!t]
    \centering
    \begin{tabular}{ l  c  c }
    \hline
      Type   &  $\Trh$ [GeV]  &  $c\tau$ [m]  \\
    \hline
      $k=2$  &  $10$  &  $1.8\times10^{-2}$\\
      $k=4$ BR &  $10$  &  $1.8\times10^{-6}$\\
      $k=4$ FR &  $10$  &  $1.7\times10^{2}$\\[3pt]
      $k=2$  &  $20$  &  $2.2 \times 10^0$\\
      $k=4$ BR &  $20$  &  $3.6\times10^{-2}$\\
      $k=4$ FR &  $20$  &  $4.7\times10^{2}$\\[3pt]
      $k=2$  &  $100$  &  $3.3\times10^{3}$\\
      $k=4$ BR &  $100$  &  $3.4\times10^{3}$\\
      $k=4$ FR &  $100$  &  $4.1\times10^{3}$\\[3pt]
      $k=2$  &  $10^4$  &  $1.3\times10^{4}$\\
      $k=4$ BR &  $10^4$  &  $1.3\times10^{4}$\\
      $k=4$ FR &  $10^4$  &  $1.3\times10^{4}$\\
    \hline
    \end{tabular}
    \caption{Values of $c\tau$ that account for the observed DM relic abundance for the benchmark point $\mdm=1\,\GeV$ and $m_P=500\,\GeV$. We show the results within BR and FR with $k=2$ and $k=4$, and $\Trh=10,\, 20, \, 100,\, 10^4\,\GeV$.}
    \label{tab:ctau_values}
\end{table}

As expected, when $\Trh>m_P$ (cf. Fig.~\ref{fig:ctaumP}, upper left panel), one recovers the classical freeze-in picture during RD, regardless of the details of the reheating phase.
In turn, the faster expansion and stronger entropy dilution when $m_P>\Trh$ imply that the rate of DM production must in general increase, and so the decay width of $P$.
This translates into smaller $P$ decay lengths than in the standard scenario (cf.~Tab.~\ref{tab:ctau_values}).
Furthermore, also the details of the reheating phase have a large impact on the predicted decay length of $P$: 
for $k>2$ and DM production deep within the reheating phase $T_\text{rh} \ll m_P$, BR scenarios require a much smaller decay length of $P$ to produce the observed DM relic density than FR scenarios. 
This is a direct consequence of the stronger dilution during BR and crucially impacts exclusion limits derived from colliders and cosmology as discussed in Section \ref{sec:Constraints}.

In Fig.~\ref{fig:yield}, we clearly observe the extended period of pure DM dilution for large ratios of $m_P/\Trh$. In addition, we find that as soon as $m_P/\Trh \gtrsim 1$, DM production proceeds mainly at larger $z$ in the $k=4$ BR scenario with respect to $k=2$ and $k=4$ FR. Indeed, as indicated in Eq. \eqref{eq:DMProductionRate}, the DM production rate (the rate of change of the comoving DM yield $\frac{d\mathcal{X}_\text{DM}}{dz}$) scales as
\begin{align}
    \frac{d\mathcal{X}_\text{DM}}{dz} =  \begin{cases} 
        z^{2+4k} K_1 \left( z \right)\,&\text{BR}\\
        z^\frac{2k+6}{k-1} K_1 \left( z \right)\;&\text{FR}
    \end{cases}. 
\end{align}
Thus, we expect to find the peak of DM production at later times in BR scenarios, i.e. $\left( \zfi \right)_{\text{BR},\,k=4}~>~\left( \zfi \right)_{k=2}~\gtrsim~\left( \zfi \right)_{\text{FR},\,k=4}$.

In combination with the dilution factor\footnote{Recall that the dilution factors estimate the dilution taking place after DM production has become inefficient. 
Since in Fig.~\ref{fig:yield} all scenarios eventually reproduce the observed relic density, a smaller dilution factor corresponds to a more pronounced peak in the yield.} given in Eq.~\eqref{eq:D(T)}, we can understand the relative heights of the peaks for the different reheating scenarios illustrated in Fig.~\ref{fig:yield}. 
The dilution factor
\begin{align} \label{eq:dilution}
    D \left( z_\text{fi} \right) = \begin{cases}
        z_\text{fi}^{1+2k}\left(\dfrac{\Trh}{m_P}\right)^{1+2k}\,\text{BR}\\
        z_\text{fi}^\frac{7-k}{k-1} \left(\dfrac{\Trh}{m_P}\right)^{\frac{7-k}{k-1}}\;\text{FR}
        \end{cases} \, ,
\end{align}
 decreases (or increases) with the largest power in $m_P/\Trh$ (or $z_\text{fi}$) for the BR scenario with $k=4$, followed by the reheating scenario with $k=2$, and then the FR scenario with $k=4$. 
As a result, the hierarchy for $m_P/\Trh~\gtrsim~1$,
\begin{align}
    D^{\text{BR},k=4} \left( T_\text{fi} \right) > D^\text{FR,k=2,k=4} \left( T_\text{fi} \right) \, ,
\end{align}
is eventually reversed for $m_P/\Trh \gg 1$. 

As a last remark, we checked whether the DM evolution lies within the freeze-in regime at \emph{every} time until the production stops.
This is ensured if the interaction rate remains subdominant with respect to the expansion rate, $\Gamma(T)<H(T)$, such that backreactions can be neglected.
Equivalently, this implies that the DM abundance remains smaller than its equilibrium abundance.
We estimate for which parameter combination this is satisfied by first fixing the DM yield to the observed one, $Y_{\text{DM}}^{0}\,\mdm=4.3\cdot10^{-10}\,\GeV$, via Eq.~\eqref{eq:Ydm_fi_dil}, which in turn determines the value of $\Gamma_P$.
We then insert it into $Y_{\text{DM}}(\zfi)\simeq\Gamma(\zfi)/H(\zfi)$ 
and compare it to the DM equilibrium comoving yield, which simply reads as $\Ydm^{\text{eq}}=45/(\pi^4\gs)$.
This implies that the ratio between the parent particle mass $m_P$ and the reheating temperature $\Trh$ needs to satisfy the following relations
\begin{align}\label{eq:FI_equilibration}
\dfrac{m_P}{\Trh}<
    \begin{cases}
       \, 160 \, \left(\dfrac{\mdm}{\GeV}\right)^{1/5}\dfrac{\zfi}{5}\quad\; k=2\\[10pt]
       \, 70 \, \left(\dfrac{\mdm}{\GeV}\right)^{1/9}\dfrac{\zfi}{10}\quad\;\; k=4\text{ BR}\\[10pt]
       \, 10^8\, \left( \dfrac{\mdm}{\GeV} \right)\dfrac{\zfi}{5}\, \qquad k=4\text{ FR}
    \end{cases}.
\end{align}
This is satisfied for the choice of parameters illustrated in Fig.~\ref{fig:yield}.
As the DM mass is lowered, however, the interaction rate needs to increase to compensate for the smaller DM energy density, such that it becomes easier to violate the freeze-in assumptions at low reheating temperatures.
For example, for $\mdm=12\,\keV$, corresponding to the lower mass bound on FIMP DM from Lyman-$\alpha$ constraints \cite{Belanger:2018sti,Heeck:2017xbu}\footnote{Notice that for FIMPs produced via two-body decays of a TeV-scale parent, a conservative (stringent) bound of $\mdm~\gtrsim~9.2\,(16)~\keV$ was recently found in Ref.~\cite{DEramo:2020gpr}, while a limit of $ 15 \, \mathrm{keV}$ was found in \cite{Decant:2021mhj} using the CLASS code \cite{Diego_Blas_2011}. These limits only apply to DM produced during radiation domination. If DM production takes place during the phase of inflationary reheating, Lyman-$\alpha$ constraints are expected to relax due to the stronger redshifting of DM during the reheating phase. For instance for $k=2$, we estimate the Lyman-$\alpha$ constraints for DM produced during reheating as $\mdm \gtrsim 10 \, \mathrm{keV} \left( \Trh / m_P  \right)^{5/3}$.},
DM equilibrates if $m_P=500\,\GeV$ and $\Trh=10\,\GeV$.
In the results presented in the following sections, we checked that the parameter space presented fulfills the freeze-in conditions.

%% file: 4ConstraintsInflation.tex
\section{\label{sec:Constraints}Results and combined constraints from LLP searches and inflation}
\begin{figure*}[!t]
    \def\sepf{0.5}
\centering
\includegraphics[scale=0.5]{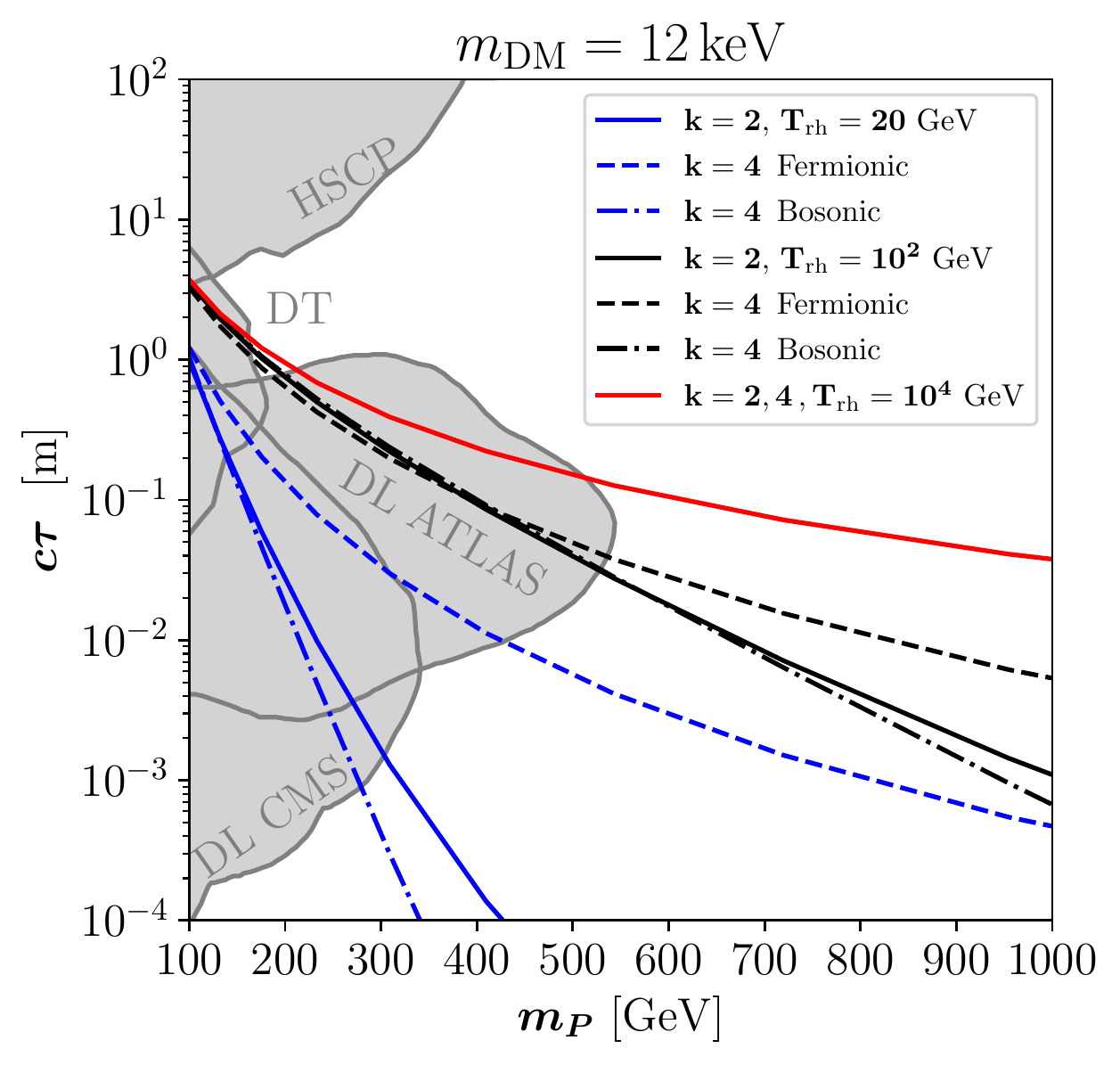}
\includegraphics[scale=0.5]{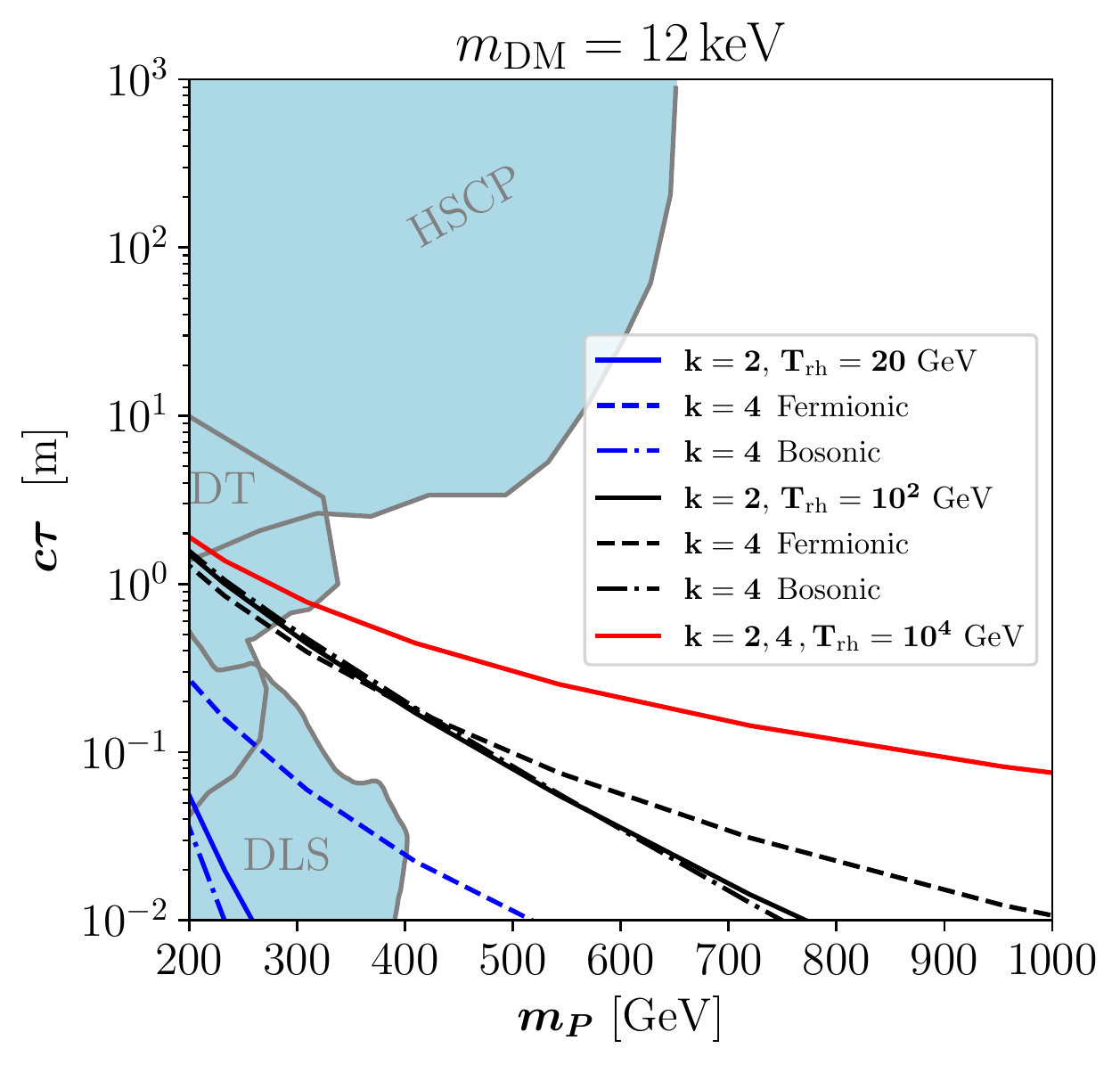}
\includegraphics[scale=0.5]{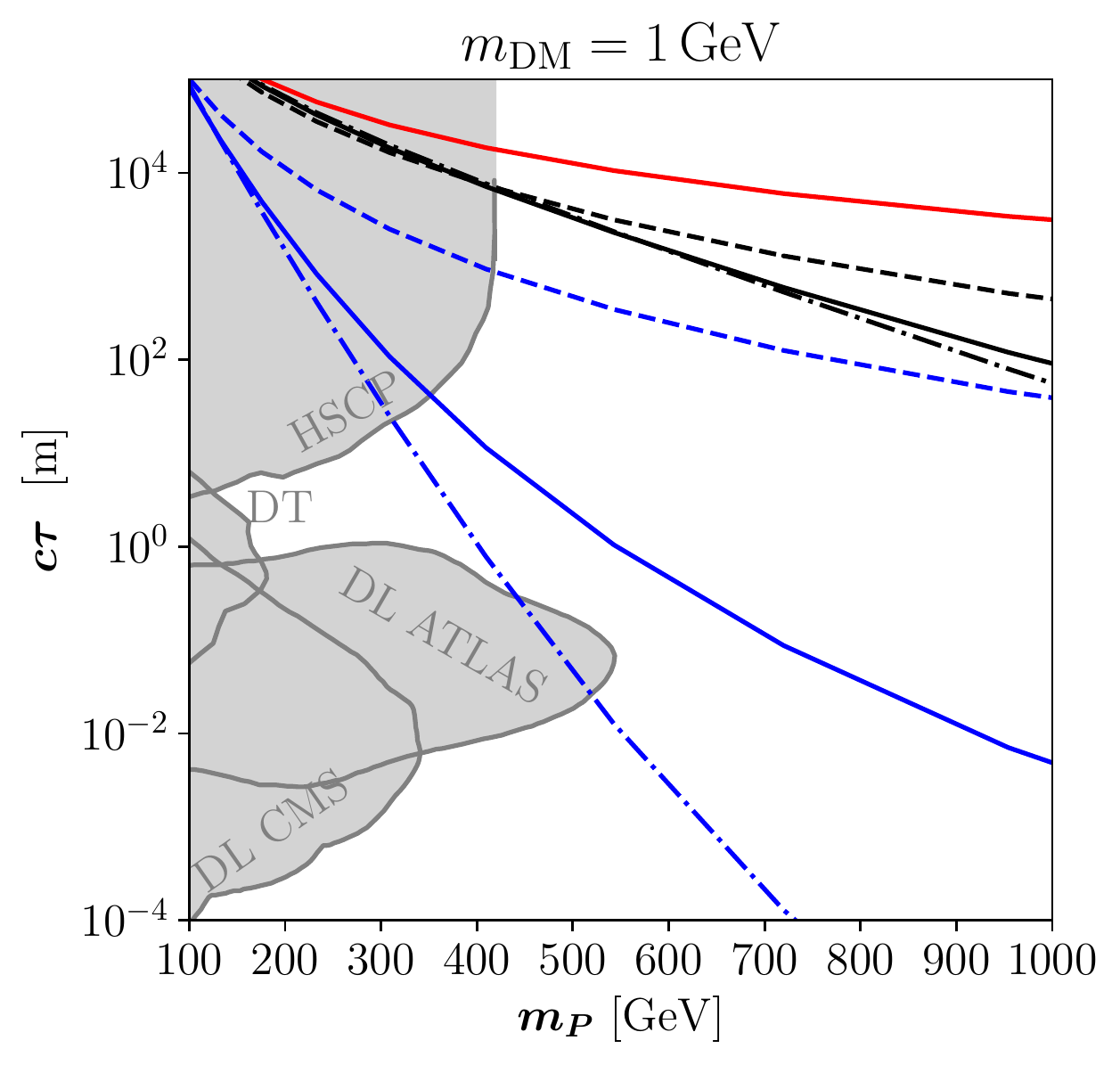}
\includegraphics[scale=0.5]{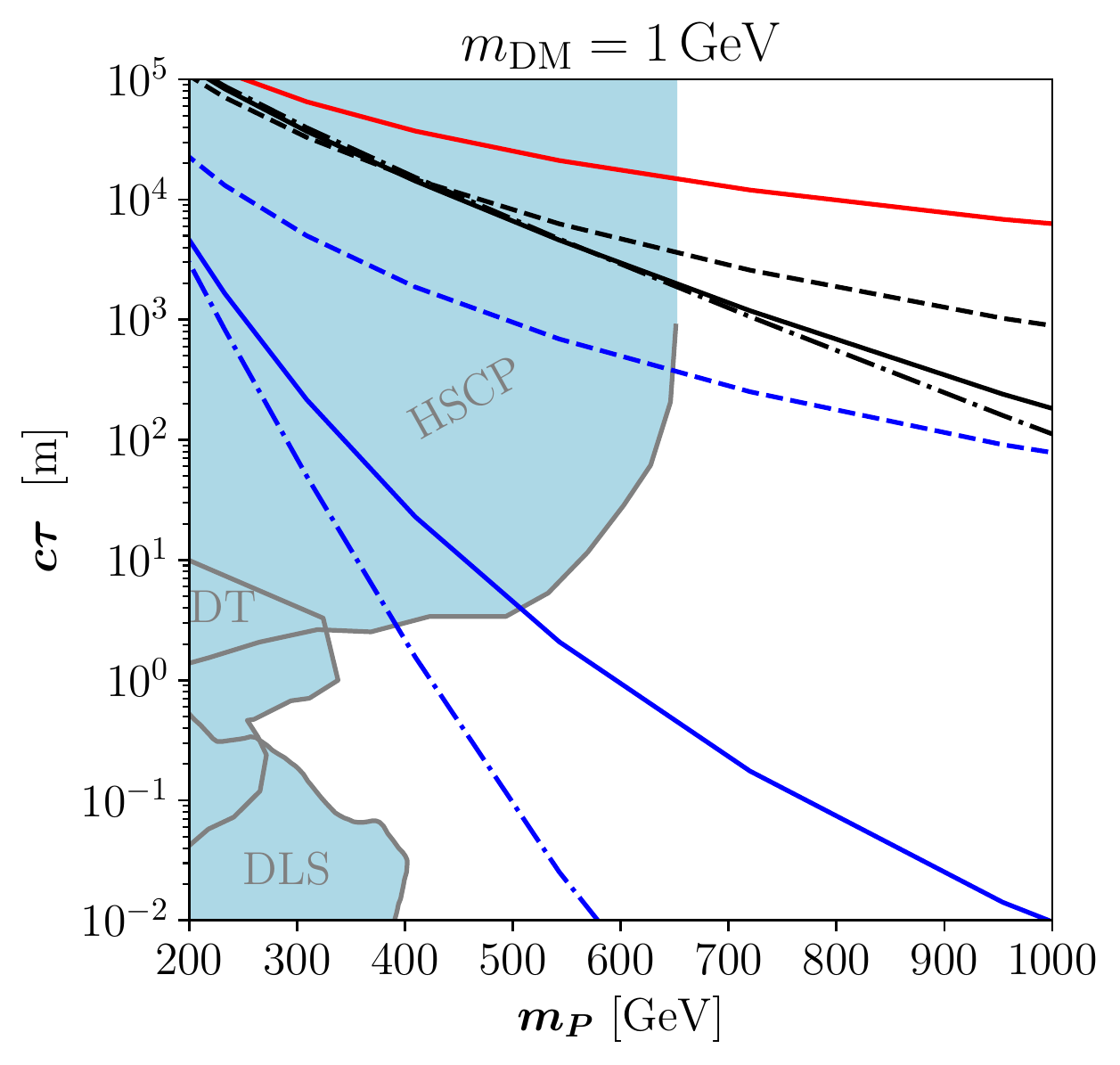}
    \caption{The contours show the values of $c\tau$ and $m_P$ accounting for the observed DM relic abundance for $\mdm=12\,\keV$ (top row) and $\mdm=1\,\GeV$ (bottom row). Blue (black) lines correspond to $\Trh=20\,(100)\,\GeV$, respectively; solid lines are valid for $k=2$, while dashed and dot-dashed lines illustrate the FR and BR scenarios for $k=4$ potentials, respectively. The red line marks $\Trh=10^4\,\GeV$, where, for the range of $m_P$ values displayed, the details of reheating are irrelevant. The constraints from LLP searches are shown with gray-shaded areas in the left column for a muonphilic Majorana DM model \cite{Calibbi:2021fld}, and with light blue colors in the right column for a leptophilic scalar singlet DM model \cite{Belanger:2018sti}.}
\label{fig:ctaumP}
\end{figure*} 
In this section, we present constraints arising from long-lived particle (LLP) searches at the Large Hadron Collider (LHC) as well as constraints on inflationary parameters from cosmological probes and discuss how they complement each other to constrain a model of FIMP DM.
As we have already mentioned in the previous sections, this analysis can be applied to FIMP models where DM is either a Majorana fermion or a real scalar singlet, interacting with a SM charged scalar or a vectorlike fermion, respectively (cf. Eq.~\eqref{eq:trilinear}).
We refer the reader to App.~\ref{sec:Model} for a brief description of these scenarios (see also \cite{Belanger:2018sti,Calibbi:2021fld}).

\subsection{Parent lifetime constraints from the relic density}

In Fig.~\ref{fig:ctaumP}, we present the parent particle decay length $c\tau$ versus its mass $m_P$ in four scenarios that reproduce the observed dark matter (DM) relic abundance. 
The left panels consist of a Majorana DM model with a scalar parent, while the right panels illustrate a scalar singlet DM model with a vectorlike fermion parent.
The upper (lower) row corresponds to $\mdm=12\,\keV$ ($1\,\GeV$).
Moreover, in both cases, we choose to couple the DM field to a SM right-handed lepton (leptophilic model) \footnote{Note that while the computed relic density and decay length hold for any leptophilic model, the LHC constrains in Fig.~\ref{fig:ctaumP} assume a muonphilic scenario for the Majorana DM model, as explained in Sec.~\ref{sec:constrB}}.
This means that the parent particle fields would be charged only under $U(1)_Y$.

DM production in the context of FR is represented by dashed lines, while the dash-dotted lines correspond to production during BR. 
When $k=2$ the FR case yields the same results as the BR scenario, as expected, hence they are both shown with the same solid lines. 
The reheating temperature is fixed to $\Trh =10^{4}\,\text{GeV}$ (red lines), $\Trh =10^{2}\,\text{GeV}$ (black lines), and $\Trh =20\,\text{GeV}$ (blue lines). 
Notably, when $\Trh \gg m_P$, freeze-in occurs mainly after reheating, rendering the details of reheating irrelevant for DM production. 
This is why all the red curves overlap, and the blue and black lines tend to merge with decreasing $m_P$.

For values of $k > 2$, the inflaton mass and decay rate are dependent on the inflaton energy density, which results in a difference in the expansion of the Universe between the BR and FR scenarios, thus affecting the production of DM during reheating, as extensively discussed in the previous sections. 
In this situation, the dilution during reheating is, in general, stronger for BR than for FR, resulting in smaller decay lengths of the parent particle for the former scenario than for the latter (cf. Tab.~\ref{tab:ctau_values}).

The DM relic abundance is proportional to the ratio of DM mass to its decay length, i.e., $\omdm~\propto~\mdm/(c\tau)$. 
Therefore, as the DM mass increases, the decay length of the parent particle should also increase to ensure that the correct relic density is achieved. 
This explains why the lines in Fig.~\ref{fig:ctaumP} move upwards when we change the DM mass from $12~\text{keV}$ to  $1~\text{GeV}$.

We stress that the predictions obtained were calculated with a full numerical evaluation of Eqs.~\eqref{eq:BoltzFriedPhi}, \eqref{eq:BoltzFriedRad}, and \eqref{eq:BEQ_chi} with the reheating temperature defined by Eq.~\eqref{eq:T_RH_Implicit}.
Our results differ from those obtained by Refs.~\cite{Belanger:2018sti} and \cite{Calibbi:2021fld}. In the first work the authors take into account only the evolution of the DM yield within RD, while in the latter, $\Trh$ is defined by the relation $\Gamma_\Phi(\Trh)=H(\Trh)$. 
However, the more accurate definition of the reheating temperature in Eq.~\eqref{eq:T_RH_Implicit} becomes especially relevant when inferring limits on the reheating temperature from CMB data.  
Generally, we find that Refs.~\cite{Belanger:2018sti} and ~\cite{Calibbi:2021fld} underestimate the relic density. 
We dedicate App.~\ref{sec:AppC} (cf. Fig.~\ref{fig:comparison}) to the comparison with the previous literature, where we explain in more detail the source of these discrepancies.

\subsection{\label{sec:constrB}Constraints from LLP searches at the LHC}
In Fig.~\ref{fig:ctaumP}, we show with shaded gray (light blue) colors the regions for the models with a scalar (vectorlike fermion) parent that are constrained by LLP searches at the LHC.
To this end, we utilize the limits provided by Ref.~\cite{Calibbi:2021fld} for the former (muonphilic model) and by Ref.~\cite{Belanger:2018sti} for the latter.
We refer the reader to these works for a deeper insight into the re-casting and re-interpretation methods of the searches employed.

Searches for ionizing tracks left by slepton-like heavy stable charged particles (HSCPs) are typically suited for probing large lifetimes of the charged parent particle: for example, in the scalar singlet DM model \cite{Belanger:2018sti}, vectorlike fermion parents with $c\tau\gtrsim 0.1\,(100)\,$m are excluded for $m_P\simeq 200\,(600)\,$GeV by the CMS \cite{CMS:2016ybj} collaboration.
The latter search, together with the ATLAS one in \cite{ATLAS:2019gqq}, have also been used for the Majorana DM model in Ref.~\cite{Calibbi:2021fld}; in this case, the limits were re-interpreted for a scalar parent and they shift the constraints to slightly smaller masses, excluding $c\tau\gtrsim 0.3\,(100)\,$m if $m_P\sim 100\,(400)\,\GeV$.

Displaced lepton (DL) searches targeting shorter decay lengths were performed by both ATLAS  \cite{ATLAS:2020wjh} and CMS \cite{CMS:2014xnn,CMS:2016isf}, with the former more efficient around $c\tau~\sim~10^{-2}\,$m, and the latter around $c\tau~\sim~10^{-3}\,$m.
Notice that, while in the first case signal regions with both displaced $e\mu$ pairs, and same-flavour ($ee$ and $\mu\mu$) pairs were analyzed, the CMS searches only include $e\mu$ pairs.
In Ref.~\cite{Calibbi:2021fld}, all the searches were employed and re-interpreted in the context of a scalar parent, while Ref.~\cite{Belanger:2018sti} utilized only the CMS analyses due to availability at that point. In the following, we denote the DL searches from Ref.~\cite{Calibbi:2021fld} as DL CMS and DL ATLAS, while the DL search from Ref.~\cite{Belanger:2018sti} is denoted by DLS.

Additionally, the ATLAS \cite{ATLAS:2013ikw,ATLAS:2017oal} and CMS \cite{CMS:2014gxa,CMS:2018rea,CMS:2020atg} collaborations have conducted searches for disappearing tracks (DT) left by charged mediators decaying within the inner tracking system of the detector. 
These searches have higher sensitivity for decay lengths between those targeted by HSCP and DL searches. 
However, the re-casted limits performed in Ref.~\cite{Calibbi:2021fld} (\cite{Belanger:2018sti}) can hardly reach masses up to $180\,(350)\,\GeV$.
Notice that, for the muonphilic scalar parent considered in Ref.~\cite{Calibbi:2021fld}, DT searches with CMS \cite{CMS:2020atg} were re-interpreted as potential kinked track (KT) signatures.
Therefore, the DT exclusion area in the left column of Fig.~\ref{fig:ctaumP} should be considered a forecast.

When comparing with Fig.~\ref{fig:ctaumP}, it becomes clear how a detailed understanding of the reheating phase is crucial to correctly interpret the constraints from LLP searches and to infer which of these searches could be relevant the most.
For example, when assuming a Majorana DM candidate with a mass of $1$ GeV (left column, bottom row in Fig.~\ref{fig:ctaumP}), a large reheating temperature (red curve) signals that only HSCP searches are currently able to probe the model.
On the contrary, for a low reheating temperature (blue curves) also DL searches could become important.

Moreover, a positive LLP signal has the potential to distinguish BR and FR scenarios, and to determine the power $k$ of the reheating potential if $\Trh \lesssim m_P$. 
This feature is illustrated by the distinct predictions for the decay length in the various reheating scenarios, which are visible in the large offset between the solid ($k=2$), dashed ($k=4$, FR) and dot-dashed ($k=4$, BR) lines in Fig.~\ref{fig:ctaumP}. 
Note that for a well defined distinction between the scenarios, we not only require knowledge about the properties of the parent particle (LLP signal) and of the DM mass but also need to infer predictions and constraints for the reheating temperature. 
These are highly dependent on the inflationary model considered and discussed in the next section. 
\begin{figure*}[!t]
    \def\sepf{0.5}
\centering
\includegraphics[scale=0.46]{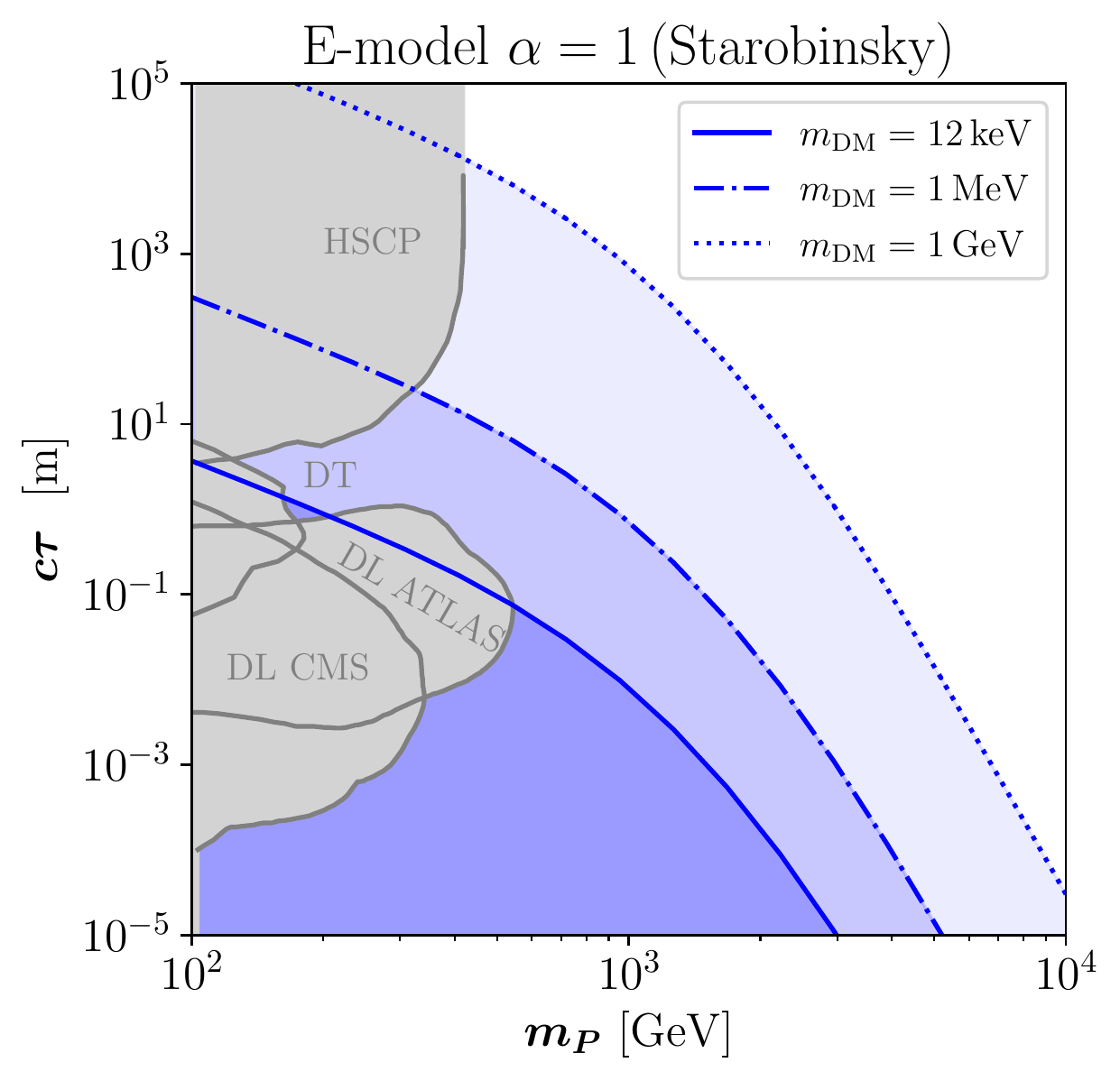}
\includegraphics[scale=0.46]{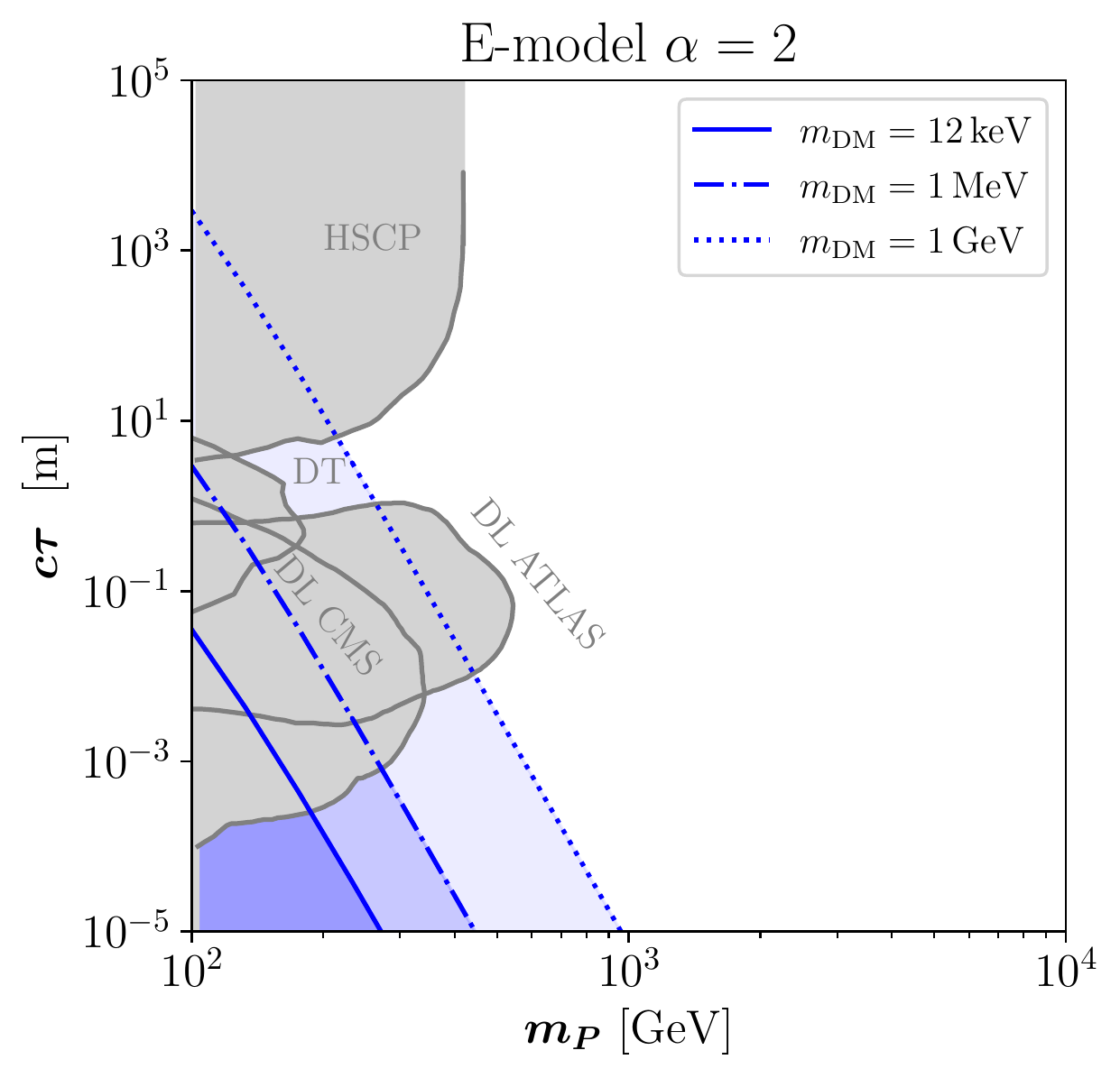}
\includegraphics[scale=0.46]{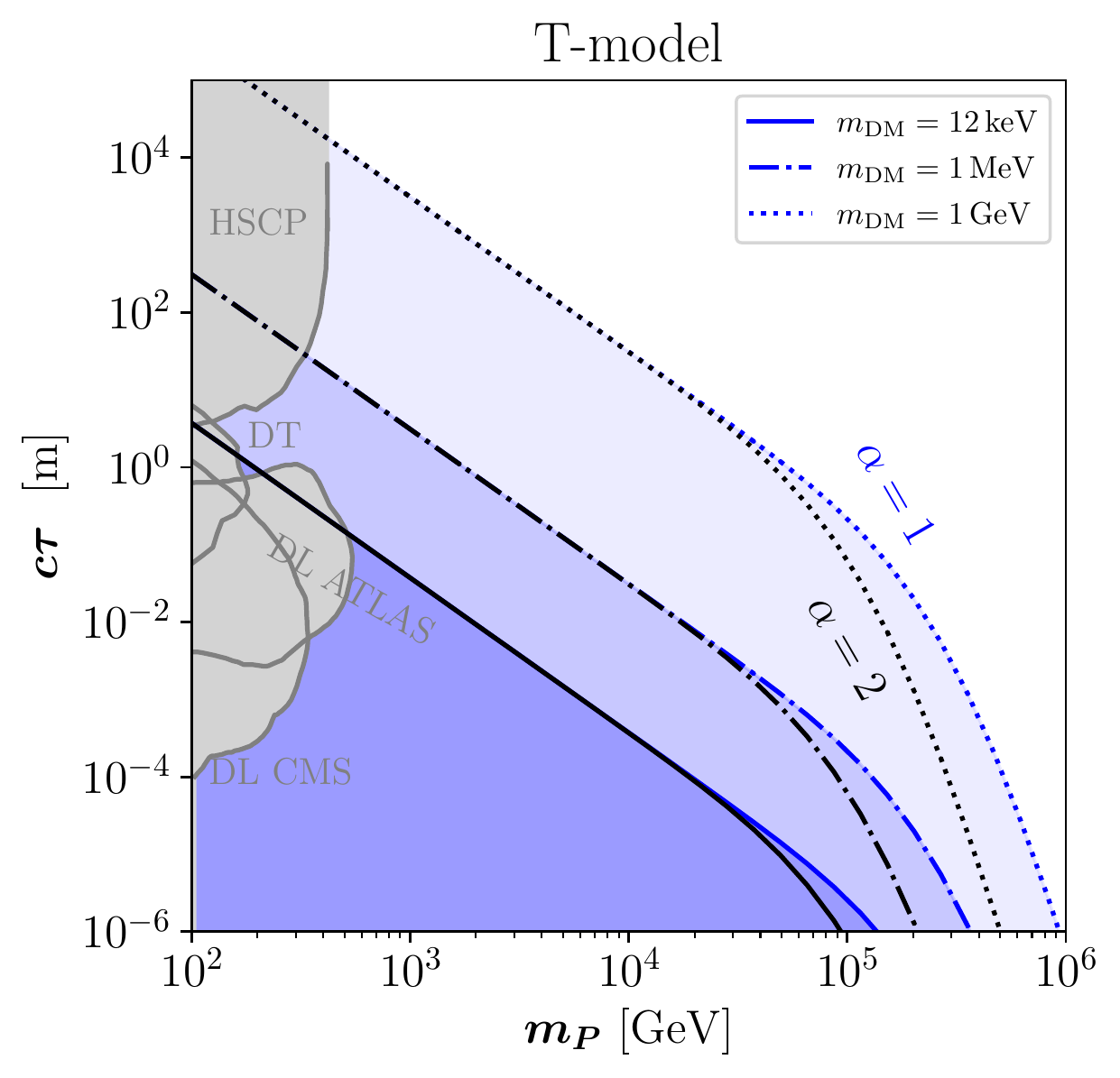}
    \caption{Collider and inflationary constraints on the leptophilic Majorana DM model with a scalar parent, by assuming reheating with $k=2$. We show scenarios for Starobinsky inflation (left panel), E-model with $\alpha=2$ (central panel), and T-model (right panel). The blue-shaded regions show where producing the observed relic abundance contradicts constraints on the reheating temperature for three choices of DM masses: $12\,\mathrm{keV}$, $1\,\mathrm{MeV}$, and $1\,\mathrm{GeV}$ in solid, dot-dashed and dotted lines, respectively. In the T-model, blue corresponds to $\alpha=1$, while black to $\alpha=2$. The gray regions indicate where different LLP searches exclude parts of the parameter space (cf. Ref.~\cite{Calibbi:2021fld} and main text).}
\label{fig:ctau_mP_inflation_scal}
\end{figure*}

\subsection{Constraints from Inflation on FIMP DM}
As pointed out above, the phase of inflationary reheating can have a substantial impact on DM production and the subsequent interpretation of collider limits. 
Thus, it is of interest to investigate the interplay of such searches with limits arising from cosmological observations, which rely on the underlying model of inflation for their interpretation.

Specifically, we consider the $\alpha-$attractor E- and T-models \cite{Kallosh:2013hoa,Kallosh:2013maa}, which reproduce the reheating potential of the form $V(\Phi)\sim\Phi^k$ (cf. Eq.~\eqref{eq:PhiPotential}) and can be described by the following potentials:
\begin{align}
    &V(\Phi)=\Lambda^4\left(1-e^{-\sqrt{\frac{2}{3\alpha}}\frac{\Phi}{\Mp}}\right)^{2n},\quad\text{E model;}
    \label{eq:V_E-mod_main}\\
    &V(\Phi)=\Lambda^4\left[\tanh\left(\frac{\Phi}{\sqrt{6\alpha}\Mp}\right)\right]^{2n},\quad\text{T model.}
    \label{eq:V_T-mod_main}
\end{align}
Here, $\Lambda$ is some energy scale, while $\alpha>0$ and $n>1$ are two dimensionless parameters.
When expanded around small field values, one can perform a matching of these quantities to the parameters in Eq.~\eqref{eq:PhiPotential} as $k=2n$, as well as $\lambda=(\frac{\Lambda}{\Mp})^4(\frac{2}{3\alpha})^n$ in the E-model or $\lambda=(\frac{\Lambda}{\Mp})^4(\frac{1}{6\alpha})^n$ for the T-model.

A more complete overview of these inflationary scenarios, their relation to the slow-roll parameters, as well as the observables constrained by the latest cosmological observations, is presented in App.~\ref{sec:AppA}.
Here, we want to mention that these models generally predict specific relations between the parameters of the inflaton potential and their observables, such as the tensor-to-scalar ratio $r$, the spectral index $n_s$, the amplitude $A_s$ of primordial scalar perturbations, and the number of $e$-folds $N_\star$ that occurred between a pivot scale $k_\star$ and the time inflation terminated (cf. Eqs.~\eqref{eq:rvsn_Emod}~-~\eqref{eq:Lambda_Tmod}).
These quantities are constrained by experiments observing the CMB \cite{Planck:2018inflation,BICEP:2021xfz}.

Moreover, it is possible to derive bounds on the reheating temperature for different scenarios by taking advantage of the fact that the dynamics of reheating affect the moment in time at which the CMB pivot scale re-enters the horizon. Consequently, we can obtain a relation between the reheating parameters and the inflationary observables \cite{Dai:2014jja} (cf. App.~\ref{sec:AppA} for the derivation) which reads as
\begin{align}\label{eq:Trh_inflation_main}
\Trh&=\bigg[\left(\frac{43}{11 g_{*s,\text{rh}}}\right)^{\frac{1}{3}} \frac{a_0 T_0}{k_\star} H_\star e^{-N_\star}  \nonumber\\
&\qquad\qquad\quad\times\left(\frac{45 V_{\text{end}} }{\pi^2 g_{*\text{rh}}}\right)^{-\frac{1}{3\left(1+w_{\text{rh}}\right)}}\bigg]^{\frac{3\left(1+w_{\text{rh}}\right)}{3 w_{\text{rh}}-1}}\,.
\end{align}
Here, $a_0$ and $T_0$ are the scale factor and the temperature at the present time, $V_\text{end}$ the potential at the end of inflation, $w_\text{rh}\neq 1/3$ the equation of state during reheating, and $H_\star$ the Hubble rate at $k_\star$. 

Note that $H_\star$, $N_\star$, and $V_\text{end}$ depend on the specific model of inflation, hence are functions of inflationary predicted observables: the spectral index $n_s$, the tensor-to-scalar-ratio $r$ as well as the scalar power spectrum $A_{s,*}$. For a given inflation model, one can then  compute  $H_\star$, $N_\star$, and $V_\text{end}$, and further use the experimental constraints on $n_s$, $r$ and $A_{s,*}$ to derive the allowed regime of $\Trh$ based on Eq.~\eqref{eq:Trh_inflation_main}. In this work we have taken $A_{s,*}=2.1\times 10^{-9}$, i.e. the central value by Planck 2018 \cite{Planck:2018inflation}. For $n_s$ and $r$, we have imposed the latest  constraints from  Planck+BICEP/Keck+WMAP analysis presented in Ref.~\cite{BICEP:2021xfz}. Note that for the two $\alpha$--attractor inflationary models examined in Eqs.~\eqref{eq:V_E-mod_main} and~\eqref{eq:V_T-mod_main}, $N_\star$ increases with $n_s$, implying that for\footnote{Note that the exponent in Eq.~\eqref{eq:Trh_inflation_main} becomes $\frac{3\left(1+w_{\text{rh}}\right)}{3 w_{\text{rh}}-1} = \frac{3k}{k-4}$ for $\Phi^k$ reheating potentials and is negative when $k<4$.} $k<4$ one could derive a lower bound on $\Trh$ by imposing the lowest allowed value of $n_s$.
For $k=4$, corresponding to $w_{\text{rh}}=1/3$, this method is unable to constrain the reheating temperature since the expansion during reheating is similar to that of a universe dominated by radiation, which makes it indistinguishable from the subsequent period of standard radiation domination \cite{Cook:2015vqa}.  We refer to App.~\ref{sec:AppA} and Fig.~\ref{fig:Trh_constrain} for more details.
\begin{figure*}[!t]
    \def\sepf{0.5}
\centering
\includegraphics[scale=0.46]{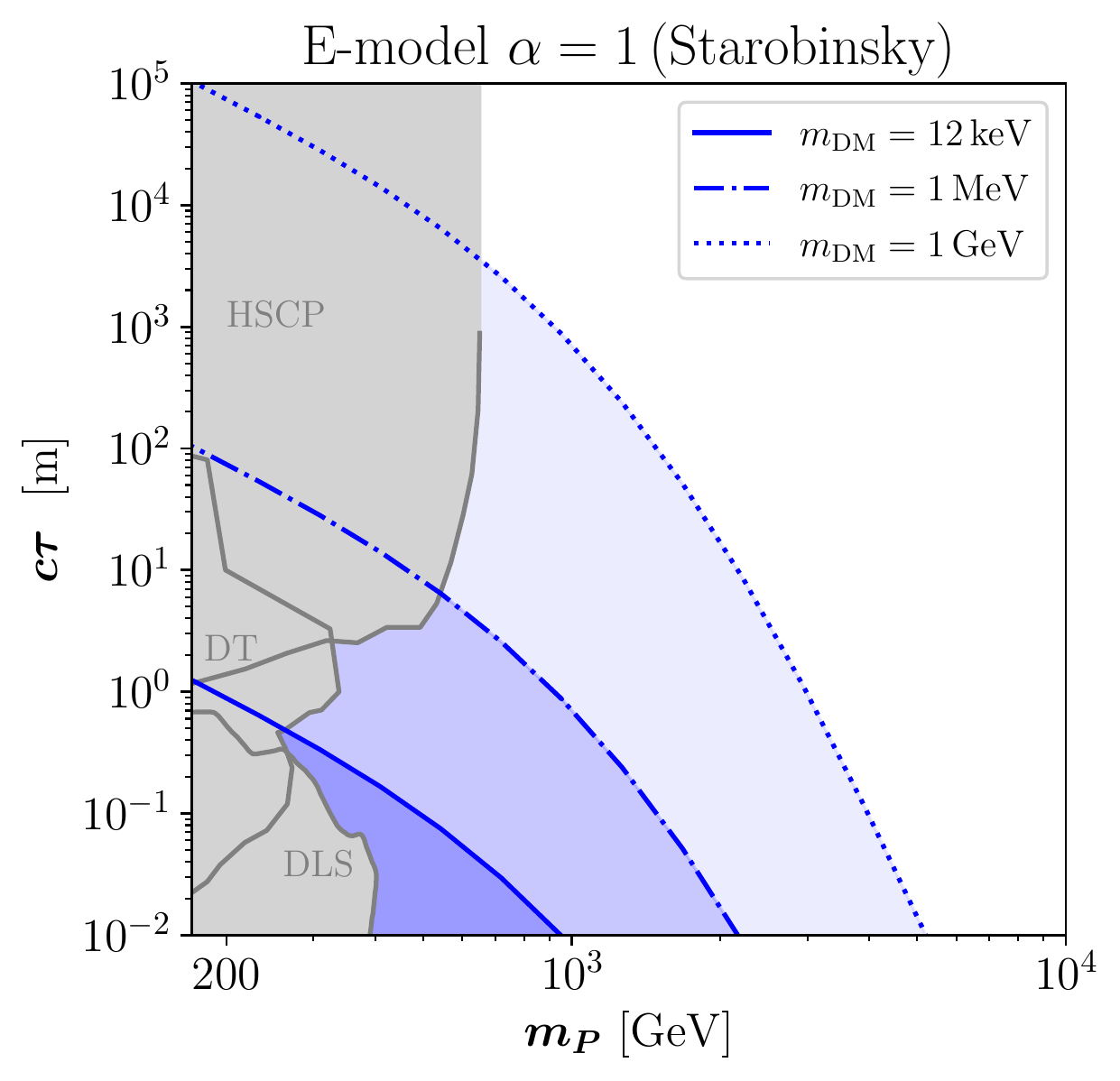}
\includegraphics[scale=0.46]{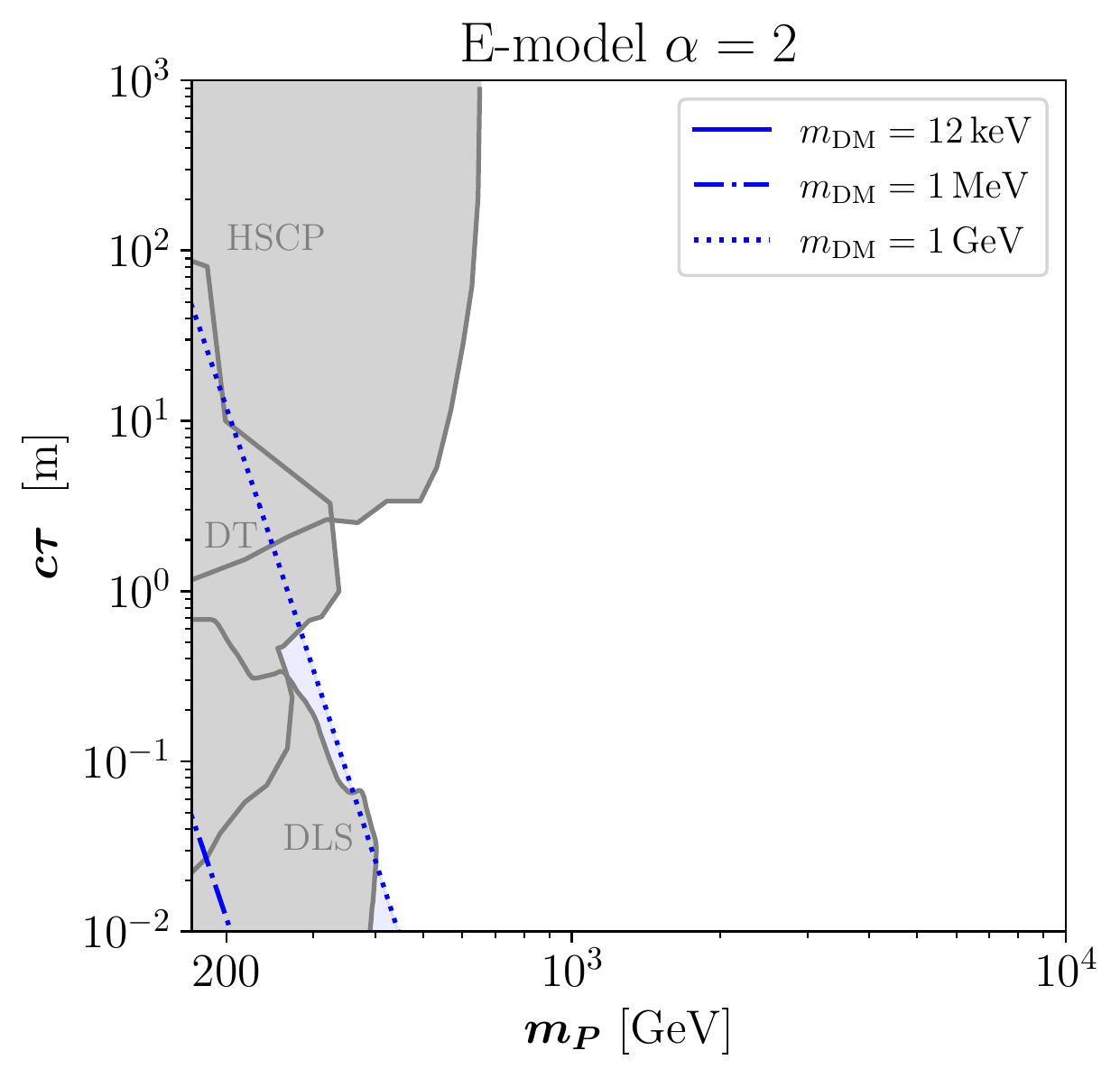}
\includegraphics[scale=0.46]{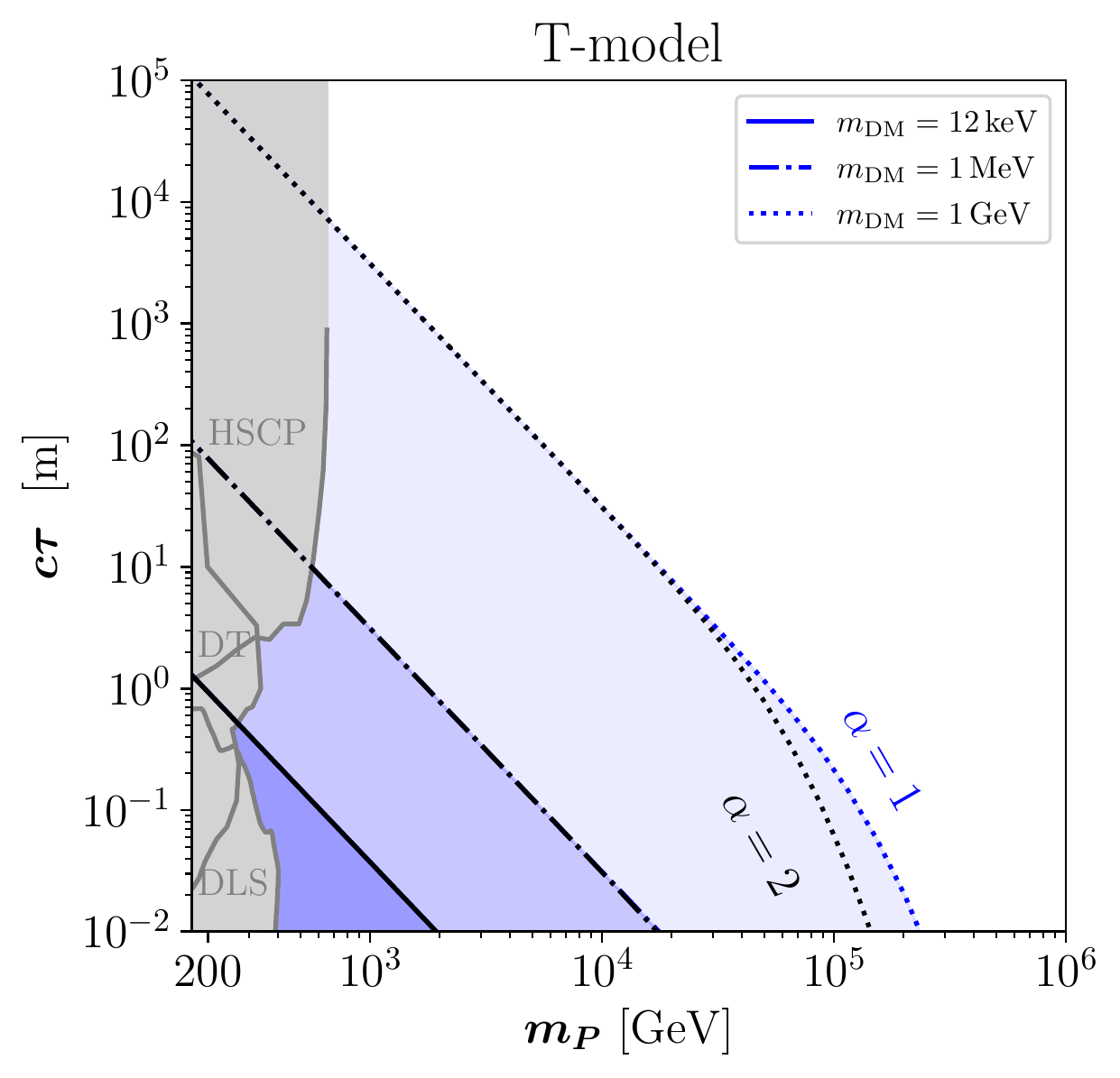}
    \caption{Same caption as in Fig.~\ref{fig:ctau_mP_inflation_scal}, but for a singlet scalar leptophilic DM model with vectorlike fermion parents and the LLP exclusion limits (gray regions) from Ref.~\cite{Belanger:2018sti} (see also main text).}
\label{fig:ctau_mP_inflation_ferm}
\end{figure*} 

In Fig.~\ref{fig:ctau_mP_inflation_scal} and in Fig.~\ref{fig:ctau_mP_inflation_ferm}, we show the combination of constraints from LLP searches and inflationary bounds for the Majorana and the scalar singlet DM models, respectively.
For the reasons explained, here we fix $k=2$.
The left and central panels correspond to the $\alpha-$attractor E-model with $\alpha=1$ (Starobinsky inflation) and $\alpha=2$, respectively, while the right panel shows both $\alpha=1$ (blue curves) and $\alpha=2$ (black curves) in a T-model inflationary scenario.
The solid, dot-dashed, and dotted curves for $\mdm=12\,\keV$, $1\,\MeV$, and $1\,\GeV$, respectively, confine the blue shaded regions where the spectral index would be $n_s< 0.9579$, a range excluded at the $2\sigma$ level by Ref.~\cite{BICEP:2021xfz}. 
For E- and T-models with $\alpha=1$ and $2$, the respective lower bounds on the reheating temperature are tabulated in Tab.~\ref{tab:reheatinglimits}.

Overall, we find more stringent constraints within the context of T-models than of E-models, as well as for smaller values of $\alpha$. 
\begin{table}[!t]
    \centering
    \begin{tabular}{ l  c    c}
    \hline
      Type   & $\alpha$  &  $\Trh$ [GeV]  \\
    \hline
      E-model  &  1  &  $1.8 \cdot 10^2$\\
      E-model &  2  &  $7.8$\\
      T-model &  1  &  $4.0 \cdot 10^4$\\
      T-model  &  2  &  $1.8 \cdot 10^4$\\
      \hline
    \end{tabular}
    \caption{Lower bounds on the reheating temperature for the models of inflation presented in Fig.~\ref{fig:ctau_mP_inflation_scal}.}
    \label{tab:reheatinglimits}
\end{table}
The lower limits on the reheating temperature can be directly converted into lower limits on the decay length $c \tau$: our analytic approximation for the relic density given in Eq.~\eqref{eq:OmegaDM_sol_main} shows that a higher reheating temperature implies a larger relic abundance, which must be compensated for by a larger decay length when fixing the masses of the parent particle and of DM. 
Thus, a lower bound on the reheating temperature also translates into a lower bound on the decay length.

Moreover, as $\omdm \sim \mdm \left( c \tau \right)^{-1}$, an increase in the DM mass results in a longer decay length, making the inflationary constraints more stringent. 
On the other hand, the Lyman-$\alpha$ bound sets a lower limit on the DM mass at $12$~keV and, thus, it follows that the constraints derived from inflationary observables for $\mdm = 12~\keV$ are unavoidable.
For example, in Starobinsky inflation (left panel in Figs.~\ref{fig:ctau_mP_inflation_ferm} and \ref{fig:ctau_mP_inflation_scal}), we find that the limits are comparable to those obtained through DT and DL searches for parent particle masses ranging from $100$~GeV to $1$~TeV. 
Besides, unlike collider searches, inflationary constraints are not limited by a kinematic threshold, and therefore, parent particle masses above 1~TeV are automatically constrained.
Additionally, these constraints have a characteristic shape for $k=2$: for $\Trh \gg m_P$, they scale as $c\tau \sim m_P^{-2}$, while for $\Trh \ll m_P$ we find $c\tau \sim m_P^{-9}$. 
This reflects the behavior found in Eqs.~\eqref{eq:Yield_prod} and \eqref{eq:Ydm_fi_dil}.

\begin{figure}[t!]
\centering
\includegraphics[scale=0.55]{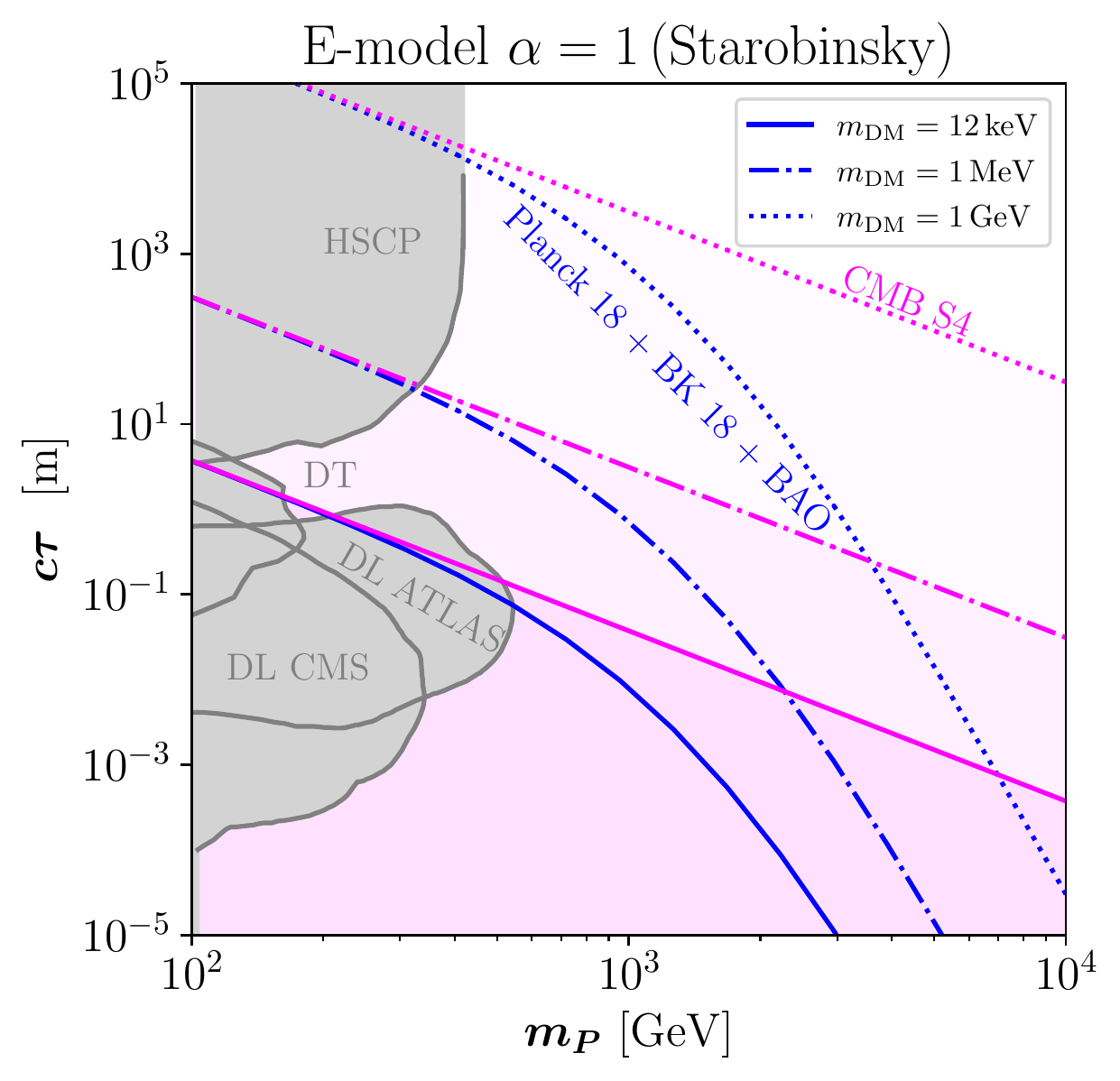}
\caption{Same as in the left panel of Fig.~\ref{fig:ctau_mP_inflation_scal} but also including the exclusion limits (magenta shaded regions) derived by the projected bounds from CMB-S4 \cite{Abazajian:2019eic}; these assume a fiducial model with $r=0.003$ (cf.~Fig.~\ref{fig:rns_plane}).}
    \label{fig:Planck_CMBs4_main}
\end{figure}

In Figure~\ref{fig:Planck_CMBs4_main}, we also show the projected limits with the future CMB-S4 experiment \cite{Abazajian:2019eic} (depicted by the magenta-filled regions), to demonstrate the potential impact on the DM model of the improved sensitivity on the inflationary parameters. 
In particular, the constraints apply to an E-model with $k=2$, and $\alpha=1$ (Starobinsky inflation) for a fiducial value tensor-to-scalar ratio value $r=0.003$; the T-model would give similar results.
Notice that CMB-S4 is in principle able to rule out the scenarios with $\alpha=2$ (cf. Fig.~\ref{fig:rns_plane}).
Since the $2\sigma$ projected limits from CMB-S4 on $n_s$ are more constraining than those from the recent combined analysis with Planck  2018, BICEP/Keck 2018 and BAO~\cite{BICEP:2021xfz}, the lower bound on $\Trh$ shifts to larger values (cf. the discussion below Eq.~\eqref{eq:Trh_inflation_main}). 
Hence, this effectively relocates the position of the kink in the inflationary constraints observed in Figs.~\ref{fig:ctau_mP_inflation_scal}~ and~\ref{fig:ctau_mP_inflation_ferm} to larger values of $m_P$, where the scaling behavior changes from $c\tau \sim m_P^{-2}$ to $c\tau \sim m_P^{-9}$, resulting in more stringent constraints at larger masses.

In summary, we have highlighted that lower reheating
temperatures imply significantly smaller decay lengths
whenever $m_P \gtrsim \Trh$. 
As a result, the minimally possible value for the decay length, implicitly set by Lyman~--~$\alpha$ constraints on the DM mass, is lifted whenever $m_P \gtrsim \Trh$.
The range of parent particle masses for which this can happen is bounded by the lowest value of the reheating temperature from CMB data constraints which strongly depends on the particular inflationary model considered.

%% file: 5Conclusions.tex
\section{\label{sec:Conclusions}Conclusions}
In this work, we have investigated the phenomenology of minimal models of feebly interacting massive particle (FIMP) dark matter (DM) produced during reheating, where the inflaton coherently oscillates around a generic potential $V(\Phi)\sim \Phi^k$, dominantly decaying either into bosons (bosonic reheating, or BR) or fermions (fermionic reheating, or FR).

We have considered a thermal parent particle with mass $m_P$ decaying to a FIMP plus a SM lepton with a decay rate $\Gamma_P$ (or equivalently decay length $c\tau \equiv 1/\Gamma_P$). 
Since freeze-in proceeds via a renormalizable operator, DM is mainly produced around $T_\text{FI} \sim m_P$. 
Consequently, DM production takes place either during radiation domination (RD) for $\Trh \gtrsim m_P$ or during reheating for $\Trh \lesssim m_P$. 
In the latter case, the produced DM is diluted until the onset of RD, such that smaller parent particle decay lengths are required to produce the observed relic density.

Additionally, we find that, depending on the value of $k$ as well as on the nature of the inflaton couplings to the daughter particle, the evolution of the DM yield shows different behaviors due to distinct entropy dilution effects. 
For reheating, we consider inflaton decays to either pairs fermions or pairs of bosons via a trilinear coupling. 
When $k=2$, the temperature evolution and hence the dilution effect in BR and FR is identical. However, for $k>2$, the inflaton mass is time-dependent, leading to a different temperature evolution in the two reheating scenarios, as depicted in Fig.~\ref{fig:rho}. 
In particular, the temperature in the bosonic case red shifts slower compared to that in a fermionic scenario, as a result of the stronger entropy injection, thus giving rise to a distinct evolution of the DM yield as shown in Fig.~\ref{fig:yield}.
Consequently, the properties of the parent $(c\tau, m_P)$ required to yield the correct DM relic abundance are different in the two reheating scenarios as illustrated in Fig.~\ref{fig:ctaumP}.
We find that, BR scenarios with $k>2$ predict shorter decay lengths than FR scenarios. 

When inferring constraints arising from LLP searches, also depicted in Fig.~\ref{fig:ctaumP}, the nature and temperature scale of reheating can have a significant impact on the interpretation of collider searches, if $\Trh \lesssim m_P$. 
For instance, displaced lepton searches are expected to probe parts of the viable parameter space for GeV DM in a BR scenario with $k=4$ and $\Trh = 20 \, \mathrm{GeV}$, whereas they are irrelevant for a scenario with a large reheating temperature.

The reheating temperature itself can be constrained from CMB data depending on the inflationary model (cf. App.~\ref{sec:AppA} and Fig.~\ref{fig:Trh_constrain}). 
In particular, we find lower bounds on $\Trh$ for all models investigated with $k=2$, whereas $k=4$ cannot be constrained with the method applied here. 
For a given DM mass, this allows us to rule out all parent particle decay lengths below a certain mass-specific threshold, as DM would be overabundant otherwise. 
These limits constrain parent decay lengths of detector size from energy scales accessible at colliders up to several orders of magnitude beyond, as shown in Figs.~\ref{fig:ctau_mP_inflation_scal}, \ref{fig:ctau_mP_inflation_ferm} and \ref{fig:Planck_CMBs4_main} crucially depending on the inflationary model.  
Since $\Omega_\text{DM} \sim \mdm (c\tau)^{-1}$, inflationary constraints on the parent decay length become more severe for large DM masses.
In turn, the constraints arising for the smallest DM mass allowed ($\mdm\gtrsim 12~\text{keV}$) are unavoidable. 

In conclusion, we find that the form of the reheating potential, the nature of the inflaton-matter coupling, as well as the magnitude of the reheating temperature, have a significant impact on the production of FIMP DM and thus on the interpretation of collider limits. 
In addition, inflationary constraints provide a measure of how strongly FIMP DM production can be altered during the reheating phase. 
At the same time, a measurement of long-lived DM parent could have the potential to shed light on the dynamics taking place during the phase of inflationary reheating.

%% file: AModels.tex
\section{\label{sec:Model}Possible FIMP models}
In what follows, we briefly comment on some particle physics models that could describe the minimal FIMP scenarios of interest for this work (we also refer to Refs.~\cite{Belanger:2018sti,Calibbi:2021fld}).
We highlight which interactions are responsible for DM production and which could be responsible for the dominant decay channel in the reheating phase.
At the same time, we comment on why we consider certain inflaton couplings.
\subsection{Majorana DM model}
In this scenario, we extend the SM with a Majorana singlet fermion $\chi$, which plays the role of DM; a charged scalar $X$ with the same quantum numbers of a right-handed SM fermion; and the inflaton real scalar $\Phi$.
The two dark sector particles, $X$ and $\chi$ could be assigned a $\mathbb{Z}_2$-symmetry under which they are odd, in order to ensure the stability of $\chi$, assumed to be lighter than $X$.
The SM fields and the inflaton are even under this transformation.
Thus, the Lagrangian for these fields reads as
\begin{align}
\Lc_\text{M} \supset \Lc_\chi + \Lc_\text{Yuk.} + \Lc_X +\Lc_\Phi +\Lc_{\Phi X}\,,
  \label{eq:LagrTOT_Majorana}
\end{align}
where
\begin{align}
    &\Lc_\chi =\frac{1}{2}i \bar{\chi}\slashed{\partial} \chi - \frac{1}{2}m_{\text{DM}}\, \bar{\chi} \chi\,,\label{eq:Lagr_chi}\\
    &\Lc_\text{Yuk.}= - \left(y_{\text{DM}}\, X  \bar{\chi}\, f_R + h.c.~\right) - y_{\chi} \Phi  \bar{\chi} \chi \label{eq:LagrYuk_chi}
\end{align}
describe the DM candidate $\chi$ and its portal couplings.
We notice that Eq.~\eqref{eq:LagrYuk_chi} would correspond to the trilinear interaction in Eq.~\eqref{eq:trilinear}.
Secondly, we note that DM could also be non-thermally generated via inflaton decays due to the trilinear interaction $y_\chi \Phi  \bar{\chi} \chi$. 
However, we assume this coupling to be negligible, focusing on the scenario in which DM is created only via the portal Yukawa interaction with SM fermions.
For the SM charged scalar we have
\begin{align}
   \mathcal{L}_{X}+\mc{L}_{\Phi X}&= \left(D_\mu X\right)^{\dagger}  \left(D^\mu X \right) -V(X) \nonumber\\
   &\quad-\mu_X\Phi |X|^2 -\frac{\sigma_X}{2}\Phi^2|X|^2\,,
   \label{eq:LagrX}
\end{align}
where we have included its interactions with gauge bosons via the covariant derivative $D_\mu=\partial_\mu+ig A_\mu$, $g$ being a gauge coupling, and with the inflaton. 
The inflaton can reheat the Universe via ``bosonic reheating'' (cf. Sec. III) through decays $\Phi\rightarrow XX^\dagger$ or through scatterings $\Phi\Phi\rightarrow XX^\dagger$.
However, Ref.~\cite{Dufaux:2006ee} showed that the former has to be dominant in order to allow the Universe to transition into a period of radiation domination. 
Thus, we neglect any contribution from the quartic interaction in the following.

Notice that, we may also consider interactions of these fields with the Higgs field, given by the following terms
\begin{align}
    &\Lc_{\Phi H}= -\mu_H \Phi |H|^2 - \frac{\lambda_{\Phi H}}{2}\Phi^2|H|^2\,, \label{eq:LagrPhiH}\\
    &\Lc_{X H} = - \lambda_{XH} |H|^2 |X|^2 \,. \label{eq:LagrXH}
\end{align}
The first two terms could play a role in reheating the Universe after inflation, while the last term mainly contributes to the thermalization of $X$, having a sub-leading role in the DM production mechanism.
Typically, one would also expect these couplings to be small in order not to spoil the flatness of the inflaton potential and the stability of the electroweak vacuum \cite{Gross:2015bea,Enqvist:2016mqj}.
Therefore, we will not consider these interactions and address the interplay between these requirements in future work.
\subsection{Scalar singlet DM model}
Another possible FIMP model arises from exchanging the particle nature of DM and of the dark parent particle.
Namely, DM can be described by a real scalar singlet $s$, while the parent is a vectorlike fermion $F$, with the same gauge quantum number of a right-handed SM fermion.
The Lagrangian reads as
\begin{align}
\Lc_\text{S} \supset \Lc_s + \Lc_\text{Yuk.}+ \Lc_F+ \Lc_\Phi + \Lc_{\Phi s} +\Lc_{\Phi F}\,,
  \label{eq:LagrTOT_scalar}
\end{align}
where
\begin{align}
    \Lc_F = \bar{F}\left(i\slashed{D}-m_F\right)F\
    \label{eq:LagrF}
\end{align}
represents the vectorlike fermion $F$ Lagrangian,
\begin{align}
    &\Lc_s = \partial_\mu s\, \partial^\mu s - \frac{\mu_s^2}{2}s^2+\frac{\lambda_s}{4}s^4 - \lambda_{sH}s^2|H|^2\,, \label{eq:Lagr_s}\\
    &\Lc_\text{Yuk.}= -\left( y_s\, s\, \Bar{F} f_R +h.c.~\right)\,,
    \label{eq:LagrYuk_s}
\end{align}
is the Lagrangian for the scalar singlet DM with its interaction to the Higgs field and the Yukawa interaction with the $F$ field; the Yukawa interaction would correspond to the trilinear coupling responsible for DM freeze-in (cf. Eq.~\eqref{eq:trilinear}).
Moreover,
\begin{align}
    &\Lc_{\Phi s} =-\mu_s s\Phi^2-\frac{\sigma_{s}}{2}s^2\Phi^2\,,\label{eq:Lagr_Phis}\\
    &\Lc_{\Phi F}= -y_F \Phi \Bar{F}F \label{eq:Lagr_PhiF}
\end{align}
are the direct couplings of $s$ and $F$ to the inflaton field.
We once again note that also in this model, DM could be non-thermally produced by direct decays and scatterings of the inflaton.
In this paper, this option is not considered and is left for future work.
In the same way, in order to focus on the trilinear Yukawa coupling in Eq.~\eqref{eq:LagrYuk_s}, we neglect DM production from Higgs scatterings or decays.
Finally, as previously discussed for the Majorana DM model, we do not to consider terms from $\Lc_{\Phi H}$ (cf. Eq.~\eqref{eq:LagrPhiH}).

%% file: AppendixA.tex
\section{\label{sec:AppA}Inflationary models and Reheating Constraints}
In this section, we revisit several well-motivated inflation models. We apply the slow-roll (SR) approximation and define the (potential) SR parameters  as \cite{Baumann:2009ds}
\begin{align}\label{eq:SR_parameter}
  \epsilon_V &\equiv \frac{\Mp^2}{2} \left( \frac{V^{\prime}}{V}\right)^2\,;
\quad  \eta_V  \equiv \Mp^2 \left(\frac{V^{\prime \prime}}{V}\right)\,.
\end{align}
During SR inflation, the SR parameters need to remain small, in particular, $\epsilon_V \ll 1$, $|\eta_V| \ll 1$. 
The end of inflation is defined via $\text{Max} \{\epsilon_V(\phiend) ,\eta_V(\phiend) \} \equiv1$.
We can find the energy density of the inflaton field at this moment of time by imposing the equation-of-state parameter to be $w_\Phi=-1/3$. 
This implies $\dot{\Phi}_\text{end}^2=V(\Phi_\text{end})$, which then yields $\rho_{\Phi,\text{end}}=\frac{3}{2}V(\Phi_\text{end})$.

The amplitude of the Gaussian curvature perturbation (or the so-called power spectrum) is \cite{Baumann:2009ds}
\begin{align}\label{eq:As}
A_{s,\star} = \frac{V}{24 \pi^2 \epsilon_V \Mp^4}\,.
\end{align}
The prediction for the tensor-to-scalar ratio and spectral index are given by \cite{Baumann:2009ds}
\begin{align}\label{eq:rns}
 r=16 \epsilon_V\,;\quad n_s= 1 - 6\epsilon_V + 2\eta_V\,.
\end{align}
The total number $N_\star$ of e--folds that occurred after the CMB pivot scale $k_{\star} = 0.05 \rm{Mpc}^{-1}$ first crossed the horizon till the end of inflation is
\begin{align} \label{eq:Nstar}
  N_\star = \int^{\Phi_\star}_{\phiend} \frac{1}{\sqrt{2 \epsilon_V}} \frac{d\phi}{\Mp}\,,
\end{align}
where the $\star$ refers to the quantities being evaluated at the Planck pivot scale, $k_\star=0.05\,\text{Mpc}^{-1}$.
The latest Planck 2018 measurements plus baryonic acoustic oscillations (BAO) give \cite{Planck:2018inflation}
\begin{align}\label{eq:Planck:2018inflatio}
    A_{s,\star}= (2.1\pm 0.1)\times10^{-9},\quad n_s=0.9659\pm0.0040\,,
\end{align}
whereas the most stringent upper bound on $r$ comes from the recent analysis by BICEP/Keck~(BK) \cite{BICEP:2021xfz}, 
\begin{align}
    r_{0.05}<0.035,\qquad 95\%\,\text{C.L.}\,.
\end{align}
These limits are used to constrain the parameters for the inflationary models considered in this work, namely the $\alpha$-attractor E- and T-models, that we introduce shortly in the following.
\subsection{$\alpha$-attractor E-model}
We first consider the $\alpha$-attractor inflation scenario described by a potential with an exponential dependence on $\Phi$ (E-model) given by \cite{Kallosh:2013hoa,Kallosh:2013maa}
\begin{align}
    V(\Phi)=\Lambda^4\left(1-e^{-\sqrt{\frac{2}{3\alpha}}\frac{\Phi}{\Mp}}\right)^{2n}\,.
    \label{eq:V_E-mod}
\end{align}
Here, $n$ is a positive integer, $\Lambda$ represents a normalization scale, while $\alpha>0$ is a parameter that classifies various possible realizations of the potential from supergravity.
Notice that for $n=1$ and $\alpha=1$, one recovers the well-known Starobinsky model \cite{Starobinsky:1980te}.

For small field values $\Phi<\Mp$ (e.g., during reheating), the potential is approximated by
\begin{align}
    V(\Phi)\simeq \Lambda^4\left(\frac{2}{3 \alpha}\right)^n \left(\frac{\Phi}{\Mp}\right)^{2n} \equiv \frac{\lambda}{\Mp^{k-4}} \Phi^{k}\,,
\end{align}
\begin{figure}[t!]
\centering
\includegraphics[scale=0.65]{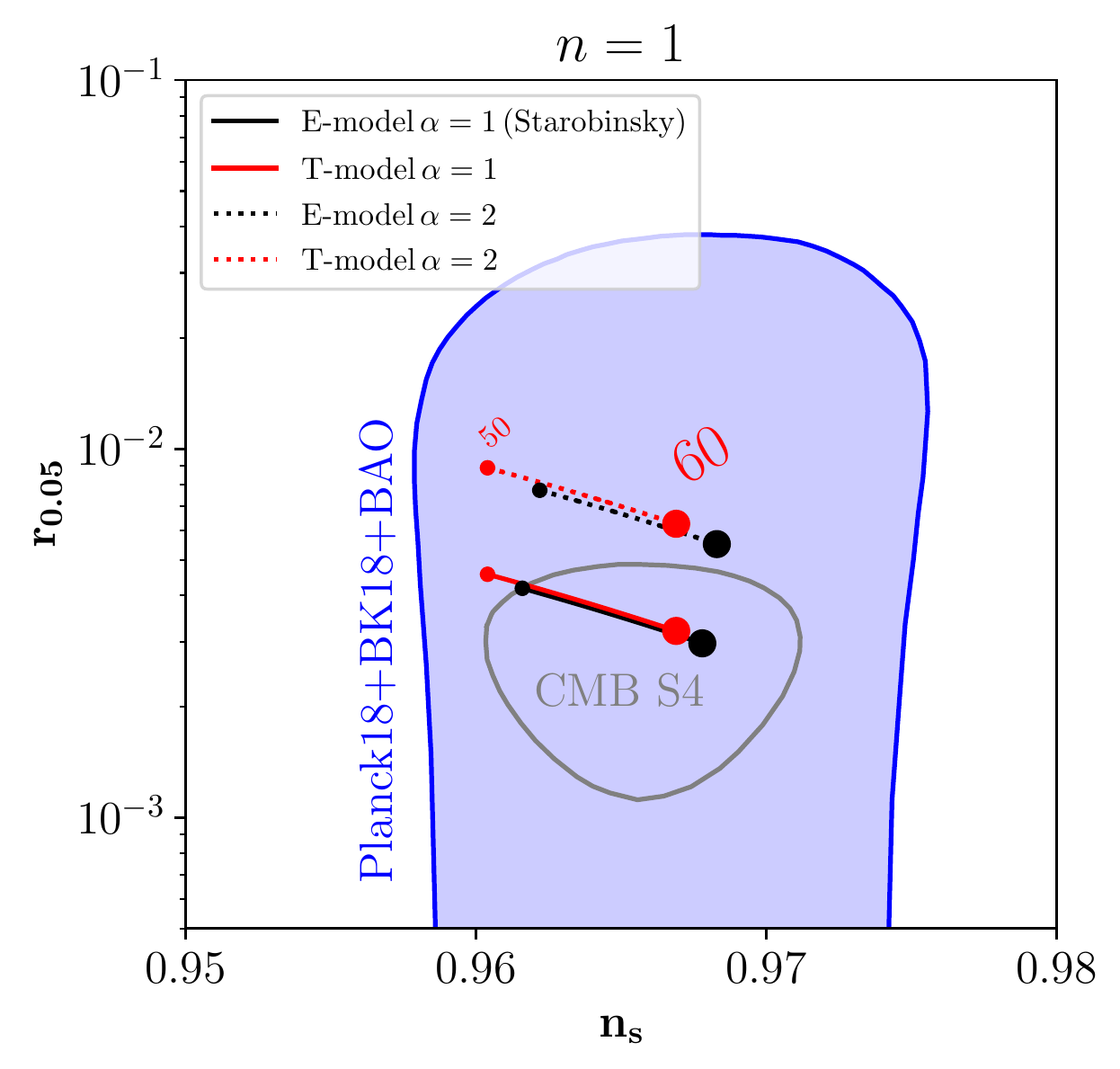}
\caption{The $r-n_s$ plane with the predictions for the $n=1$ T- (red lines) and E-models (black lines), for $\alpha=1$ (solid lines) and $\alpha=2$ (dotted lines). 
The curves lie within 50 (small circles) and 60 (big circles) number of $e$-folds.
The blue-shaded region indicates the allowed parameter space at the $95\%$ C.L. at the pivot scale $k_\star=0.05\mathrm{Mpc}^{-1}$ \cite{BICEP:2021xfz}. Future CMB-S4 projections with fiducial $r=0.003$ are indicated with a gray contour \cite{Abazajian:2019eic}.}
\label{fig:rns_plane}
\end{figure}
where we defined $k=2n$, and $\lambda\equiv (\frac{\Lambda}{\Mp})^4 (\frac{2}{3\alpha})^n$.
Using Eqs.~\eqref{eq:SR_parameter} and \eqref{eq:rns}, the tensor-to-scalar ratio and spectral index for the potential in Eq.~\eqref{eq:V_E-mod} are
\begin{align} \label{eq:r_Emod}
 r& =\frac{64 n^2}{3 \alpha\left(e^{\sqrt{\frac{2}{3 \alpha}} \frac{\Phi_k}{\Mp}}-1\right)^2}  \,,
\end{align}
\begin{align} \label{eq:ns_Emod}
 n_s&=1-\frac{8n 
 \left(e^{\sqrt{\frac{2}{3 \alpha}} \frac{\Phi_k}{\Mp}}+n\right)}{3\alpha \left( e^{\sqrt{\frac{2}{3 \alpha}} \frac{\Phi_k}{\Mp}}-1\right)}\,.
\end{align}
Based on Eq.~\eqref{eq:ns_Emod}, one can solve for $\Phi_k$ to find
\begin{align}\label{eq:phik_Emod}
\Phi_k = \Mp\sqrt{\frac{3\alpha}{2}} \ln \left[1+\frac{4n + 2\sqrt{4n^2 +6n\,\alpha(1+n)(1-n_s)}}{3\alpha(1-n_s)}\right]\,,
\end{align}
which, with  Eq.~\eqref{eq:r_Emod}, yields
\begin{align}
    r = 48\alpha \left[ \frac{ n (1-n_s)}{2n + \sqrt{4n^2 +6n(1+n)(1-n_s)} }\right]^2\,.
    \label{eq:rvsn_Emod}
\end{align}
The total number of $e$-folds $N_\star$ after inflation that occurred after the CMB pivot scale $k_\star$ has crossed the horizon for the first time follows from Eq.~\eqref{eq:Nstar} as
\begin{align}
      N_\star = \int^{\Phi_\star}_{\phiend} \frac{1}{\sqrt{2 \epsilon_V}} \frac{d\Phi}{\Mp}= \frac{3 \alpha}{4 n}\left[e^{\sqrt{\frac{2}{3 \alpha} }\frac{\Phi_\star}{\Mp}}-e^{\sqrt{\frac{2}{3 \alpha} }\frac{\Phi_{\text {end }}}{\Mp}}-\sqrt{\frac{2}{3 \alpha}} \frac{\left(\Phi_\star- \phiend \right)}{\Mp} \right]\,,
      \label{eq:Nk_Emod}
\end{align}
where $\phiend$ denotes the field value at the end of inflation and is defined via $\text{Max} \{\epsilon_V(\phiend) ,\eta_V(\phiend) \} \equiv1$, such that
    \begin{align}\label{eq:phiend_Emod}
\phiend = \Mp \sqrt{\frac{3\alpha}{2}} \ln\left(1+\frac{2n}{\sqrt{3\alpha}}\right)\,.
\end{align}
With Eq.~\eqref{eq:phik_Emod} and Eq.~\eqref{eq:phiend_Emod}, we can further rewrite the number of $e$-folds in Eq.~\eqref{eq:Nk_Emod} as
\begin{align}
 N_\star =
 \frac{3 \alpha}{4 n}
 \Bigg[&
\frac{2\left( n \left(2-\sqrt{3}(1-n_s) \sqrt{\alpha} \right) +\sqrt{4 n^2+6 n(1+n)(1-n_s) \alpha}\right)}{3(1-n_s) \alpha}  \\
 &-\ln \left(
\frac{4 n+3(1-n_s) \alpha+2 \sqrt{4 n^2+6 n(1+n)(1-n_s) \alpha}}{(1-n_s)(2 \sqrt{3} n+3 \sqrt{\alpha}) \sqrt{\alpha}}
\right)
 \Bigg].
\end{align}
We show in Fig.~\ref{fig:rns_plane} the $(r,n_s)-$plane in the $n=1$ case ($k=2$ during reheating) with some benchmark lines from Eq.~\eqref{eq:r_Emod} and Eq.~\eqref{eq:Nk_Emod} with $50\leq N_\star\leq 60$, together with the current $95\,\%$ C.L. limits from \cite{BICEP:2021xfz}.

Finally, the overall normalization of the potential and hence $\Lambda$ can be fixed by the amplitude of the scalar perturbations in Eq.~\eqref{eq:As} as follows:
\begin{align}
\Lambda
&= \Mp \left(\frac{3}{2}\pi^2 r A_{s,*}\right)^{1/4}
\left[\frac{4n + 2\sqrt{4n^2 +6n\alpha(1+n)(1-n_s)}}{4n + 2\sqrt{4n^2 +6n\alpha(1+n)(1-n_s)}+3\alpha(1-n_s)} \right]^{-n/2}\,.
\end{align}

\subsection{$\alpha$-attractor T-model}
The second model of $\alpha$-attractors we consider comprises a class of potentials with the following form \cite{Kallosh:2013hoa,Kallosh:2013maa}
\begin{align}
    V(\Phi)=\Lambda^4\left[\tanh\left(\frac{\Phi}{\sqrt{6\alpha}\Mp}\right)\right]^{2n}\,,
    \label{eq:V_T-mod}
\end{align}
where $\alpha$, $n$, and $\Lambda$ are understood as before.
For small-field values, the potential in Eq.~\eqref{eq:PhiPotential} is recovered with $k=2n$ and $\lambda=(\frac{\Lambda}{\Mp})^4 (\frac{1}{6\alpha})^n$.
Following the same arguments as just explained for the E-model, the derivation of all the relevant quantities can be straightforwardly computed, resulting in:
\begin{align}
n_s = 1 - \frac{8n}{3\alpha} \left[ n+ \text{Cosh}\left(\sqrt{\frac{2}{3\alpha}}\frac{\Phi_k}{\Mp}\right) \right] \left[\text{Csch}\left(\sqrt{\frac{2}{3\alpha}}\frac{\Phi_k}{\Mp}\right)\right]^2\,,
\label{eq:ns_Tmod}
\end{align}
\begin{align}
    \Phi_k =\Mp \sqrt{\frac{3\alpha}{2}}\times\text{arcosh}\left(\frac{4n+\sqrt{9 \alpha^2 (1-n_s)^2 + 8n^2 (2+3\alpha -3\alpha n_s)}}{3\alpha(1-n_s)}\right)\,,
\end{align}

\begin{align}
    r=\frac{24n\alpha\,(1-n_s)^2}{n\left[4 + 3\alpha (1-n_s)\right]+\sqrt{9 \alpha^2 (1-n_s)^2 + 8n^2 (2+3\alpha -3\alpha n_s)}}\,,
    \label{eq:r_Tmod}
\end{align}
\begin{align}
    \phiend= \Mp \sqrt{\frac{3\alpha}{2}} \times \text{arcosh}\left(1+\frac{4n^2}{3\alpha}\right)\,,
\end{align}
\begin{align}
    N_\star=\frac{3\alpha}{4n^2}\left[ \frac{4n+\sqrt{9\alpha^2(1-n_s)^2 + 8n^2 (2+3\alpha -3\alpha n_s)}}{3\alpha (1-n_s)}  -\sqrt{1+\frac{4n^2}{3\alpha}} \right]\,,
    \label{eq:Nk_T-mod}
\end{align}
\begin{align}
    \Lambda= \Mp\left(\frac{3}{2}\pi r A_{s,*}\right)^{1/4}\left[\frac{4n+\sqrt{9 \alpha^2 (1-n_s)^2 + 8n^2 (2+3\alpha -3\alpha n_s)}-3\alpha(1-n_s)}{4n+\sqrt{9 \alpha^2 (1-n_s)^2 + 8n^2 (2+3\alpha -3\alpha n_s)}+3\alpha(1-n_s)}\right]^{-n/4}\,.
    \label{eq:Lambda_Tmod}
\end{align}
We show in Fig.~\ref{fig:rns_plane} the $(r,n_s)-$plane in the $n=1$ case ($k=2$ during reheating) with some benchmark scenarios for the E-model (red line) using Eq.~\eqref{eq:r_Emod} and  the T-model (black line) using Eq.~\eqref{eq:r_Tmod}  with $50\leq N_\star\leq 60$, together with the current $95\,\%$ C.L. limits from \cite{BICEP:2021xfz}.

Note that the E-model with $\alpha =1$ reproduces Starobinsky inflation \cite{Starobinsky:1980te}, with its prediction $r \sim 10^{-3}$ being potentially testable with the next-generation CMB experiments, such as 
CORE \cite{COrE:2011bfs}, LiteBIRD
\cite{Matsumura:2013aja}, and
CMB-S4 \cite{Abazajian:2019eic}.

\subsection{Matching Reheating to Inflationary Parameters}\label{App:CMBReheatingLimits}
\begin{figure}[t!]
    \centering
    \includegraphics[scale=0.65]{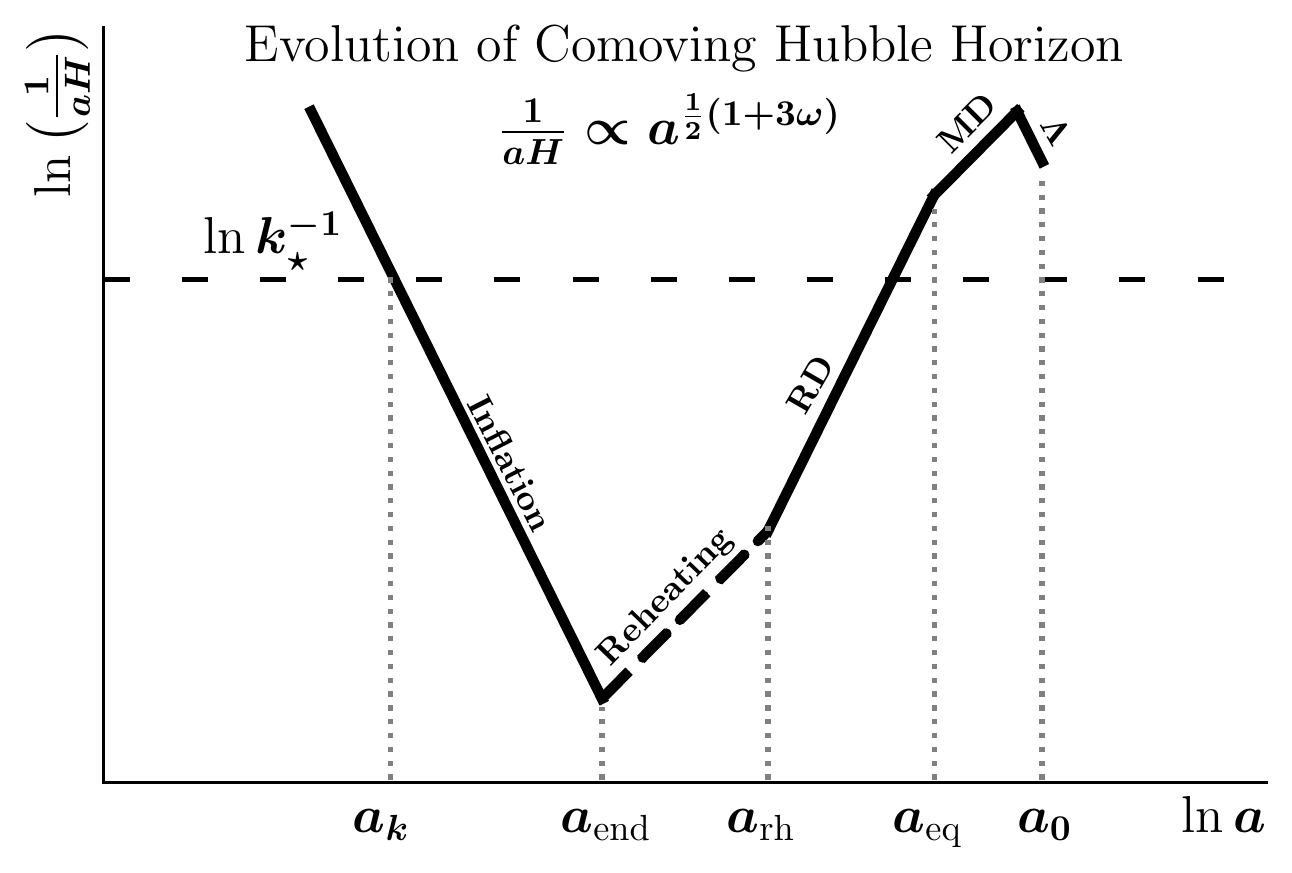}
    \caption{Evolution of the comoving Hubble horizon in different epochs from inflation to the present era. The horizon redshifts as $a^{\frac{1+3w}{2}}$, hence during reheating, RD, and MD, its evolution can differ, as the steepness of the curve reveals.}
    \label{fig:history}
\end{figure}
The period of reheating affects the moment at which the CMB pivot scale re-enters the horizon, allowing one to relate CMB predictions with the reheating parameters, e.g. the equation of state $w_{\text{rh}}\equiv w_{\Phi}(a<\arh)$, the duration of reheating, as well as the reheating temperature \cite{Dai:2014jja, Cook:2015vqa,Drewes:2023bbs}.
In Fig.~\ref{fig:history}, we show the evolution of the comoving Hubble horizon $(aH)^{-1}$ in different cosmological epochs. 
To show the correlation between inflationary predictions and reheating parameters, we consider the horizon crossing for a given mode, e.g. the pivot scale $k\equiv k_\star=a_\star H_\star$, so that we can write
\begin{align}\label{eq:aHk}
\frac{k_\star}{a_0 H_0}=\frac{a_\star}{a_{\text {end }}} \frac{a_{\text {end }}}{a_{\text {rh}}} \frac{a_{\text{rh}}}{a_{\text{eq}}} \frac{a_{\text{eq}} }{a_0} \frac{H_\star}{H_0}\,,
\end{align}
where a subscript `0' stems from ``present day'' values.
This implies
\begin{align}
\frac{a_0 } {a_{\text{eq}} } = \frac{a_\star}{a_{\text {end }}} \frac{a_{\text {end }}}{a_{\text {rh }}} \frac{a_{\text{rh}}}{a_{\text{eq}}} \frac{a_0 H_\star}{k_\star}\,.
\end{align}
Using $\rho  = \rho_{\text{end}} \left(\frac{a}{a_{\text{end}}}\right)^{-3(1+w_{\text{rh}})} $ during reheating (cf. Eq.~\eqref{eq:Rho_Phi}), one has
\begin{align}\label{eq:Nrh1}
N_{\text{rh}}\equiv \ln\left(\frac{\arh}{a_\text{end}}\right)= \frac{1}{3(1+w_{\text{rh}})} \ln \left(\frac{\rho_{\text{end}}}{\rho_{\text{rh}}}\right)\,,
\end{align}
where $\rho_{\text{rh}} =\frac{g_{*,\text{rh}} \pi^2}{30} \Trh^4$ and $\rho_{\text{end}} = \frac{3}{2}V_\text{end}$ with $V_\text{end}$ denoting the inflaton potential energy at the end of inflation.

\begin{figure*}[t!]
    \def\sepf{0.5}
\centering
\includegraphics[scale=0.55]{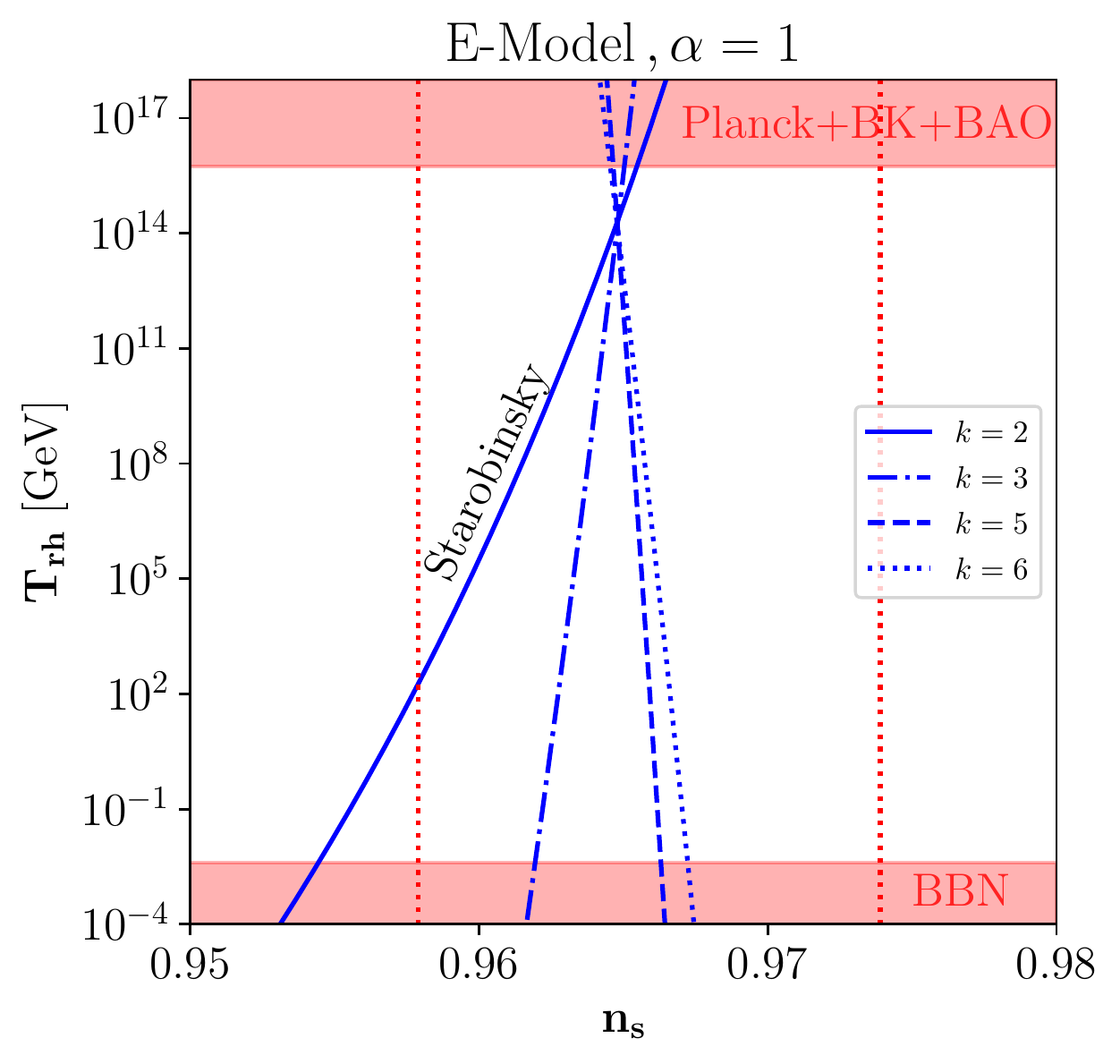}
\includegraphics[scale=0.55]{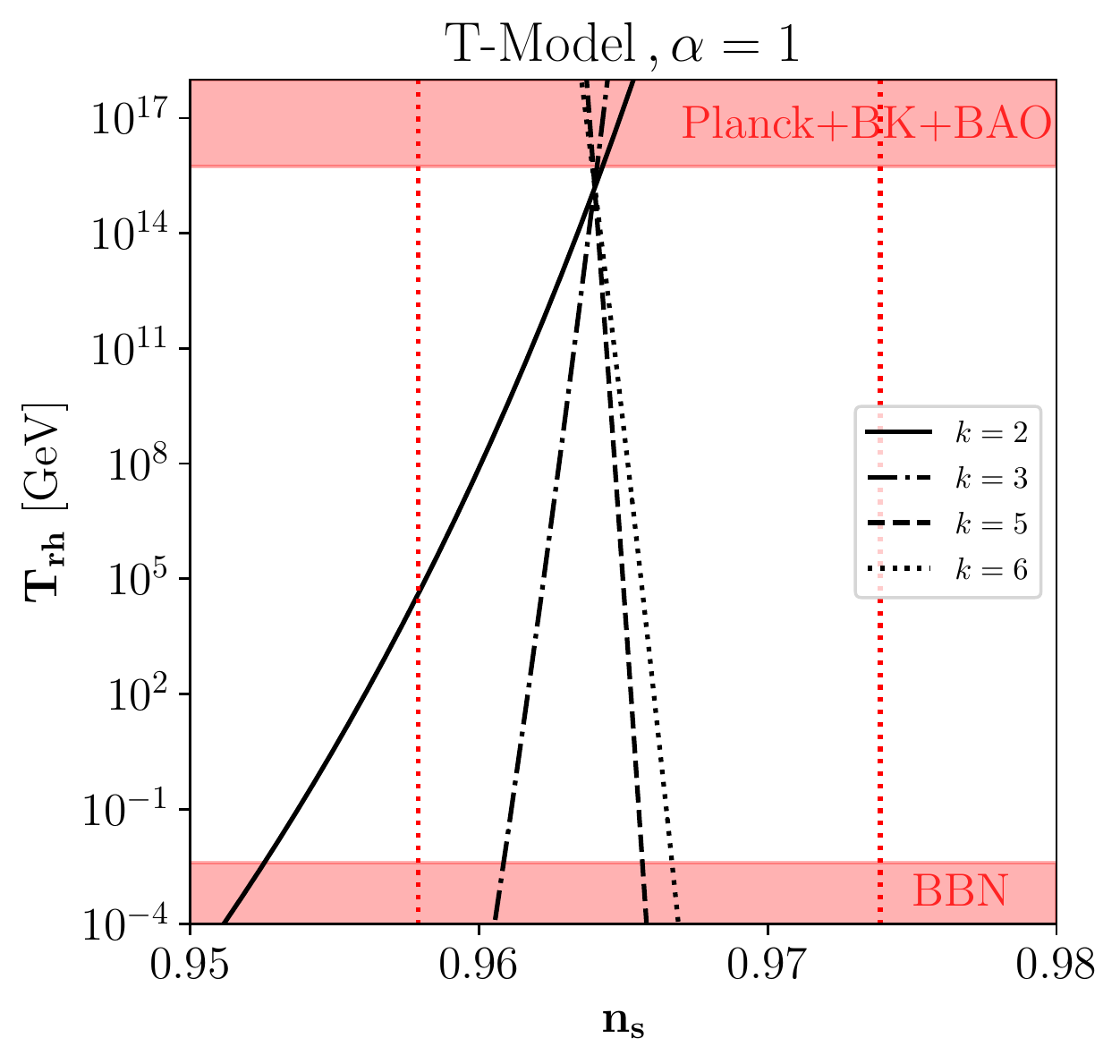}\\
\includegraphics[scale=0.55]{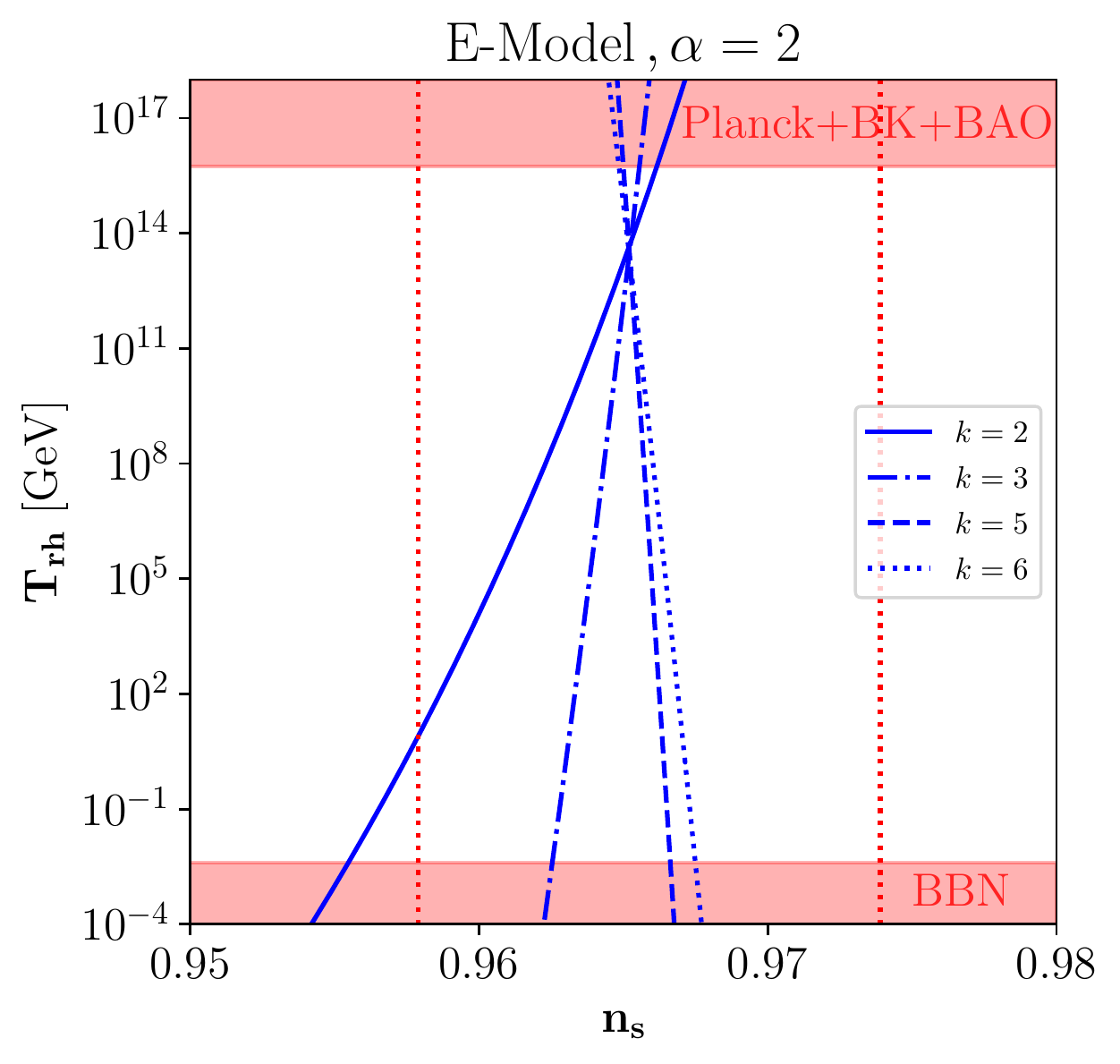}
\includegraphics[scale=0.55]{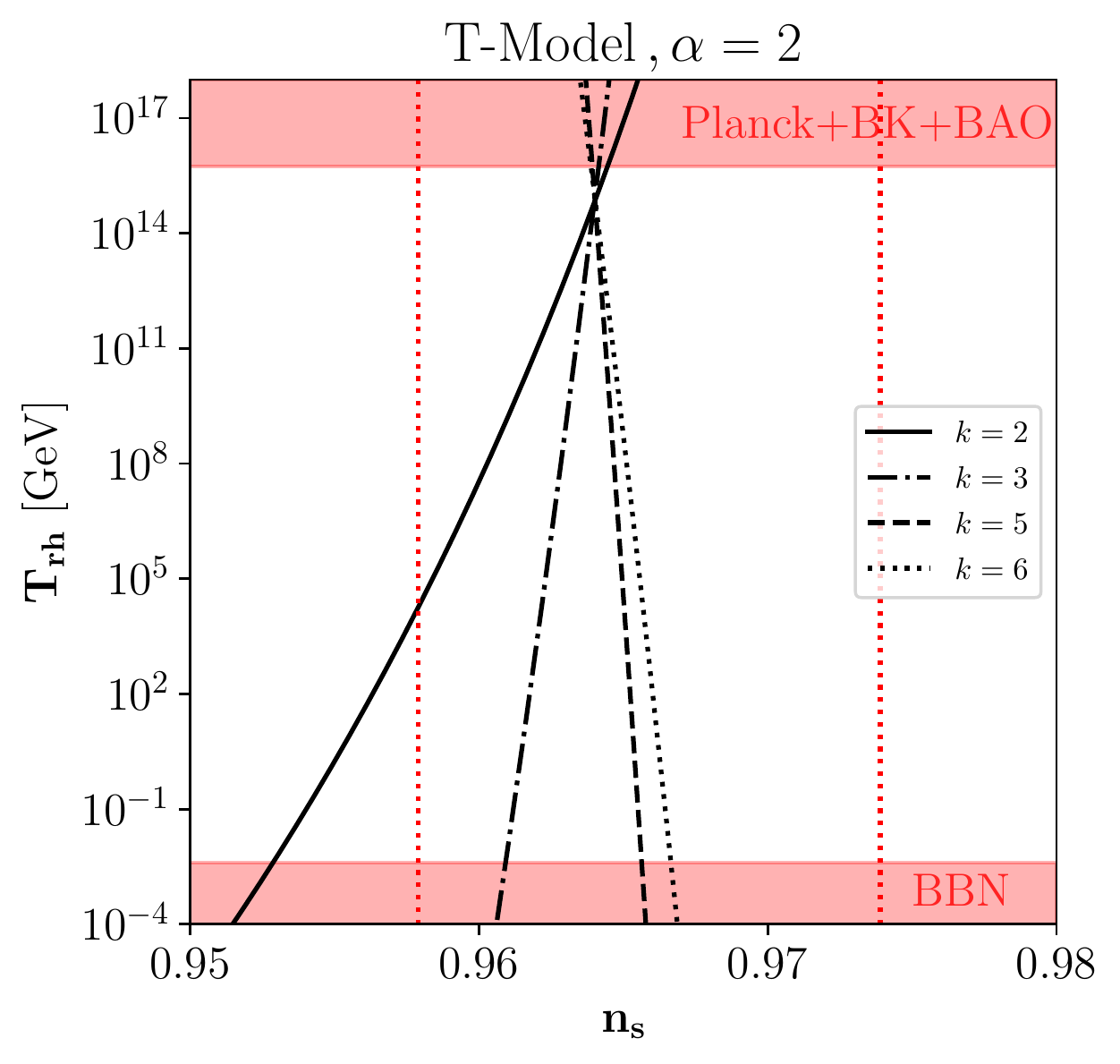}
\caption{Reheating temperature values as functions of $n_s$ for $\alpha$-attractor E- (left column, blue curves) and T-models (right column, black curves) with $\alpha=1$ (upper row) and $\alpha=2$ (lower row). We show with solid, dot-dashed, dashed, and dotted lines the predictions for reheating scenarios with $k=2$, 3, 5, and 6, respectively. Values of $\Trh$ within the $2\sigma$ constraints from Ref.~\cite{BICEP:2021xfz} are allowed. The case for $k=4$ would only give a vertical line, as explained in the text (cf. Eq.~\eqref{eq:Trh_inflation}), and hence cannot be constrained. The upper red-shaded area is excluded by limits on $r$ from Ref~\cite{BICEP:2021xfz}, while the lower one is excluded by BBN \cite{Kawasaki:2000en,ParticleDataGroup:2022pth}.}
\label{fig:Trh_constrain}
\end{figure*}

After reheating, entropy is conserved, which implies
\begin{align}
g_{*s,\text{rh}}\, T_{\text{rh}}^3=\left(\frac{a_0}{a_{\text{rh}}}\right)^3\left(2\, T_0^3+6 \cdot \frac{7}{8} T_{\nu 0}^3\right)\,,
\end{align}
where $T_{\nu 0} =\left( \frac{4}{11} \right)^{1/3} T_0$ denotes the neutrino decoupling temperature.
Thereafter, we find
\begin{align} \label{eq:Trh}
\frac{T_{\text{rh}}}{T_0} &=\left(\frac{43}{11 g_{*s,\text{rh}}}\right)^{1 / 3} \frac{a_0}{a_{\text{eq}}} \frac{a_{\text{eq}}}{a_{\text{rh}}} = \left(\frac{43}{11 g_{*s, \text{rh}}}\right)^{1 / 3}  e^{-N_{\text{rh}}} 
 e^{-N_\star} \frac{a_0H_\star}{k_\star}\,,
\end{align}
where we have used Eq.~\eqref{eq:aHk} in the second step. Using Eq.~\eqref{eq:Trh} and Eq.~\eqref{eq:Nrh1}, one finds~\cite{Cook:2015vqa}
\begin{align}\label{eq:Nrh2}
N_{\text{rh}} = &\frac{4}{3(1+w_{\text{rh}})} \Bigg[\frac{1}{4}\ln \left(\frac{45}{\pi^2 
 {g_{*,\text{rh}}}}\right) + \ln\left(\frac{V^{1/4}_{\text{end}}}{H_\star}\right)   \frac{1}{3} \ln\left(\frac{11\, g_{*s, \text{rh}}}{43}\right) + \ln \frac{k_\star}{a_0 T_0} + N_\star + N_{\text{rh}}\Bigg]\,.
\end{align}
If $w_{\text{rh}} \neq 1/3$, we have
\begin{align} \label{eq:Nrh3}
N_{\text{rh}}=\frac{4}{1-3 w_{\text{rh}}}\Bigg[&-N_\star-\ln\left( \frac{k_\star}{a_0 T_0}\right)-\frac{1}{4} \ln\left(\frac{45}{g_{*\text{rh}} \pi^2} \right) -\frac{1}{3} \ln \left( \frac{11 g_{*s, \text{rh}}}{43} \right)+\ln\left(\frac{V^{1/4}_{\text{end}}}{H_\star}\right)\Bigg]\,,
\end{align}
where  $H_\star =\pi M_P \sqrt{\frac{r A_{s,\star}}{2} }\,.$ Finally, by using Eq.~\eqref{eq:Nrh2} and Eq.~\eqref{eq:Trh}, we obtain~\cite{Cook:2015vqa}
\begin{align}\label{eq:Trh_inflation}
\Trh=\left[\left(\frac{43}{11 g_{*s,\text{rh}}}\right)^{\frac{1}{3}} \frac{a_0 T_0}{k_\star} H_\star e^{-N_\star}\left(\frac{45 V_{\text{end}} }{\pi^2 g_{*\text{rh}}}\right)^{-\frac{1}{3\left(1+w_{\text{rh}}\right)}}\right]^{\frac{3\left(1+w_{\text{rh}}\right)}{3 w_{\text{rh}}-1}}\,,
\end{align}
where $H_\star$, $N_\star$ and $V_{\text{end}}$ are inflation model dependent. 
We would like to comment on the special case when $w_{\text{rh}} =1/3$. 
In such case, one cannot solve Eq.~\eqref{eq:Nrh2} and hence cannot make a prediction for $N_{\text{rh}}$, since this quantity cancels on each side of the equation. 
Note that one defines the RD phase when $\rho_\text{R}\geq\rho_\Phi$, i.e., when the EoS of the Universe approaches $1/3$; however, if it is already $1/3$ during reheating, then one cannot 
differentiate between the two regimes, and hence the value of $N_{\text{rh}}$ is left undefined \cite{Cook:2015vqa}.

The reheating constraints for the $\alpha$ attractor E-models are shown in Fig.~\ref{fig:Trh_constrain}, where the two vertical red dotted lines correspond to the $2 \sigma$ range of $n_s$ (cf. Eq.~\eqref{eq:Planck:2018inflatio}). For the E-model with $\alpha=1$ and $k=2n=2$ (or Starobinsky model), the minimal allowed reheating temperature is $\Trh^{\text{min}} \simeq 180.7~\text{GeV}$. 
With larger $\alpha$, the lower bound on $\Trh$ decreases, e.g. $\Trh^{\text{min}} \simeq 7.8~\text{GeV}$ with $\alpha =2$. 
For T-models, the minimal allowed reheating temperature is   $\Trh^{\text{min}} \simeq 4.0\times 10^{4}~\text{GeV}$ for $\alpha =1$ and $\Trh^{\text{min}} \simeq 1.8 \times 10^{4}~\text{GeV}$ $\alpha =2$. 
For larger values of $k$, the equation of state $w_{\text{rh}}$ increases, implying a change of slope on $\Trh$ with respect to $n_s$.

%% file: AppendixC.tex
\section{\label{sec:AppC}DM yield and parent lifetime}
In this appendix, we set in context the results of this work and the findings of Refs.~\cite{Belanger:2018sti,Calibbi:2021fld}.
We start by solving the Boltzmann equation in Eq.~\eqref{eq:BEQ_chi} for the comoving yield of DM $\Xdm(a)=\ndm(a)a^3$ as follows:
\begin{align}
    \Xc_\text{DM}(a)\simeq n_\text{DM}(\aend)+\int_{\ln\aend}^{\ln\arh}\dd\ln a'\,\left(\frac{a'}{\aend}\right)^3\frac{\mc{C}_\text{rh}}{H_\text{rh}} + \int_{\ln\arh}^{\ln a}\dd\ln a'\,\left(\frac{a'}{\aend}\right)^3\frac{\mc{C}_\text{RD}}{H_\text{RD}}\,,
    \label{eq:Chi(a)}
\end{align}
where $n_\text{DM}(\aend)\simeq 0$ from the freeze-in assumptions and where $\mc{C}_\text{rh}$, $H_\text{rh}$, $\mc{C}_\text{RD}$, and $H_\text{RD}$ are the collision operator in Eq.~\eqref{eq:CollOp_DM} and the Hubble rate in Eq.~\eqref{eq:HubblePhiRad} during inflaton-dominated reheating and during RD, respectively.
In Eq.~\eqref{eq:Chi(a)}, DM production has been broken down into two parts: the contribution from the reheating phase (first integral) and the contribution from the RD phase (second integral). 
This separation helps us to understand that if the majority of DM production occurs \emph{after} the reheating phase, then the first integral becomes zero, and all the DM production is accounted for in the second integral.

This is precisely the approach that was discussed in Equation (3.2) of Ref.~\cite{Belanger:2018sti}, even in scenarios where the mass of the parent particle, $m_P$, exceeds the reheating temperature $\Trh$. 
However, in this situation, it is crucial to note that this method has two limitations. Firstly, the yield would be significantly suppressed by the Boltzmann tail of the parent particle distribution function at temperatures lower than $m_P$. Secondly, by adopting this approach, one would neglect the bulk of the production time, which predominantly occurs in the first integral that is not considered.

\begin{figure}[!t]
\centering
\includegraphics[scale=0.65]{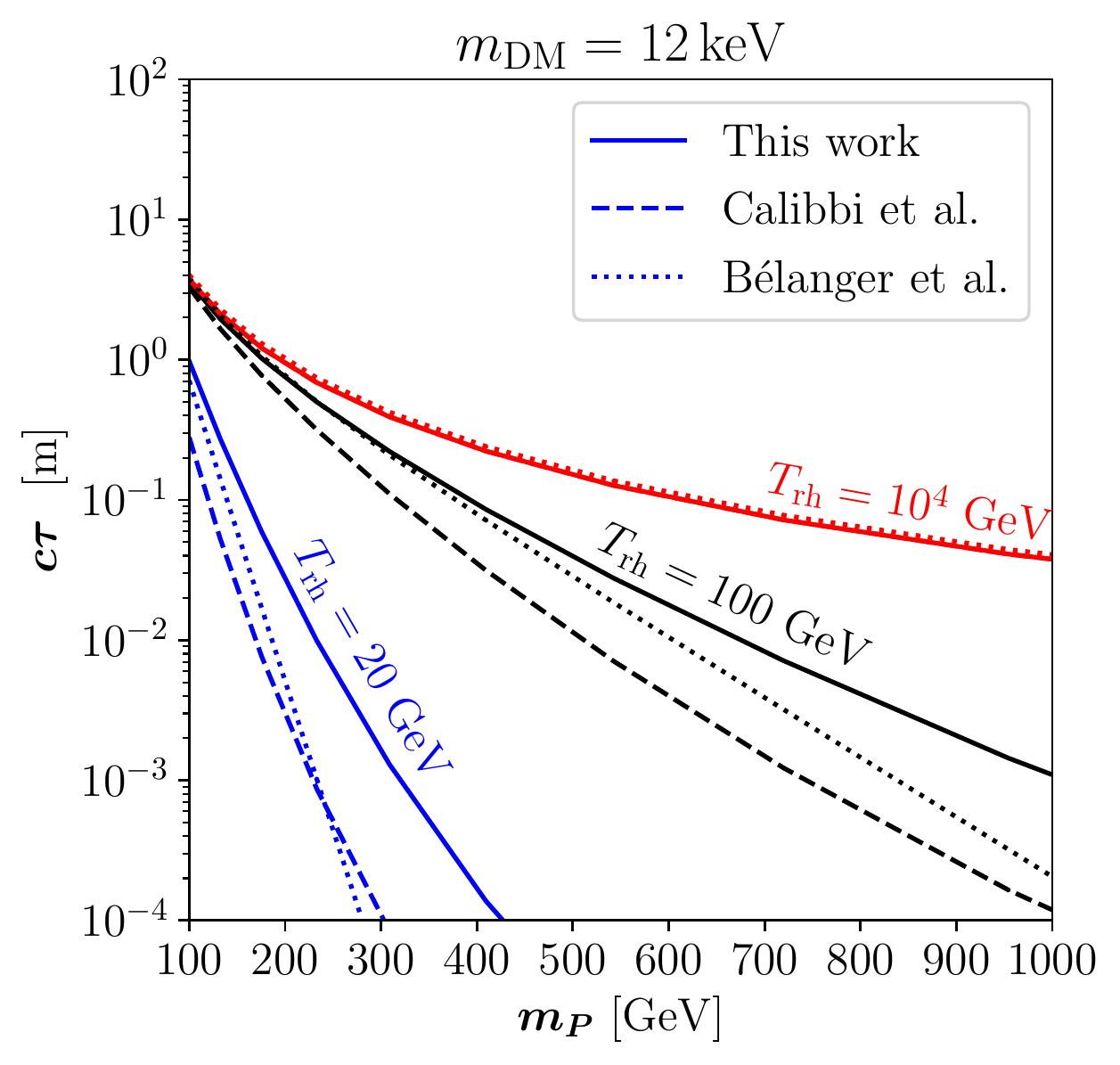}
\caption{Values of the parent's lifetime vs. its mass accounting for the observed DM relic density for $\mdm=12\,\keV$. We compare the approach utilized in this work in the $k=2$ scenario (solid line) and the methods employed in Bélanger et al. \cite{Belanger:2018sti} (dotted lines) and Calibbi et al. \cite{Calibbi:2021fld} (dashed line). 
The predictions coincide when DM production is well within RD (red lines, $\Trh=10^4\,\GeV$), while they differ in LRT scenarios depicted by the lines in black ($\Trh=100\,\GeV$) and blue ($\Trh=20\,\GeV$), as explained in the text.
}
    \label{fig:comparison}
\end{figure}

In this sense, for fixed values of $\Trh$ and $m_P$, the yield computed in \cite{Belanger:2018sti} would \emph{always} be smaller or equal to the yield calculated using Eq.~\eqref{eq:Chi(a)}.
Consequently, for fixed $\mdm$, the required $\Gamma_P$ values to obtain the observed $\Omega_\text{DM}h^2$ have to be always \emph{larger} in \cite{Belanger:2018sti} than in this study, irrespective of the considered reheating epoch. 
As a result, we can conclude that the parent particle's lifetimes in \cite{Belanger:2018sti} would be smaller, leading to lower yields. 
This fully elucidates the behavior of the curves in Fig.~\ref{fig:comparison}.

In order to quantify this correction, we solve Eq.~\eqref{eq:Chi(a)} and we evaluate it at the present time $a_0$, in such a way that the relic abundance from Eq.~\eqref{eq:OmegaDM_def} reads as
\begin{align}
    \omdm\simeq 0.12\,g_P\left(\frac{1.5\,\text{m}}{c\tau}\right)\left(\frac{106.75}{\gs}\right)^{3/2}\left(\frac{\mdm}{100\,\keV}\right)\left(\frac{200\,\GeV}{m_P}\right)^2\times
    \begin{cases}
        &\dfrac{2k+4}{3k-3}\left(\dfrac{\Trh}{m_P}\right)^{\frac{9-k}{k-1}}\mc{I}_\text{rh,f}+\mc{I}_\text{RD}^0\qquad\text{in FR}\\[10pt]
        &\dfrac{2k+4}{3}\left(\dfrac{\Trh}{m_P}\right)^{4k-1}\mc{I}_\text{rh,b}+\mc{I}_\text{RD}^0\qquad\text{in BR}\
    \end{cases}\,,
    \label{eq:OmegaDM_sol}
\end{align}
or, accordingly, the lifetime of the parent particle reads as
\begin{align}
    c\tau[\text{m}] \simeq 1.5\,g_P\left(\frac{0.12}{\omdm}\right)\left(\frac{106.75}{\gs}\right)^{3/2}\left(\frac{\mdm}{100\,\keV}\right)\left(\frac{200\,\GeV}{m_P}\right)^2\times
    \begin{cases}
        &\dfrac{2k+4}{3k-3}\left(\dfrac{\Trh}{m_P}\right)^{\frac{9-k}{k-1}}\mc{I}_\text{rh,f}+\mc{I}_\text{RD}^0\qquad\text{in FR}\\[10pt]
        &\dfrac{2k+4}{3}\left(\dfrac{\Trh}{m_P}\right)^{4k-1}\mc{I}_\text{rh,b}+\mc{I}_\text{RD}^0\qquad\text{in BR}
    \end{cases}
    \label{eq:ctau_ours}\,,
\end{align}
where
\begin{align}
    \mc{I}_\text{rh,f}=\int_{z_\text{end}}^{z_\text{rh}}\dd z' z'^{\frac{2k+6}{k-1}}K_1(z')\,,\quad \mc{I}_\text{rh,b}=\int_{z_\text{end}}^{z_\text{rh}}\dd z' z'^{2+4k}K_1(z')\,,\quad\mc{I}_\text{RD}^0=\int_{z_\text{rh}}^{z_0}\dd z' z'^3 K_1(z')\,, \label{eq:DMProductionRate}
\end{align}
with $z=m_P/T$.
For comparison, the lifetime calculated in \cite{Belanger:2018sti} is
\begin{align}
c\tau[\text{m}] \simeq 4.5\,\xi\,g_P\,\left(\frac{\omdm}{0.12}\right)\left(\frac{\mdm}{100\,\keV}\right)\left(\frac{200\,\GeV}{m_P}\right)^2\left(\frac{102}{\gs}\right)^{3/2}\left(\frac{\mc{I}_\text{RD}}{3\pi/2}\right)\,,
    \label{eq:ctau_Belanger}
\end{align}
where the parameter $\xi=1 (2)$ if the parent particles are self-conjugate (not self-conjugate).
From Eq.~\eqref{eq:ctau_Belanger} we can clearly see that the considered integrated yield is missing the $\mc{I}_\text{rh}$ part.

This issue has been previously discussed in \cite{Calibbi:2021fld}, where the authors argued that, in an early matter-dominated epoch resulting from the decaying inflaton (i.e., our $k=2$ case), the lifetime of the parent particle is always smaller than that found in \cite{Belanger:2018sti}, due to the suppression from the non-adiabatic evolution. At first glance, this may seem to contradict the argument we have just presented.

However, there are two subtle caveats that need to be considered. First, in \cite{Calibbi:2021fld}, the initial inflaton energy density is fixed at $(10^{16}\,\GeV)^4$, whereas we use $\rhoend=\frac{3}{2}V(\phiend)$, derived from the slow-roll conditions (cf. App.~\ref{sec:AppA}). 
Although this choice does not lead to significant differences, it implies different $\Tmax$ values, which affect the duration of the reheating phase and are model-dependent for fixed $\Trh$. Nevertheless, as long as $\Tfi\ll \Tmax$, these effects are negligible.

The second and more important difference is in the definition of the reheating temperature. In \cite{Calibbi:2021fld}, $\Trh$ is defined by the time $t_\text{decay}\simeq t_\text{universe}$ or equivalently, $\Gamma_\Phi= H$, with $H$ determined solely by $\rho_\text{R}(\Trh)$. This means that the decay width is fixed by the following equation
\begin{align}
\Gamma_\Phi=\sqrt{\dfrac{\pi^2\gs}{90}}\dfrac{\Trh^2}{\Mp}\,.
\label{eq:GammaCal}
\end{align}

Instead, we define the reheating temperature by requiring the equality $\rho_\Phi=\rho_\text{R}$ (cf. Eq.~\eqref{eq:T_RH_Implicit}), utilizing the model-dependent definitions in Eq.~\eqref{eq:GammaBoson} and Eq.~\eqref{eq:GammaFermion} for the decay width.
The exact moment when the equality happens is determined numerically, while an analytical approximation can be found in Eq.~\eqref{eq:TRH_boson} and Eq.~\eqref{eq:TRH_fermion}.
The two approaches have different consequences. 
Requiring $\Gamma_\Phi(\Trh)=H(\Trh)$ implies that the period immediately after $\Trh$ is still dominated by the inflaton, and not by radiation. Thus, the evolution is non-adiabatic and the effects of faster expansion and entropy dilution persist during this period. In contrast, imposing $\rho_\Phi(\Trh)=\rho_\text{R}(\Trh)$ leads to the Universe being radiation-dominated \emph{immediately} after $\Trh$, with the effects of $\rho_\Phi$ and non-adiabaticity becoming quickly less important.
Moreover, the definition of the reheating temperature must provide the properties relevant for the deduction of the reheating temperature from CMB observations as discussed in Appendix \ref{App:CMBReheatingLimits}, namely that the equation of state changes at $\Trh$.

This implies that the approach of \cite{Calibbi:2021fld} predicts larger reheating temperatures $\Trh^\star$ and larger dilution factors than what we have presented in this work.
We can verify this by computing the ratio between $\Trh^\star$ and our analytical approximations in Eq.~\eqref{eq:TRH_boson} and Eq.~\eqref{eq:TRH_fermion} for the $k=2$ scenario (equivalent to the MD considered in \cite{Calibbi:2021fld}). 
This amounts to $\Trh^\star~\simeq~1.58~\Trh$.

We illustrate the differences between the two approaches by showing the parameter space predictions in Fig.~\ref{fig:comparison}, where we fix $\mdm=12\,\keV$ and take three benchmark choices for $\Trh=20$, $100$, and $10^4$ GeV, which lead to the observed relic abundance.
For DM produced well within RD (red lines), all definitions agree.
However, in low reheating temperature scenarios (black and blue lines), the assumptions in \cite{Calibbi:2021fld} necessarily lead to larger DM couplings and hence to smaller values of the parent decay length than with the method discussed in the present work.

%% file: AppendixB.tex
\section{\label{sec:AppB}Pre-thermalization effects}
In this appendix, we discuss in which scenario pre-thermalization effects could be of relevance for the cosmological evolution and give an estimate why we can safely neglect them. 
Hereby, we follow Ref.~\cite{Garcia:2018wtq}, which itself bases its discussion on a series of papers \cite{Harigaya:2014waa,Harigaya:2013vwa,Kurkela:2011ti,Mukaida:2015ria}. 

During our calculations presented in the main text, we assume that the parent particle is in thermal equilibrium with the SM particles with a temperature $T$ at all times via gauge interactions, i.e. its distribution function is given by $f_P \left( p, T \right) = f_P^\text{eq} \left( p, T \right)$.
The temperature is calculated from the energy density that is transferred from inflaton decays, described by the Eqs.~\eqref{eq:BoltzFriedPhi}-\eqref{eq:1stFriedmann}. 
However, the SM sector, including the parent particle, is not thermalized instantly. 
If thermalization is mediated by a gauge interaction of strength $\alpha_g$, the timescale of thermalization $t_\text{th}$ can be estimated and translated to a temperature scale of thermalization \cite{Garcia:2018wtq}
\begin{align}
     T_\text{th} = \alpha_g^\frac{4}{5} m_\Phi \left( \frac{24}{\pi^2 g_{\star\,\text{rh} }} \right)^\frac{1}{4} \left( \frac{\Gamma_\Phi M_{Pl}^2}{m_\Phi^3} \right)^\frac{2}{5} \,.
\end{align}
For times $t<t_\text{th}$, the parent particle does not necessarily follow its equilibrium distribution such that 
\begin{align}
    f_P \left( p, T \right) =\theta\left( t_\text{th} - t\right) f_P^\text{non-eq} \left( p, T \right)
    +\,\theta\left( t - t_\text{th} \right) f_P^\text{eq} \left( p, T \right) \, .
\end{align}
We expect the most drastic pre-thermalization effects on the DM abundance when the inflaton does not directly decay to the parent particle and one can (over)estimate these effects by setting $f_P^\text{non-eq} \left( p, T \right) = 0$, since we expect $\int d^3 p f_P^\text{non-eq} \left( p, T \right) < \int d^3 p f_P^\text{eq} \left( p, T \right)$ at all times, as the inflaton decay products are highly relativistic. 
Thus, for our estimate we effectively neglect any DM production from parent decays during this phase. 

Inspecting the temperature evolution depicted in Fig.~\ref{fig:rho}, reveals that the temperature initially increases from $0$ to $T_\text{max}$ and then redshifts depending on the reheating potential. 
This implies that we find $T= T_\text{th}$ before and after $T=T_\text{max}$, since $T_\text{th}<T_\text{max}$. We identify $T_\text{th}$ with the time \emph{after} we have reached $T_\text{max}$. 
In this way, we find an estimate for the largest pre-thermalization effects on the relic abundance of DM.  

We use an analytic approximation for the contribution of DM to $\omdm$ during the reheating phase, which is given in Eq.~\eqref{eq:OmegaDM_sol}.
This solution assumes the relation between the temperature and the scale factor given in Eq.~\eqref{eq:Temp_BR} or Eq.~\eqref{eq:Temp_FR}, which hold for $z \gtrsim z_\text{max}$.
Thus, when deriving an analytic result, we replace $z_\text{end}$ by $z_\text{max}$, which is only an accurate substitution if $z_\text{max} \ll 1$, a scenario realized for all parameters analyzed in this work.
As we neglect any contribution to the production of DM before the time of thermalization, we simply need to modify the integral $\mc{I}_\text{rh}$ (cf. Eq.~\eqref{eq:DMProductionRate})
\begin{align}
    \int_{z_\text{max}}^{z_\text{rh}}\dd z' z'^{10}K_1(z') \rightarrow \int_{z_\text{th}}^{z_\text{rh}}\dd z' z'^{10}K_1(z') \, .
\end{align}
In the equations above, we have assumed $k=2$. 
Then, the estimate of the largest pre-thermalization effect on our results, which were derived assuming a thermalized parent, is given by
\begin{align}
    \frac{ \int_{z_\text{max}}^{z_\text{th}}\dd z' z'^{10}K_1(z')}{ \int_{z_\text{max}}^{z_\text{rh}}\dd z' z'^{10}K_1(z')} \, ,
\end{align}
and can be evaluated numerically for a given set of model parameters. 
To illustrate why these effects are typically negligible in our setup, let us assume $z_{rh} \gg 1$, meaning DM production proceeds during reheating. Notice that we do not expect any pre-thermalization effects when DM is produced during RD, as all the SM-charged fields, including the parent particle, are expected to have already reached thermal equilibrium.
Furthermore, as our parameters fulfill $z_\text{max} \ll 1$ and $z_\text{th} < 1$, we set $z_\text{max} = 0$ and expand in $z_\text{th} \ll 1$ and find 
\begin{align}
    \frac{ \int_{z_\text{max}}^{z_\text{th}}\dd z' z'^{10}K_1(z')}{ \int_{z_\text{max}}^{z_\text{rh}}\dd z' z'^{10}K_1(z')} \approx 7 \cdot 10^{-8} z_\text{th}^{10} \, .
\end{align}
Thus, we expect the reduction of the DM relic density due to pre-thermalization effects  to be negligible if $z_\text{th} \ll 1$.
Using the analytic estimate for the reheating temperature, given in Eq.~\eqref{eq:Temp_FR}, for $k=2$, we find 
\begin{align}
    z_{\text{th}}\equiv \frac{m_P}{T_\text{th}} &=  3 \times 10^{-3} \left(\frac{\gs (\Trh)}{106.75}\right)^{\frac{1}{20}} \left(\frac{m_P}{1~\text{TeV}}\right)^\frac{1}{5} \left(\frac{m_P}{\Trh}\right)^{\frac{4}{5}} \left(\frac{\lambda}{10^{-12}}\right)^{\frac{1}{10}} \left(\frac{\alpha_g}{10^{-2}}\right)^{-\frac{4}{5}} \, ,
\end{align}
where $\lambda$ is the dimensionless coefficient of the leading order term in the  reheating potential given in Eq.~\eqref{eq:PhiPotential}. As a consequence, we expect pre-thermalization effects are small as long as the ratio $\frac{m_P}{\Trh}$ is not too large. For instance for $\alpha_g = 10^{-2}$, $\lambda=10^{-12}$ and $\gs (\Trh)= 106.75$, we find $\frac{m_p}{\Trh} \lesssim 10^{3} \left(\frac{1~\text{TeV}}{m_P}\right)^{1/5}$.
Note that all of the estimates in this section were derived for $k=2$ but we expect qualitatively similar results to hold for larger $k$ and a first step in this direction was taken recently in \cite{Chowdhury:2023jft}.

In summary, we expect pre-thermalization effects to be negligible for almost all of the parameter space presented analyzed and presented within this paper. 
However, for DM production taking place deep in the reheating phase $m_P \gtrsim 10^3 \Trh$, pre-thermalization could reduce the amount of DM produced and thus the model would require a smaller decay length to produce the observed relic density. 
A more dedicated analysis of these effects and their interplay with constraints from colliders and the CMB is left for future work.